\documentclass[12pt,journal,compsoc]{IEEEtran}

\ifCLASSOPTIONcompsoc
  \usepackage{cite}
\else
  \usepackage{cite}
\fi

\ifCLASSINFOpdf
  \usepackage[pdftex]{graphicx}
  \graphicspath{{./Figures/}}
  \DeclareGraphicsExtensions{.pdf,.png}
\else
  \usepackage[dvips]{graphicx}
  \graphicspath{{./Figures/}}
  \DeclareGraphicsExtensions{.pdf,.png}
\fi

\usepackage[cmex10]{amsmath}
\usepackage{amssymb}

\ifCLASSOPTIONcompsoc
 \usepackage[caption=false,font=normalsize,labelfont=sf,textfont=sf]{subfig}
\else
 \usepackage[caption=false,font=footnotesize]{subfig}
\fi

\usepackage{url}

\usepackage[symbol]{footmisc}

\newcommand{\Z}{\mathbb{Z}}
\newcommand\given[1][]{\:#1\vert\:}

\begin{document}

\title{Artifacts of Quantization in Distance Transforms
\thanks{Funding for this research was provided by Alberta Innovates Health Solutions (BAB) and the Natural Sciences and Engineering Research Council (BAB, CGS-D; TDK, CGS-M).}}

\author{
    \IEEEauthorblockN{
      Bryce~A~Besler\IEEEauthorrefmark{1}, 
      Tannis~D~Kemp\IEEEauthorrefmark{1}, 
      Nils~D~Forkert\IEEEauthorrefmark{2}\IEEEauthorrefmark{3}, 
      Steven~K~Boyd\IEEEauthorrefmark{1}\IEEEauthorrefmark{3},
    } \\
    \IEEEauthorblockA{
      \IEEEauthorrefmark{1}McCaig Institute for Bone and Joint Health
    } \\
    \IEEEauthorblockA{
      \IEEEauthorrefmark{2}Hotchkiss Brain Institute
    } \\
    \IEEEauthorblockA{
      \IEEEauthorrefmark{3}Department of Radiology
    }\\
    University of Calgary, Canada
}

\IEEEtitleabstractindextext{%
\begin{abstract}
Distance transforms are a central tool in shape analysis, morphometry, and curve evolution problems.
This work describes and investigates an artifact present in distance maps computed from sampled signals.
Namely, sampling reflects through the distance transform causing quantization in the resulting distance map.
Gradients of the quantized distance map show banding, affecting the quality of subsequence processing.
Furthermore, this error is independent of the sampling period of the signal and cannot be removed by modifying the number of samples across an objects width.
Where needed, distance maps should be computed from representations other than binary images.
In the case where exact representations are needed, a dithering and noise removal algorithm is proposed.
\end{abstract}

}

\maketitle

\IEEEdisplaynontitleabstractindextext

%
\IEEEpeerreviewmaketitle

\section{Introduction}
The distance transform is a fundamental tool of image processing, used extensively in shape analysis~\cite{blum1967transformation,kimmel1995skeletonization,siddiqi2002hamilton,niblack1992generating}, morphometry~\cite{hildebrand1997new}, and curve evolution~\cite{osher1988fronts, caselles1993geometric, malladi1995shape}.
The distance signal is a single, non-parametric representation of shape and volume, able to represent objects of any dimensions, of arbitrary topology, and of arbitrary definitions of distance.
These features make signed distance transforms appealing as a digital representation of spatial objects.

Many methods exist to digitally embed an object.
For binary images, the algorithm is simply called the distance transform~\cite{rosenfeld1966sequential,rosenfeld1968distance,danielsson1980euclidean,borgefors1986distance}.
Representation-specific methods are available for meshes~\cite{baerentzen2005signed} and parametric curves~\cite{pottmann2003geometry}.
This article is concerned with the fidelity of the distance transform of binary signals.

This study demonstrates that the distance transform of a sampled signal is equivalent to a quantized distance transform.
This leads to errors and artifacts in the gradients of the distance transform signal.
An algorithm is presented for removing the quantization and artifacts.

\section{Preliminaries}
\subsection{Distance Transform}
\label{sec:distance_transforms}
Define a metric space $(X, g)$ where $X$ is a set and $g$ a metric on the set.
Consider a subset $A \subset X$.
The distance transform assigns every $x \in X$ a positive real number corresponding to the minimum distance between $x$ and $A$.
\begin{equation}
  d(x, A) = \inf_{a \in A} g(x, a)
\end{equation}

\begin{figure}[t]
  \centering
  \begin{tabular}{cc}
    \subfloat[$I(x)$]{
      \includegraphics[width=0.4\linewidth]{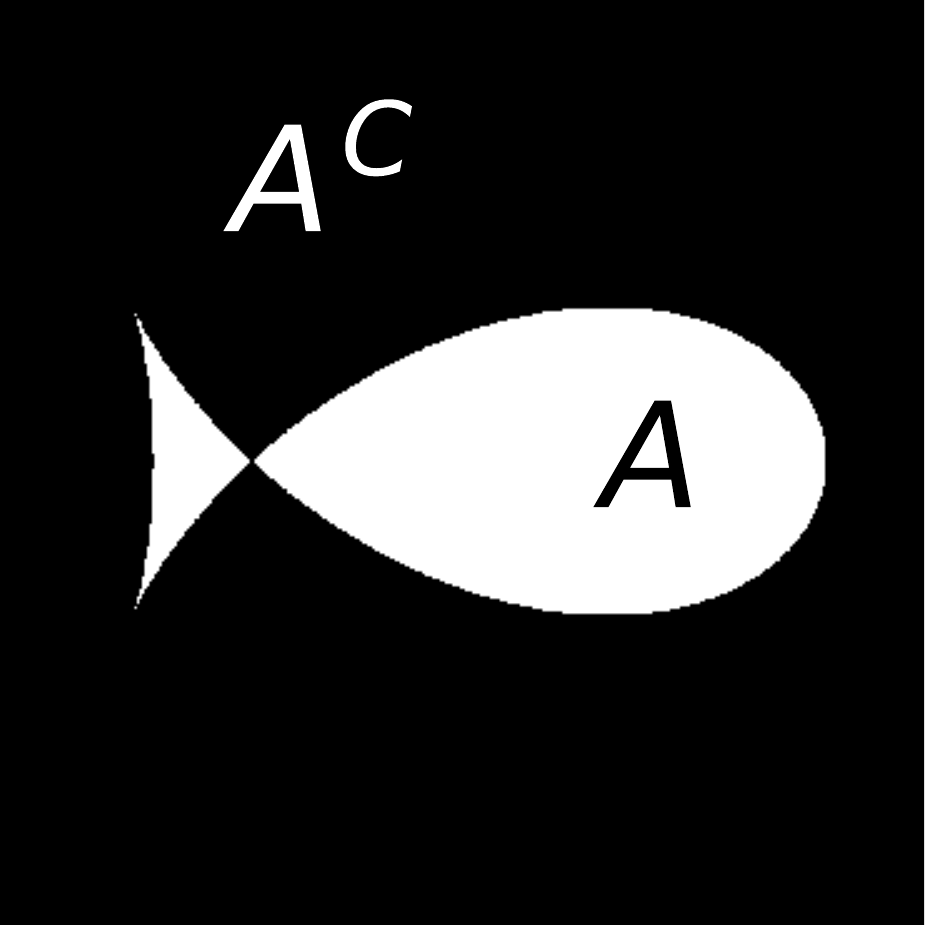}%
      \label{fig:fish:fish}
    } &
    \subfloat[{$d(x, A)$}]{
      \includegraphics[width=0.4\linewidth]{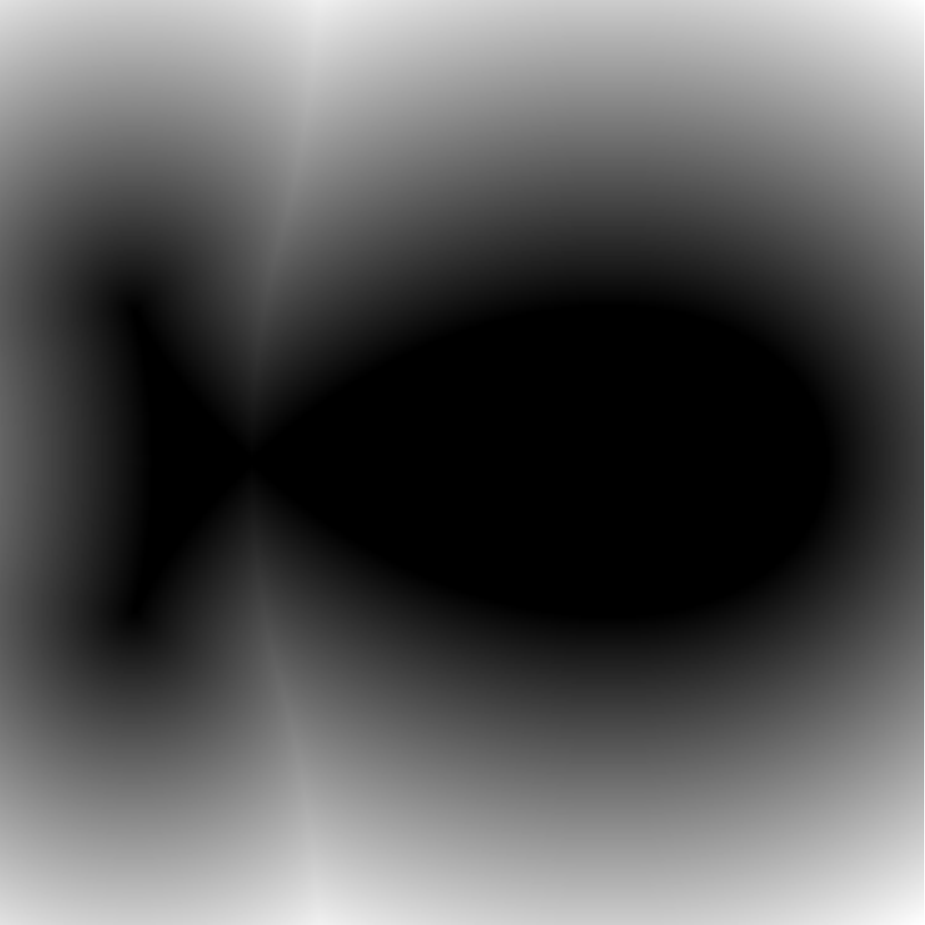}%
      \label{fig:fish:backward}
    } \\
    \subfloat[$d(x, A^C)$]{
      \includegraphics[width=0.4\linewidth]{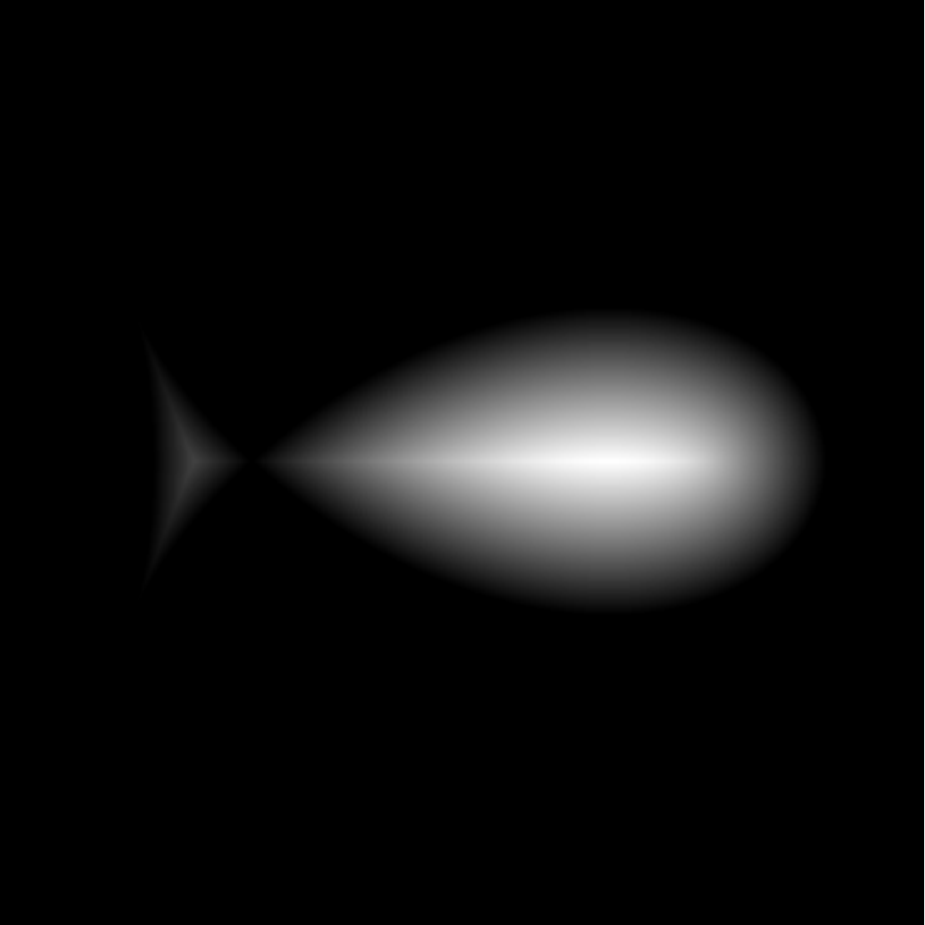}%
      \label{fig:fish:forward}
    } &
    \subfloat[{$\phi(x)$}]{
      \includegraphics[width=0.4\linewidth]{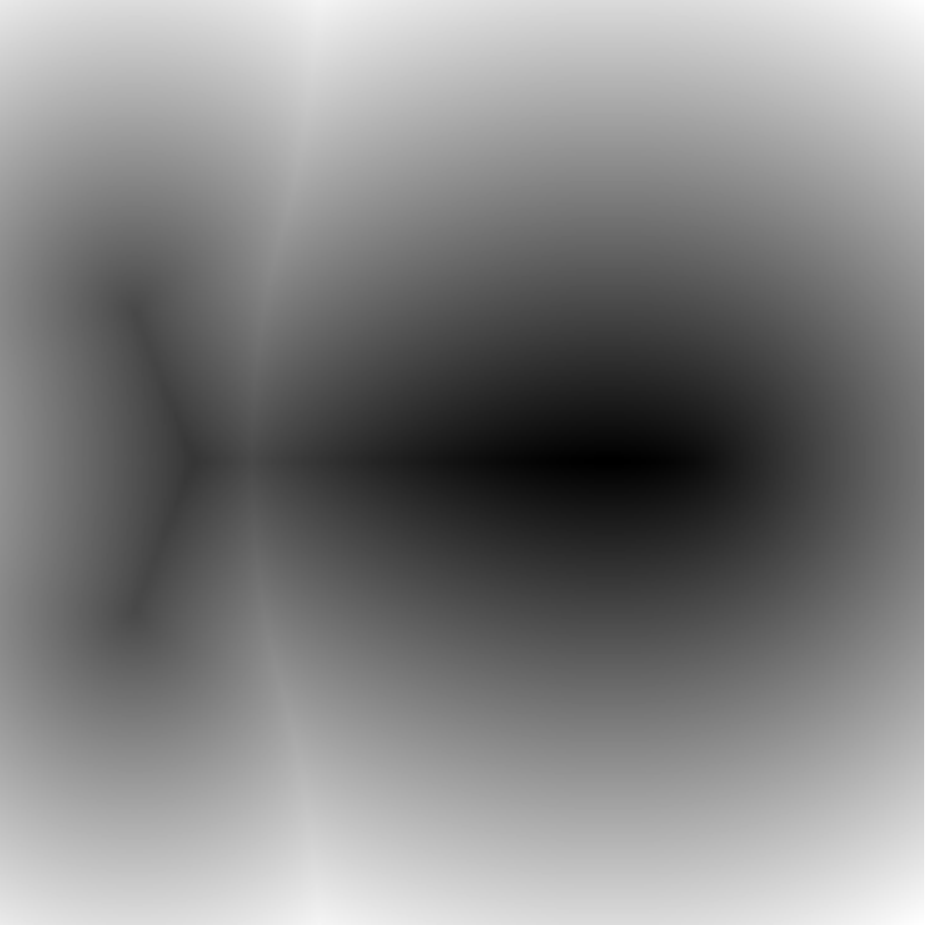}%
      \label{fig:fish:signed}
    }
  \end{tabular}
  \caption{For the fish curve (\ref{fig:fish:fish}), demonstration of the background distance transform (\ref{fig:fish:backward}), foreground distance transform (\ref{fig:fish:forward}), and signed distance transform (\ref{fig:fish:signed}).}
  \label{fig:fish}
\end{figure}

In many cases, an extension of the distance transform is used called the signed distance transform.
The boundary of the set $\partial A$ is defined as the set of points belonging to the closure of $A$ but not the interior.
The signed distance transform assigns every $x \in X$ the distance to the boundary with the added condition that the sign indicates if the point is inside or outside the set.
\begin{equation}
  \label{eqn:sdt}
  \phi(x) = \left\{
  \begin{matrix}
    + d(x, A) & \text{if } x \in A^C \\
    - d(x, A^C) & \text{if } x \in A
  \end{matrix}
  \right.
\end{equation}
This work uses the convention that the inside of the curve is negative.
The distance transform and signed distance transform of binary images are demonstrated in Figure~\ref{fig:fish} for the fish curve~\cite{lockwood1967book}.

\subsection{Sampling and Quantization}
\begin{figure}[h]
  \centering
  \begin{tabular}{cc}
    \subfloat[$y(x)$]{
      \includegraphics[width=0.44\linewidth]{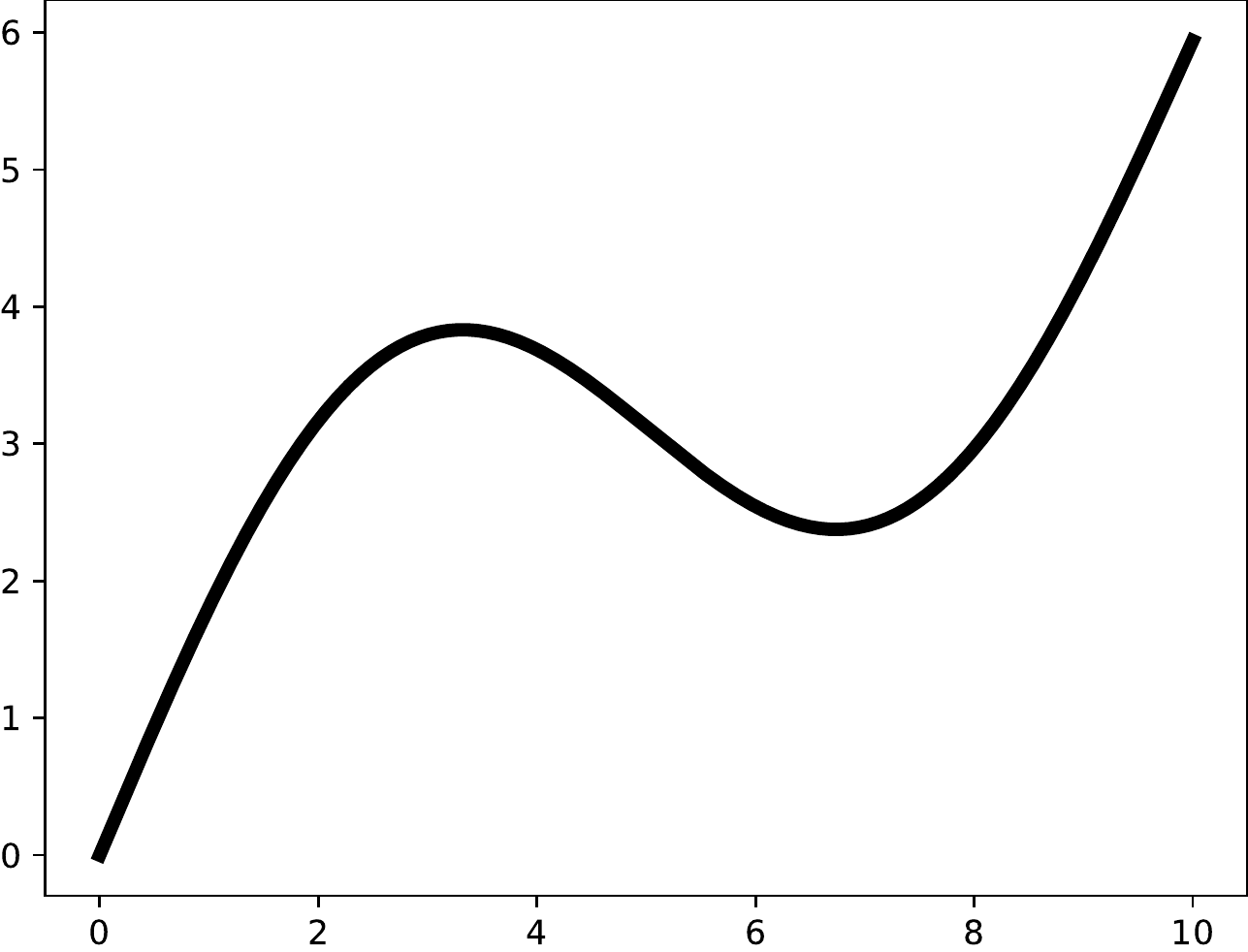}%
      \label{fig:sampling-quant:real}
    } &
    \subfloat[{$y[nh]$}]{
      \includegraphics[width=0.44\linewidth]{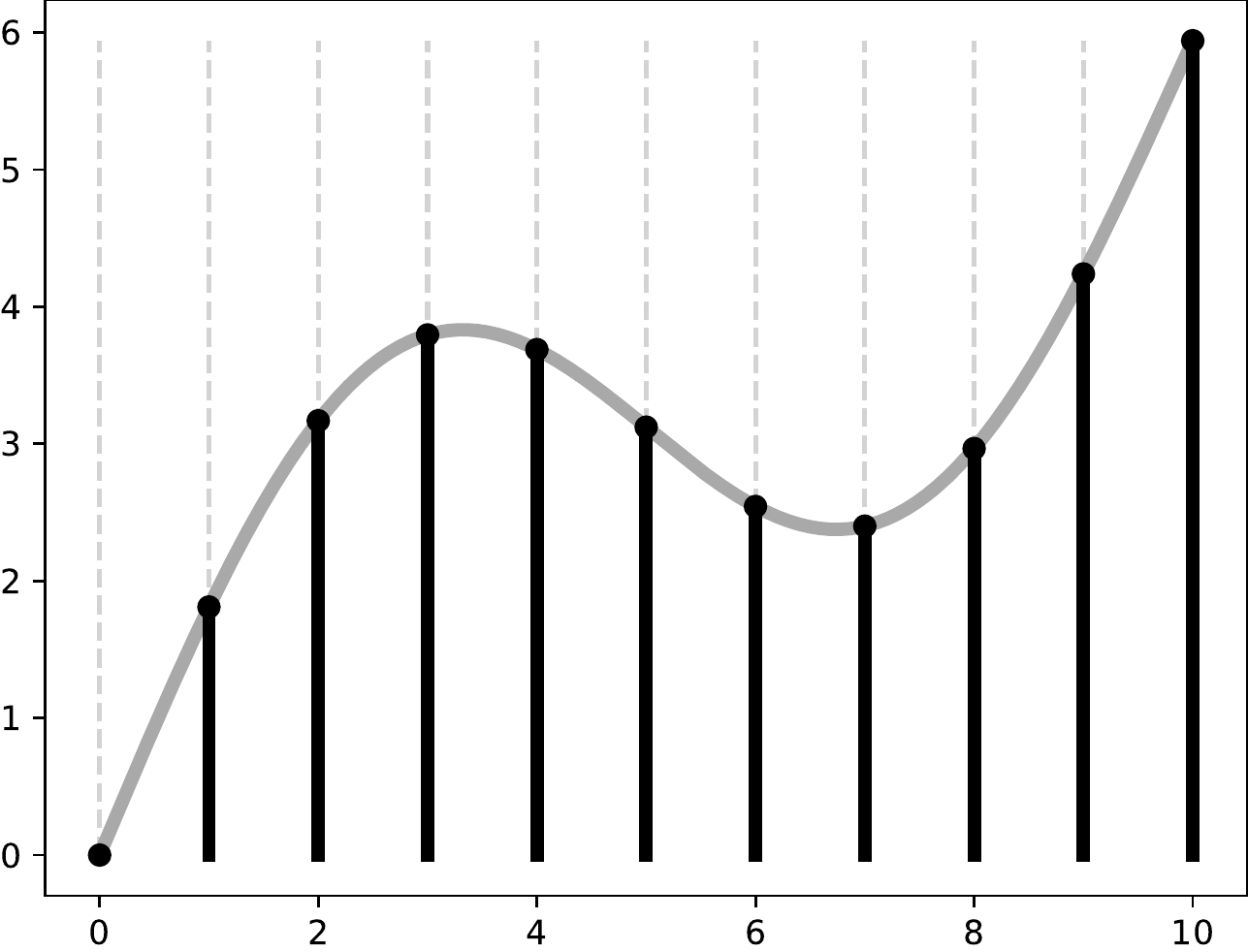}%
      \label{fig:sampling-quant:sampled}
    } \\
    \subfloat[$\tilde{y}(x)$]{
      \includegraphics[width=0.44\linewidth]{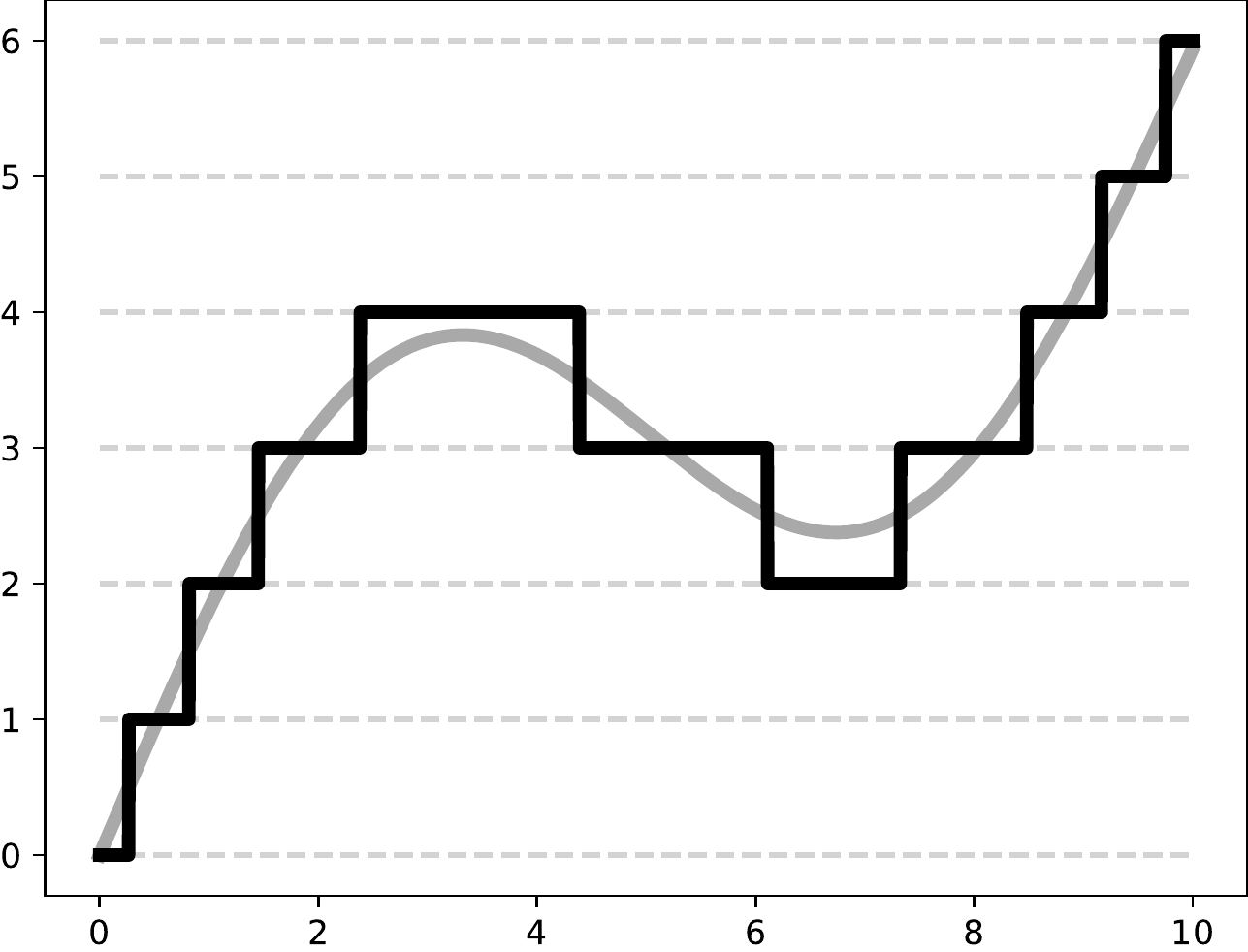}%
      \label{fig:sampling-quant:quantized}
    } &
    \subfloat[{$\tilde{y}[nh]$}]{
      \includegraphics[width=0.44\linewidth]{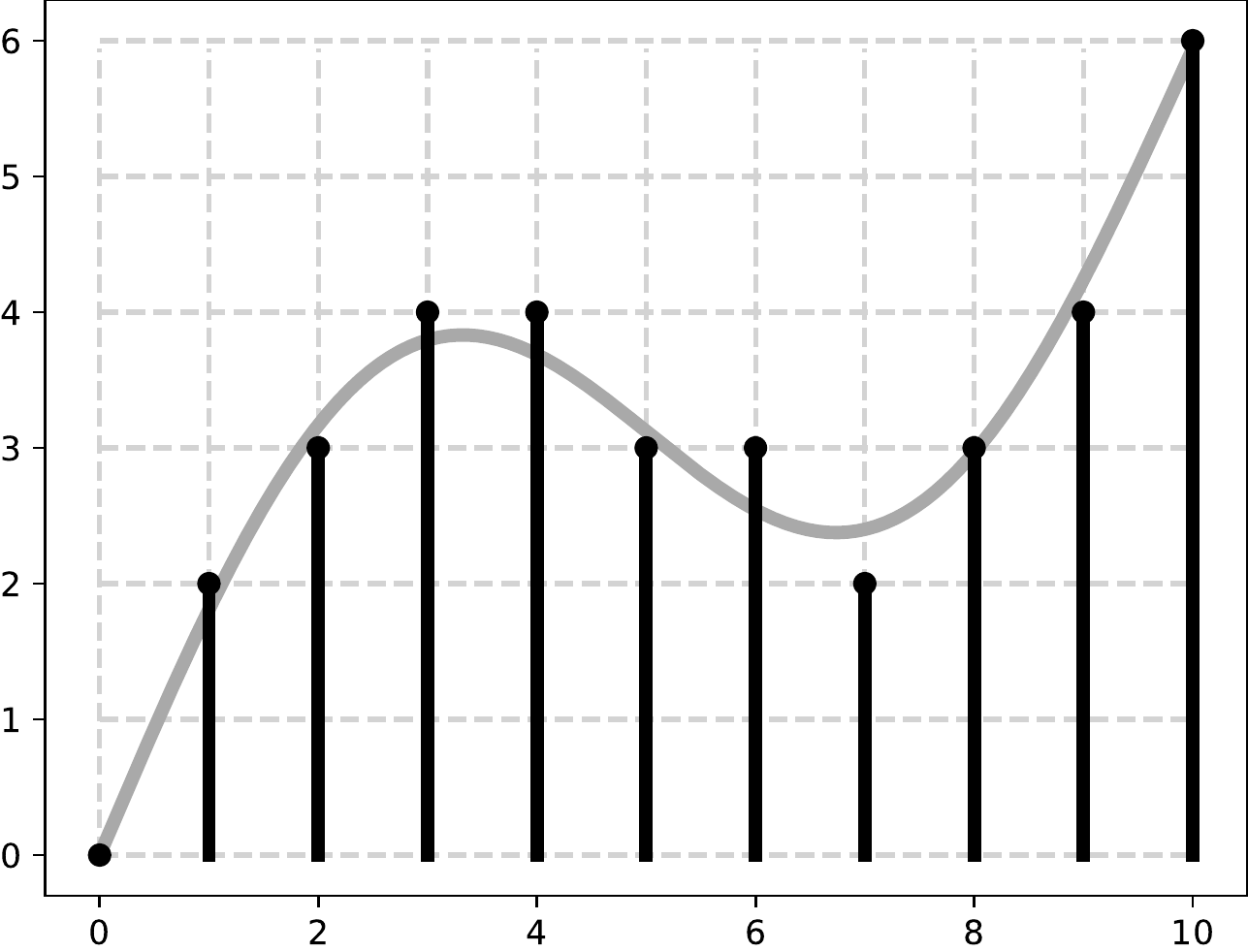}%
      \label{fig:sampling-quant:sampled-and-quantized}
    }
  \end{tabular}
  \caption{(\ref{fig:sampling-quant:real}) Continuous signal, (\ref{fig:sampling-quant:sampled}) sampled ($h = 1.0$), (\ref{fig:sampling-quant:quantized}) quantized, (\ref{fig:sampling-quant:sampled-and-quantized}) and both sampled and quantized .}
  \label{fig:sampling-quant}
\end{figure}

From signal processing theory, digitization refers to the discretization of both the domain (termed sampling) and intensity (termed quantization) of a continuous signal.
The distinction is demonstrated in Figure~\ref{fig:sampling-quant}.
In this work, circular brackets are used to represent continuous signals and square brackets are used to represent sampled signals:
\begin{equation}
  y(x) = y[nh]
\end{equation}
where $h$ is the sampling period and $n$ the sample number.
Similarly, $\tilde{y}(x)$ is used to denote a quantized signal.
A signal $\tilde{y}[nh]$ can be both sampled and quantized.
It will be seen that the distance transform of a sampled signal leads to quantization errors in the distance signal that produces structured artifacts in subsequent processing of the distance transform.

\subsection{Finite Differences}
In many applications, gradients need to be computed in the image.
These require finite approximations to the infinitesimal calculus.
Here, we use the following notation for forward, backward, and central difference:
\begin{eqnarray}
  D^{+x}\phi &=& \frac{\phi(x + h) - \phi(x)}{h} \\
  D^{-x}\phi &=& \frac{\phi(x) - \phi(x - h)}{h} \\
  D^{0x}\phi  &=& \frac{\phi(x + h) - \phi(x - h)}{2h}
\end{eqnarray}
One important property of signed distance transforms is that their magnitude gradient is $+1$ everywhere except at singularities (the medial axis).
Choosing a numerical approximation to the gradients, $D\phi$, the magnitude gradient can be computed at each point in an n-dimensional image:
\begin{equation}
  \lvert\nabla\phi\rvert \approx |D\phi| = \sqrt{\sum_i^n \left(D^i \phi\right)^2}
\end{equation}
Similar approximations can be made for Laplacian, Hessian, and curvature by substituting finite difference approximations for infinitesimal differences.
These derivatives are used in specific image processing tasks such as active contours~\cite{caselles1993geometric, malladi1995shape} and skeletonization~\cite{kimmel1996sub,siddiqi2002hamilton}.

\section{Quantization in the \\ Distance Transform}
It is demonstrated that the distance transform of a sampled signal produces a quantized distance signal.
Pedagogical examples are given in one and two dimensions.
Finally, the problem is synthesized based on discretizing the distance metric.

\label{section:quantization}
\subsection{1D Example}
For motivation, we consider the signed distance transform of a one-dimensional rectangular function.
The one-dimensional sphere defined by its radius $r$ and center $x_0$ is given by:
\begin{equation}
  \phi_{1D}(x) = \lvert x - x_0\rvert - r
\end{equation}
where $\lvert \cdot \rvert$ is the absolute value function.
The binary signal can be reconstructed using the Heaviside function, $H$
\begin{equation}
  I(x) = H(-\phi(x))
\end{equation}
and embedded numerically to produce the estimation to the signed distance transform, $\tilde{\phi}_{1D}$.
A tilde is used as it will be seen that $\tilde{\phi}_{1D}$ is a quantized representation of $\phi_{1D}$.

\begin{figure}[h]
  \centering
  \begin{tabular}{cc}
    \subfloat[$\phi(x)$]{
      \includegraphics[width=0.44\linewidth]{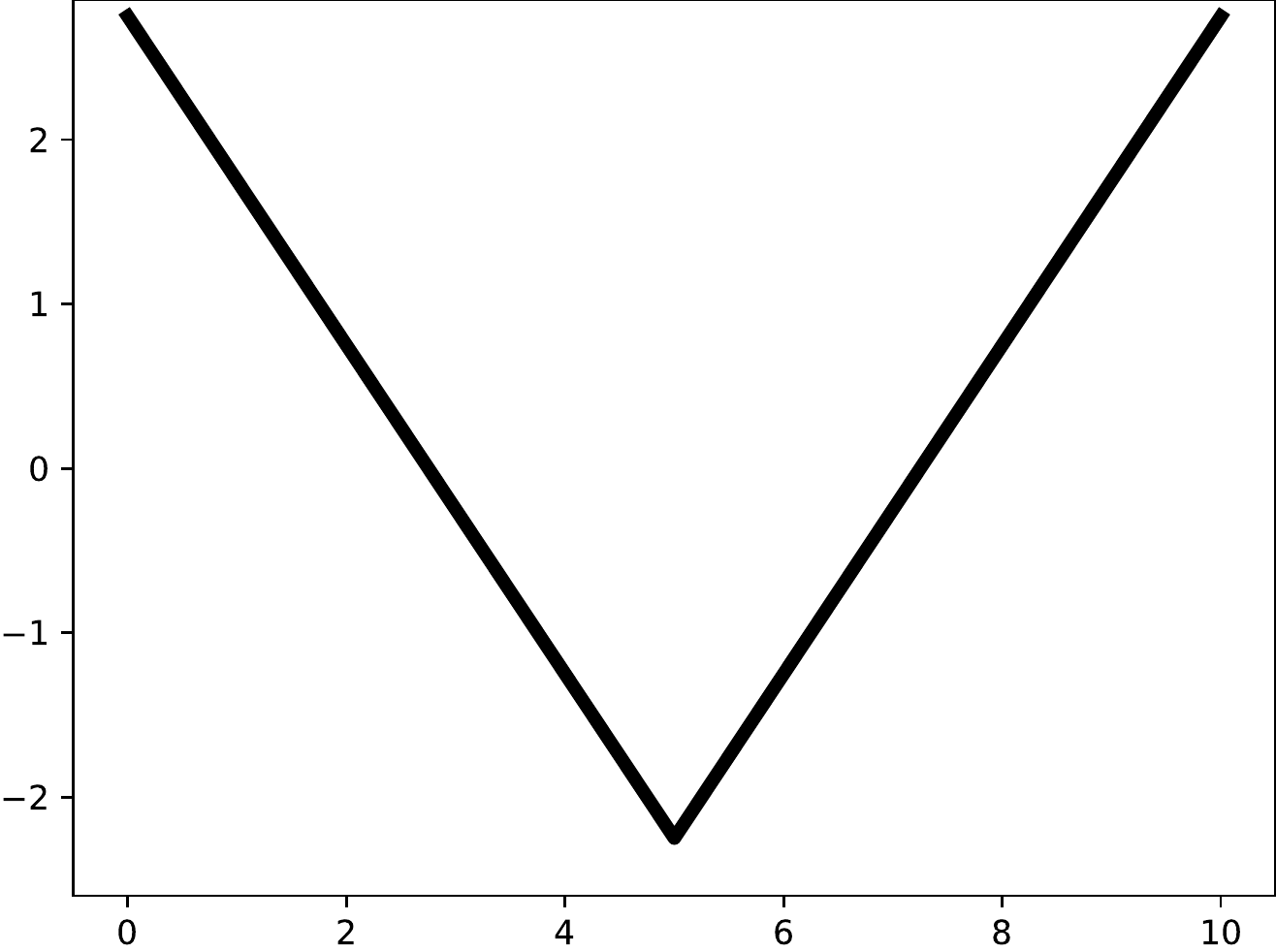}%
      \label{fig:example1d:real}
    } &
    \subfloat[{$I(x) = H(-\phi(x))$}]{
      \includegraphics[width=0.44\linewidth]{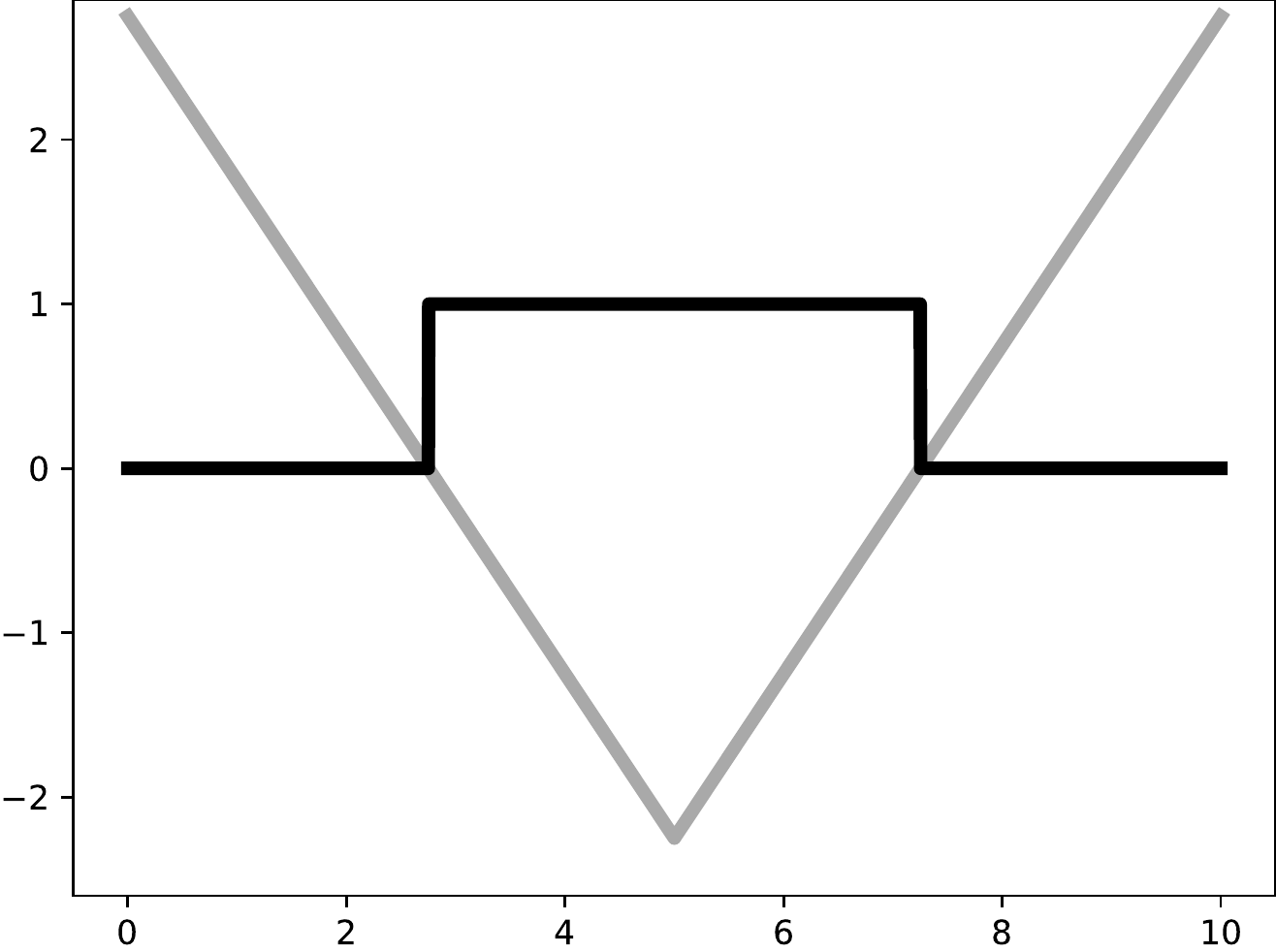}%
      \label{fig:example1d:Heaviside}
    } \\
    \subfloat[{$I[nh]$}]{
      \includegraphics[width=0.44\linewidth]{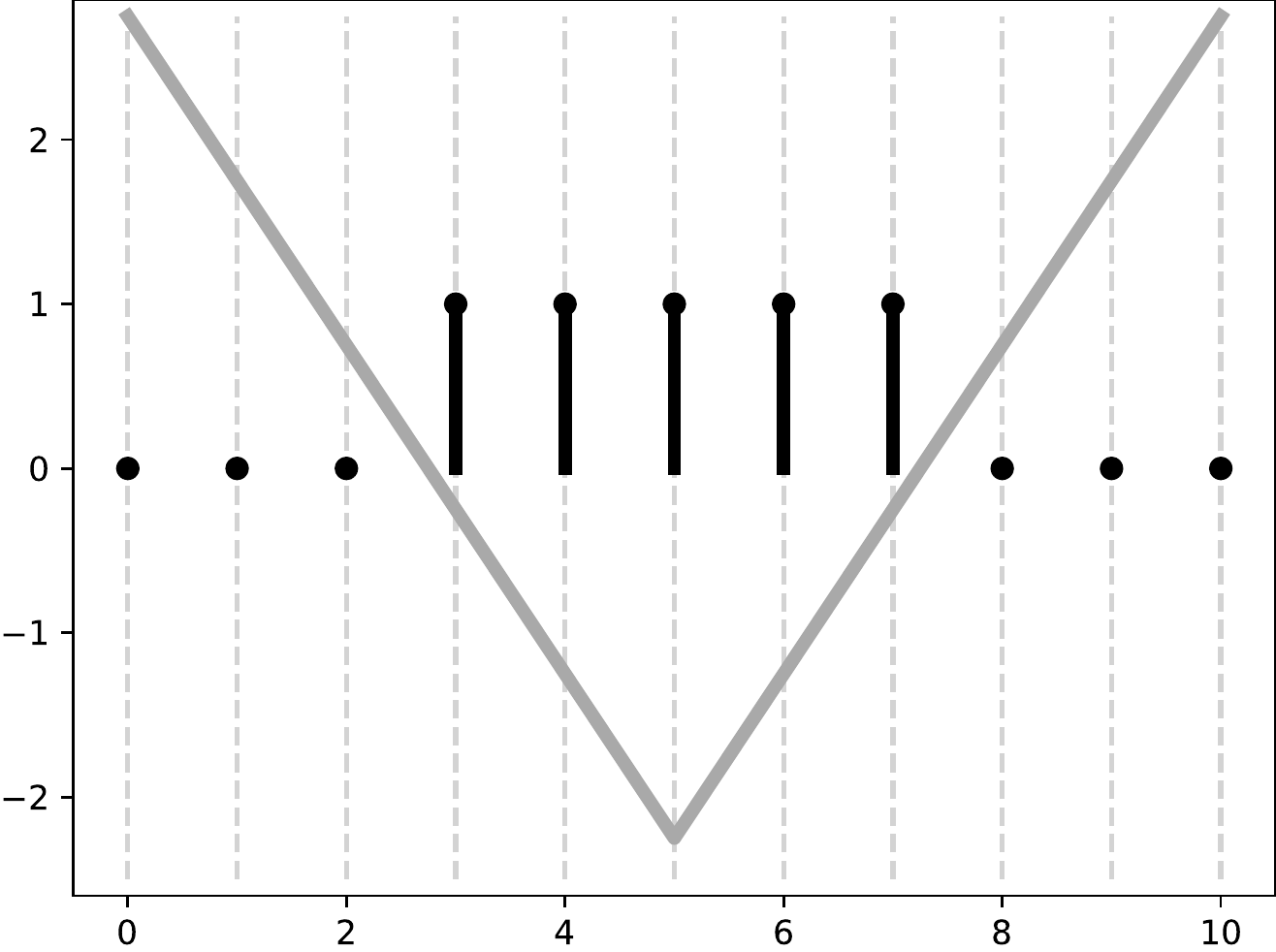}%
      \label{fig:example1d:sampled}
    } &
    \subfloat[{$\tilde{\phi}[nh] = sdt(I[nh])$}]{
      \includegraphics[width=0.44\linewidth]{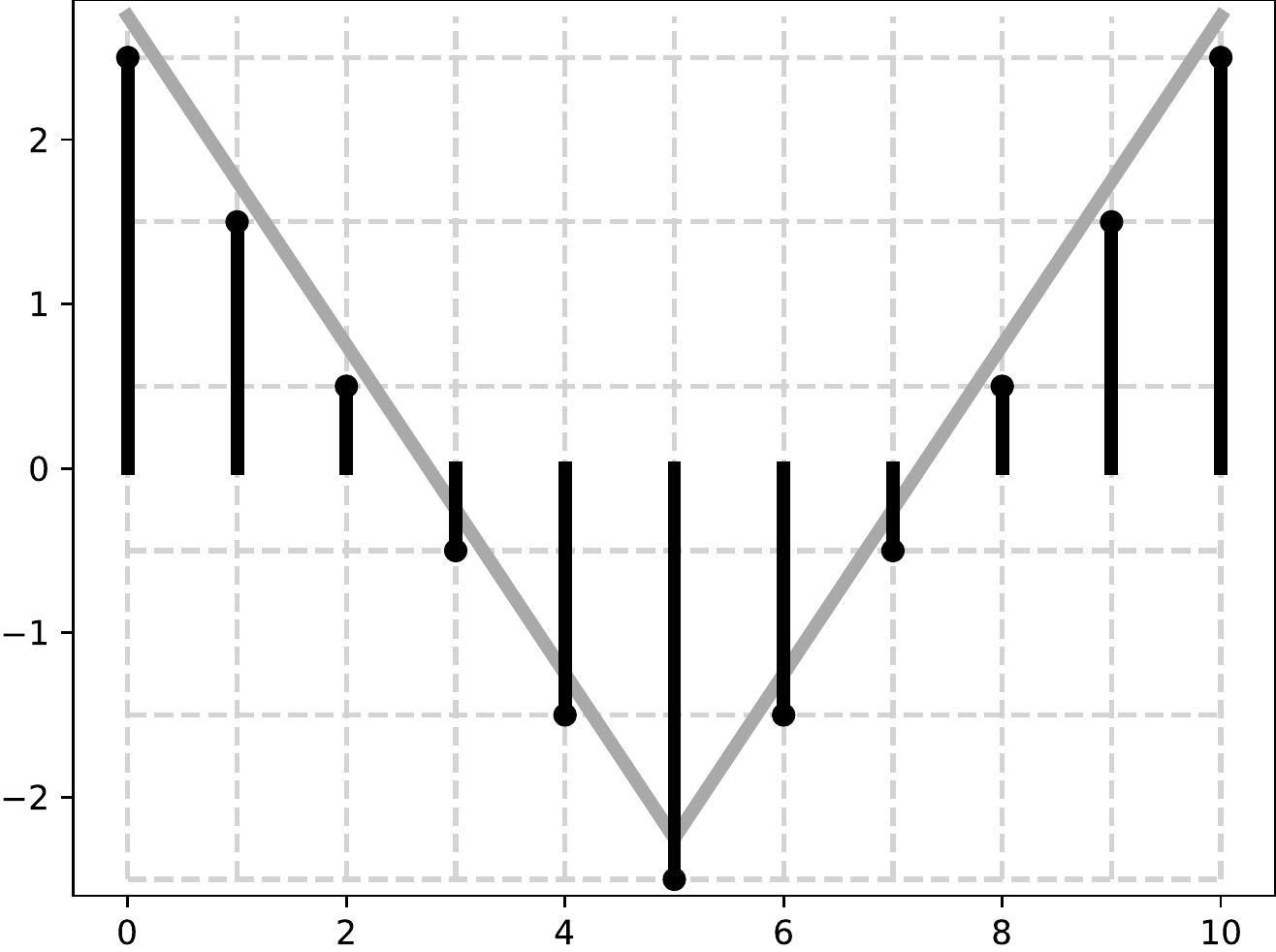}%
      \label{fig:example1d:dt_Heaviside}
    }
  \end{tabular}
  \caption{(\ref{fig:example1d:real}) The signed distance transform of a 1D sphere ($x_0=5.0$, $r=2.25$), (\ref{fig:example1d:Heaviside}) binarized, (\ref{fig:example1d:sampled}) sampled ($h=1.0$), and (\ref{fig:example1d:dt_Heaviside}) the computed signed distance transform. The computed signed distance transform is quantized and biased compared to the continuous distance map.}
  \label{fig:example1d}
\end{figure}

Figure~\ref{fig:example1d} plots the continuous signal, the sampled signal, the Heaviside of the sampled signal, and the resulting distance transform.
This is synonymous with the standard method of representing a continuous object in a digital distance transform image.
As seen in Figure~\ref{fig:example1d:dt_Heaviside}, the distance transform of a sampled signal produces a quantized representation of the underlying distance transform.
Additionally, the quantized signal is a constant offset from the true signal.
Appendix~\ref{app:sdt} details why these values are at half integers and not whole integers.

\subsection{2D Example}
\begin{figure}[h]
  \centering
  \begin{tabular}{cc}
    \subfloat[{$\phi_{2D}[ih, jh]$}]{
      \includegraphics[width=0.44\linewidth]{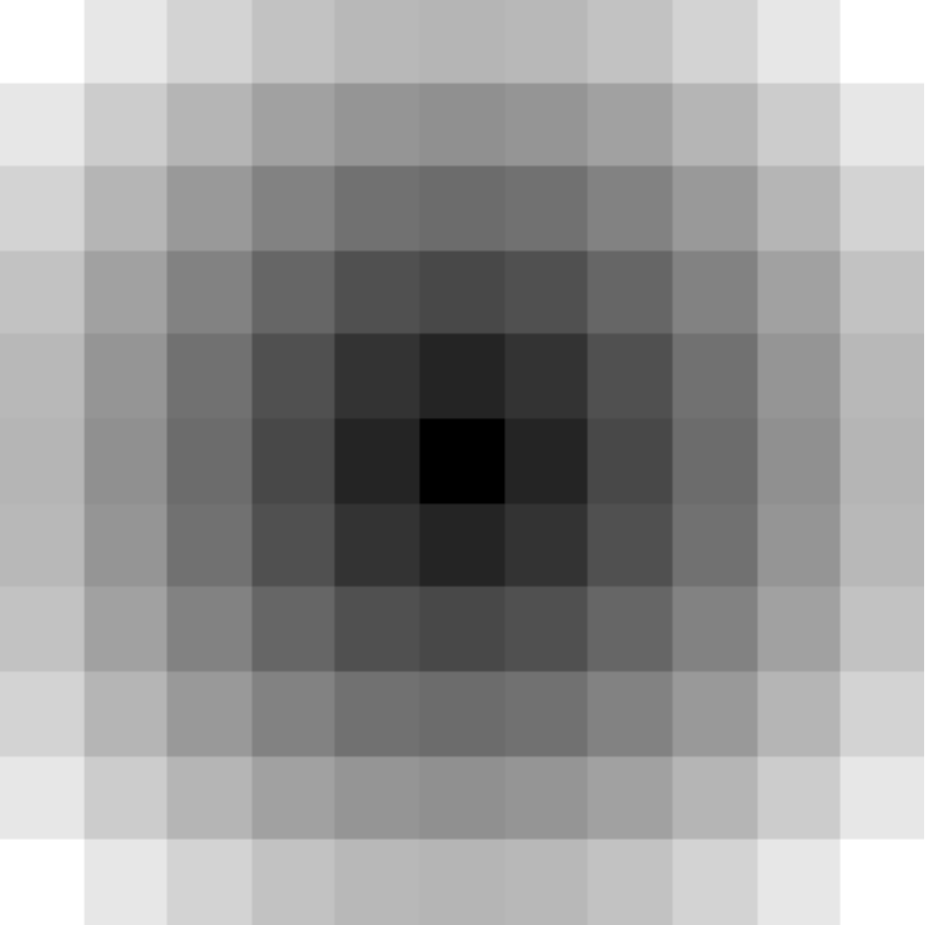}%
      \label{fig:example2d:phi}
    } &
    \subfloat[{$I = -H(\phi_{2D})$}]{
      \includegraphics[width=0.44\linewidth]{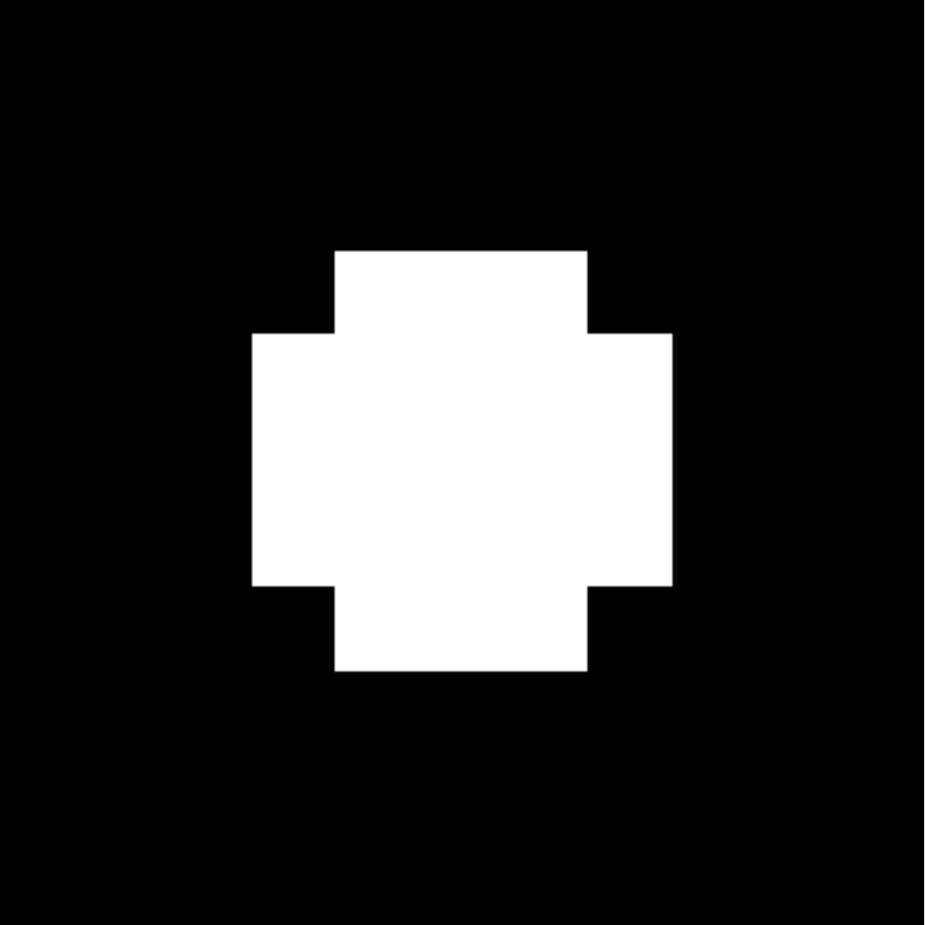}%
      \label{fig:example2d:In}
    } \\
    \subfloat[{$\tilde{\phi}_{2D}[ih, jh] = sdt(I)$}]{
      \includegraphics[width=0.44\linewidth]{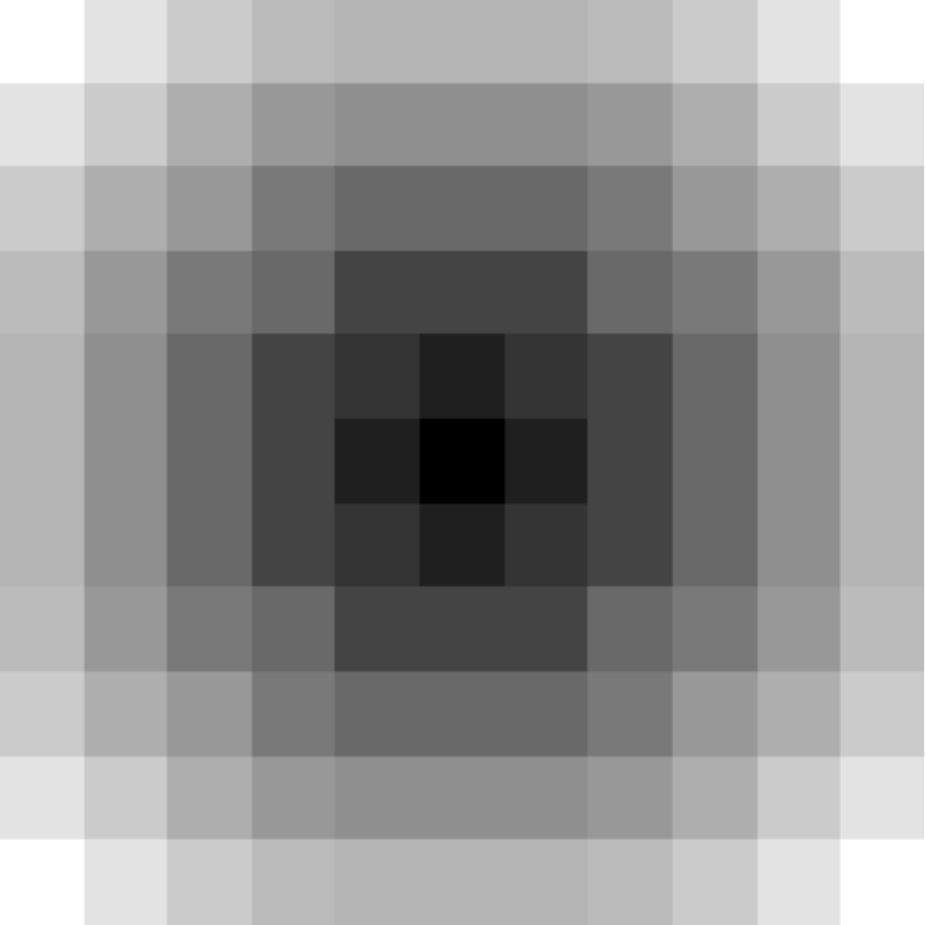}%
      \label{fig:example2d:d}
    } &
    \subfloat[Quantization]{
      \includegraphics[width=0.44\linewidth]{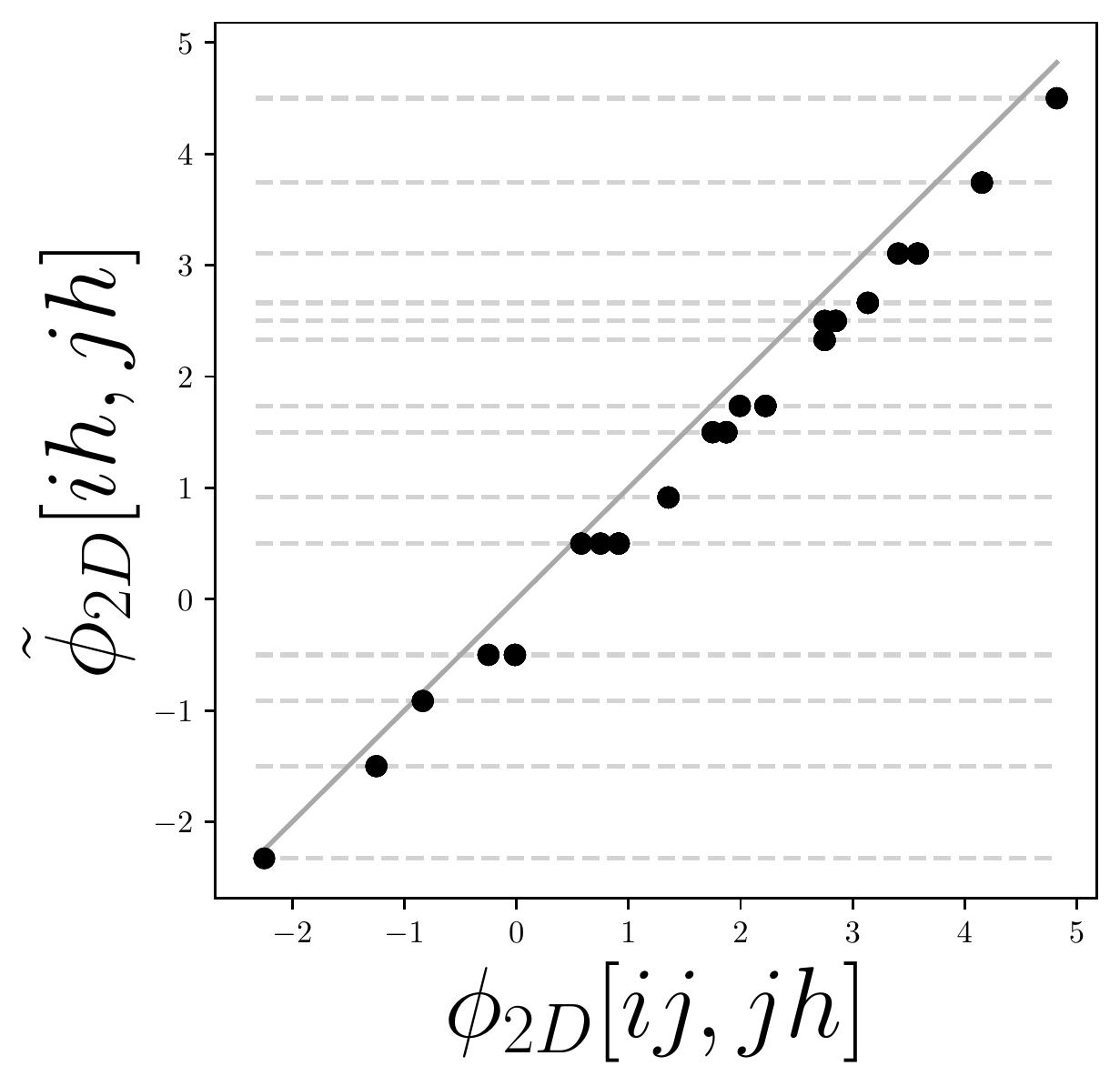}%
      \label{fig:example2d:regression}
    }
  \end{tabular}
  \caption{Quantization in the signed distance transform of a sampled 2D sphere ($x_0=\left(5.0, 5.0\right)^T$, $r=2.25$, $h=1.0$). \ref{fig:example2d:phi} is the ideal signed distance transform, \ref{fig:example2d:In} the binarized image, and \ref{fig:example2d:d} the computed signed distance transform demonstrating quantization. \ref{fig:example2d:phi} is plotted against \ref{fig:example2d:d} in \ref{fig:example2d:regression} where dashed lines represent quantized levels and the solid line demonstrates the ideal relationship $y=x$.}
  \label{fig:example2d}
\end{figure}

The problem is now explored in two dimensions.
A circular signal is again used with the absolute value replaced with the $\ell^2$ norm and $x$, $x_0$ understood as 2D vectors.
\begin{equation}
  \label{eqn:phi_2d_sphere}
  \phi_{2D}(x) = \lVert x - x_0 \rVert_2 - r
\end{equation}
In Figure~\ref{fig:example2d}, the signed distance transform of the Heaviside of the sampled signal is compared to the original signal.
To demonstrate quantization, the numerical signal $\tilde{\phi}_{2D}[ih, jh]$ is plotted against $\phi_{2D}[ih, jh]$.

Comparing Figure~\ref{fig:example2d:phi} to Figure~\ref{fig:example2d:d}, the signed distance transform of the sampled signal has fewer gray levels.
From Figure~\ref{fig:example2d:regression}, quantization is seen as intensities aligning along discrete horizontal lines.
However, the quantization is not necessarily at integers, taking on a finite set of non-integer multiples of $h$.
Furthermore, as seen in the 1D case, the quantized signal is slightly lower (or biased) compared to the ideal signal.

\subsection{Analysis and Synthesis}
\begin{figure}[h]
  \centering
  \includegraphics[width=0.5\linewidth]{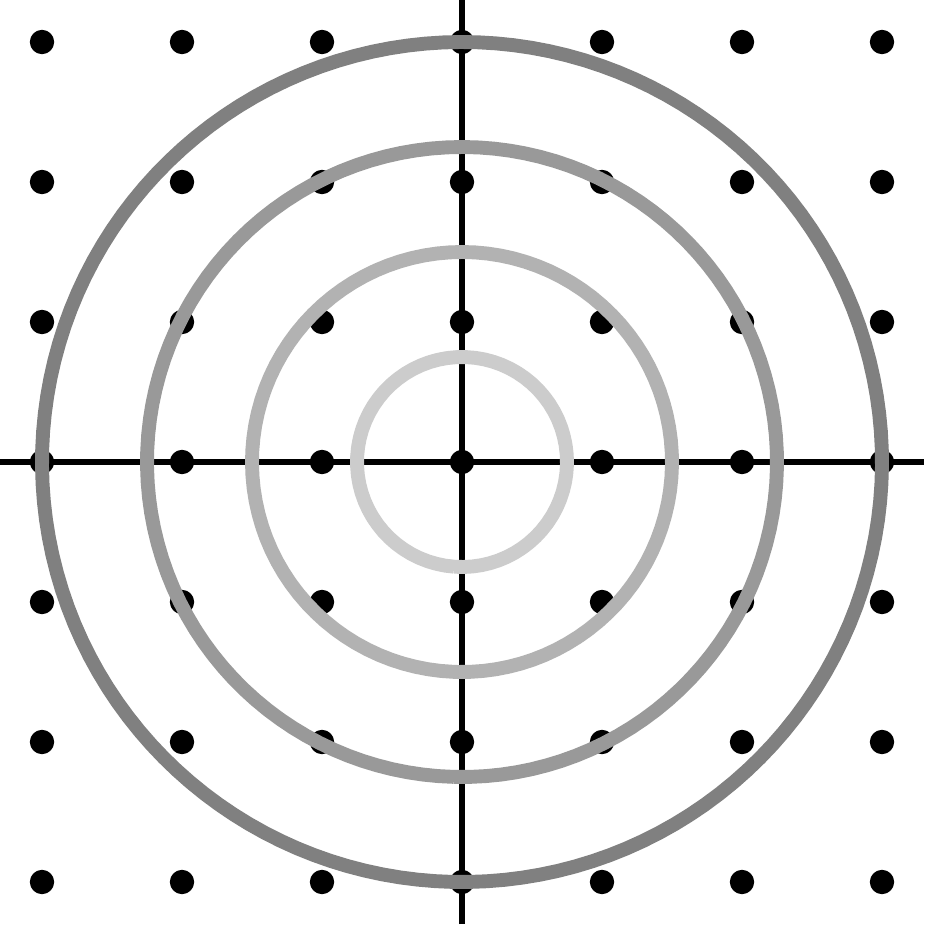}
  \caption{Discretization of the $\ell^2$ metric. Grey lines denote contours of the continuous metric while points denote possible samples from quantization.}
  \label{fig:discretized_metric}
\end{figure}

The problem is now generalized to vectors of arbitrary dimension.
Consider any element of the set $x \in X$ that can be represented as an integer vector multiplied by the sampling period.
\begin{equation}
  x[i, j] = h \left(i, j\right)^T
\end{equation}
The Euclidean distance between any two points will be of the form
\begin{equation}
  d(x, p) = h \sqrt{(i_x-i_p)^2 + (j_x - j_p)^2}
\end{equation}
The consequence is that $d$ can only take a finite set of values.
\begin{equation}
  d \in \left\{h l \given l = \sqrt{i^2 + j^2}, i,j \in \Z \right\}
\end{equation}
This can be extended to arbitrary dimensions by considering the metric $g$ between all discrete samples of the space $X$ (see Section~\ref{sec:distance_transforms}).
For vectors, this set is the metric between the integer vectors and the zero vector.
\begin{equation}
  \label{eqn:quant-metric}
  d \in \left\{ hl \given l = g(z, 0) , z \in \Z^n \right\}
\end{equation}
Discretization of the metric is visualized in Figure~\ref{fig:discretized_metric}.

In summary, the distance transform of a sampled signal is a sampled and quantized representation of the true distance transform.
The samples are a finite set of integer and non-integer multiples of the sampling period.
For a Euclidean metric, these values will be square roots of the sum of square integers.

\section{Artifacts of Quantization}
The consequences of quantization are now explored.
It will be seen that quantization leads to artifacts in the gradients of the distance transform.
These artifacts appear from the distance transform being flat from quantization.

\subsection{Gradient of a 2D Sphere}
From Equation~\ref{eqn:phi_2d_sphere}, the gradient of the image can be computed.
\begin{equation}
  \nabla \phi_{sphere} = \frac{x - x_0}{\lVert x - x_0 \rVert_2}
\end{equation}
As this equation holds in arbitrary dimensions, the reference to dimensions is dropped.
Note that the magnitude gradient is $+1$ as is expected of a signed distance transform.

\begin{figure*}[h]
  \centering
  \begin{tabular}{cccc}
    \multicolumn{4}{c}{ \begin{tabular}{ccc}
    \subfloat[{$\phi[ih, jh]$}]{
      \includegraphics[width=0.2\linewidth]{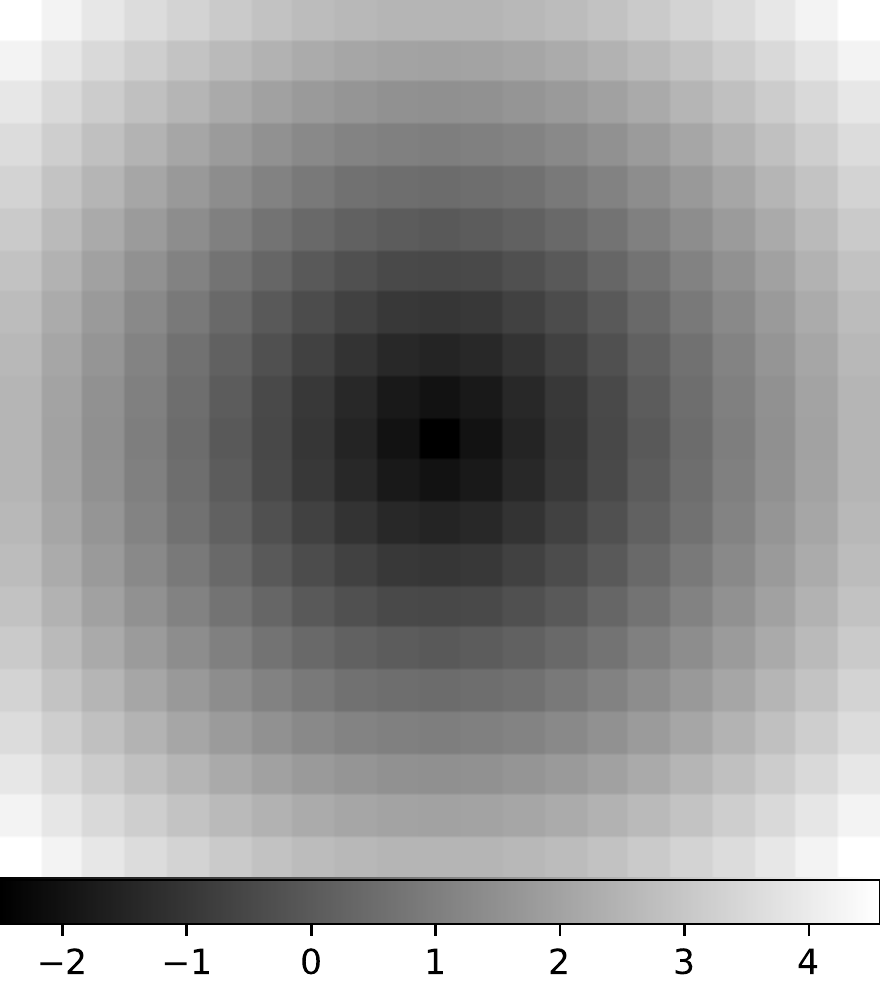}%
      \label{fig:gradients:phin}
    } &
    \subfloat[{$I = H(-\phi)$}]{
      \includegraphics[width=0.22\linewidth]{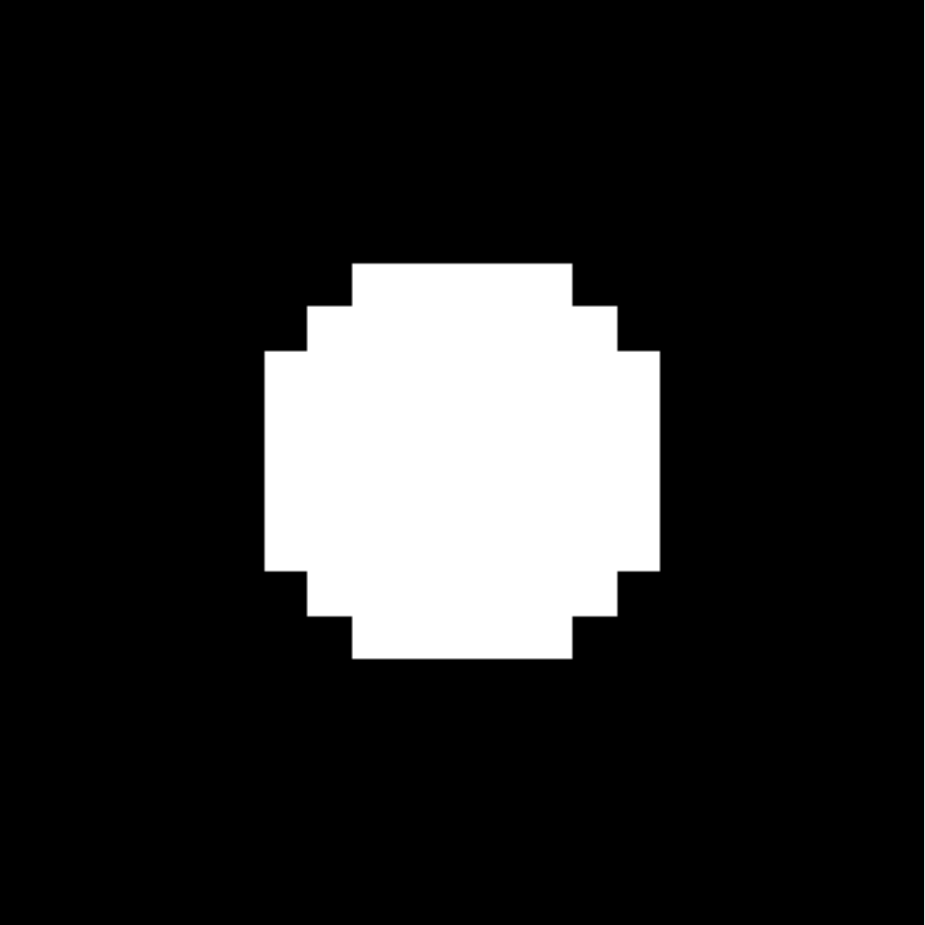}%
      \label{fig:gradients:In}
    } &
    \subfloat[{$\tilde{\phi}[ih, jh]$}]{
      \includegraphics[width=0.2\linewidth]{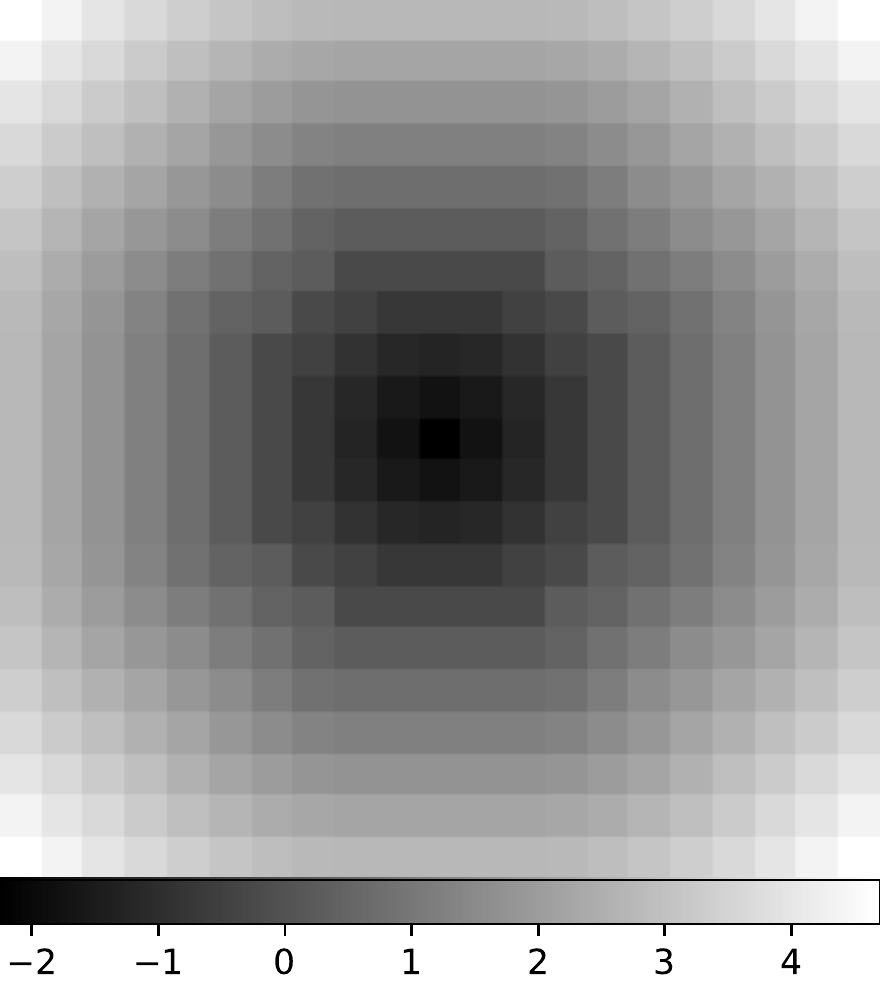}%
      \label{fig:gradients:approx_phi}
    } \end{tabular}} \\
    \subfloat[{$\nabla_x \phi$}]{
      \includegraphics[width=0.2\linewidth]{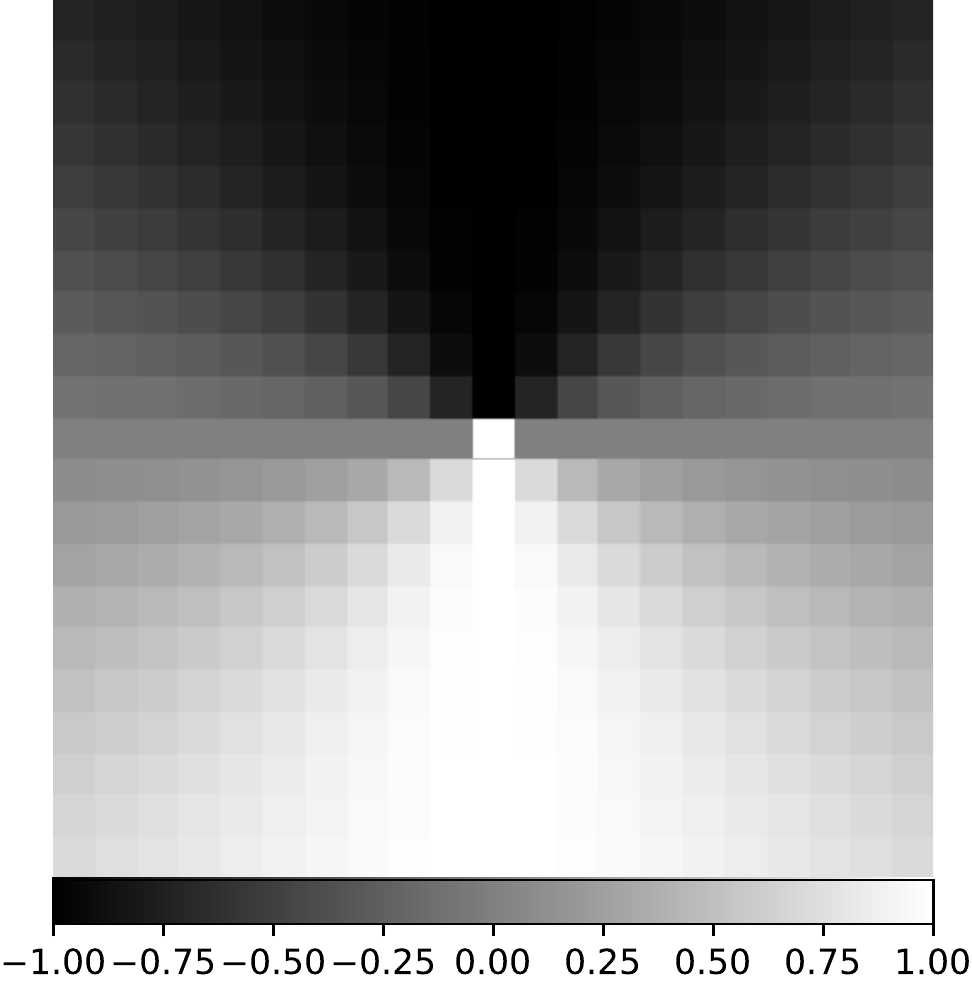}%
      \label{fig:gradients:grad}
    } &
    \subfloat[{$D^{0x}\tilde{\phi}$}]{
      \includegraphics[width=0.2\linewidth]{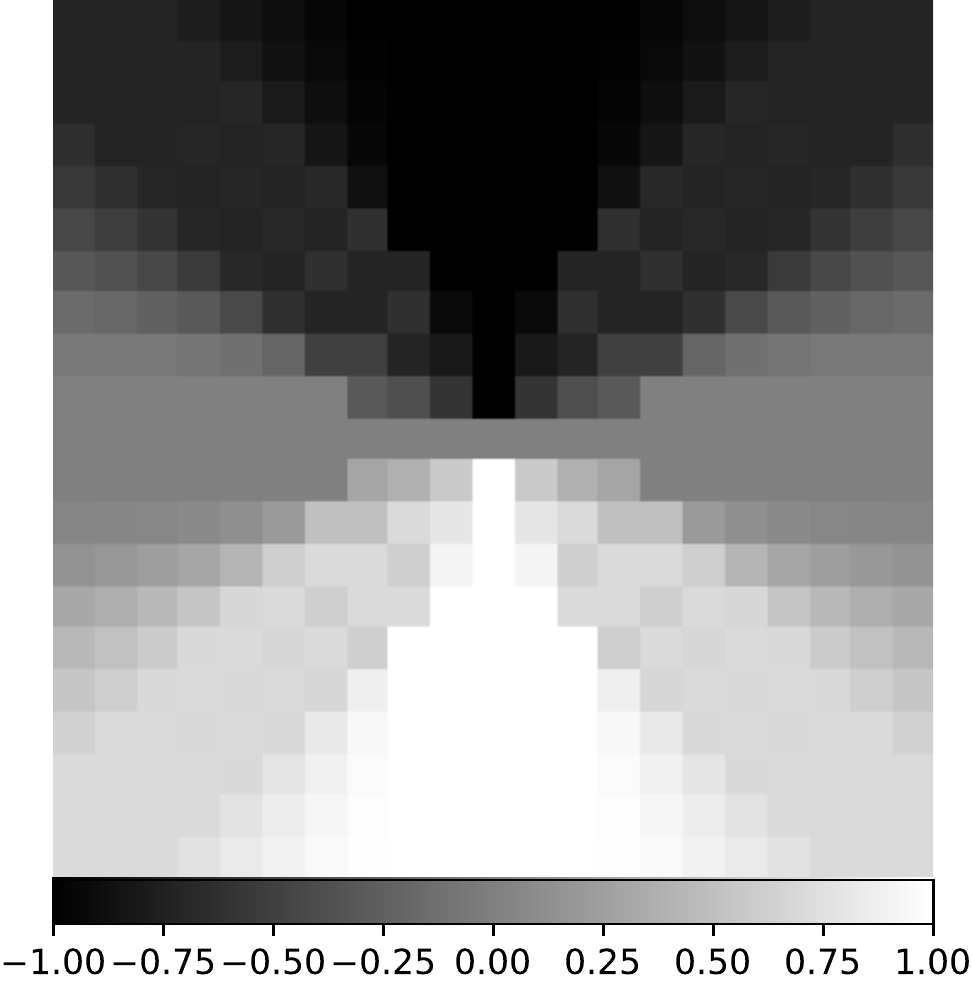}%
      \label{fig:gradients:dc}
    } &
    \subfloat[{$\lVert\nabla\phi\rVert$}]{
      \includegraphics[width=0.2\linewidth]{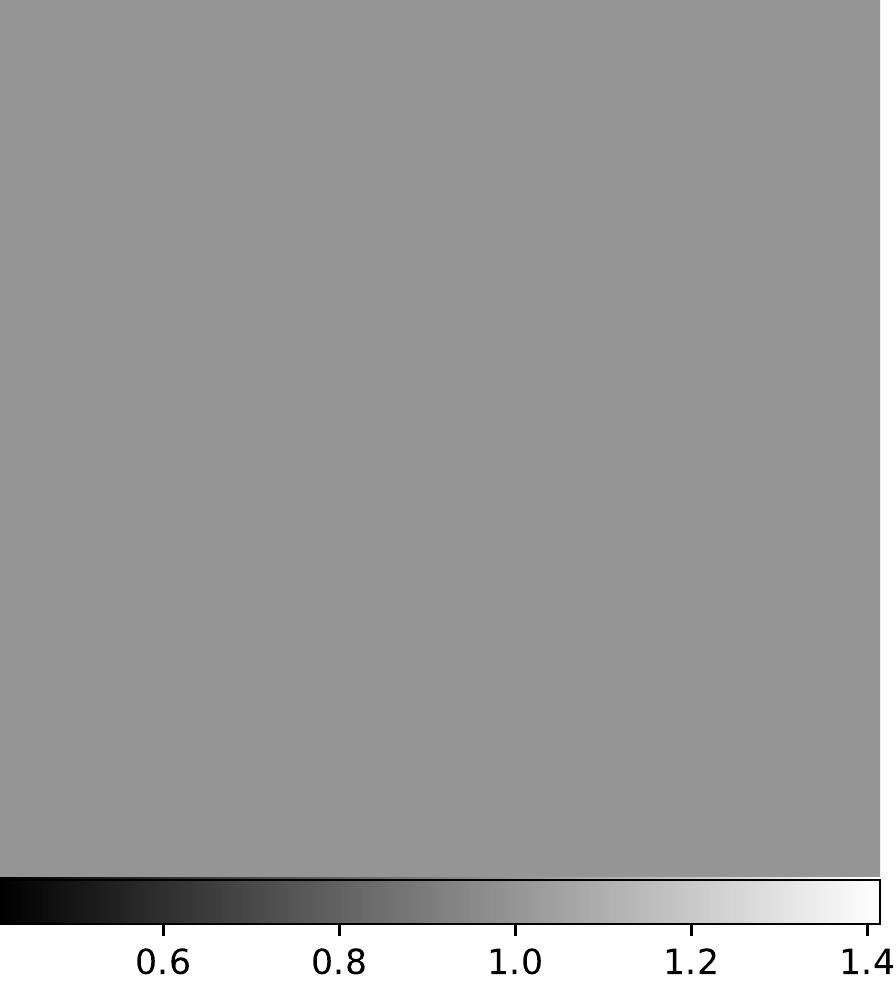}%
      \label{fig:gradients:mag_grad}
    } &
    \subfloat[{$\lVert D^0\tilde{\phi}\rVert$}]{
      \includegraphics[width=0.2\linewidth]{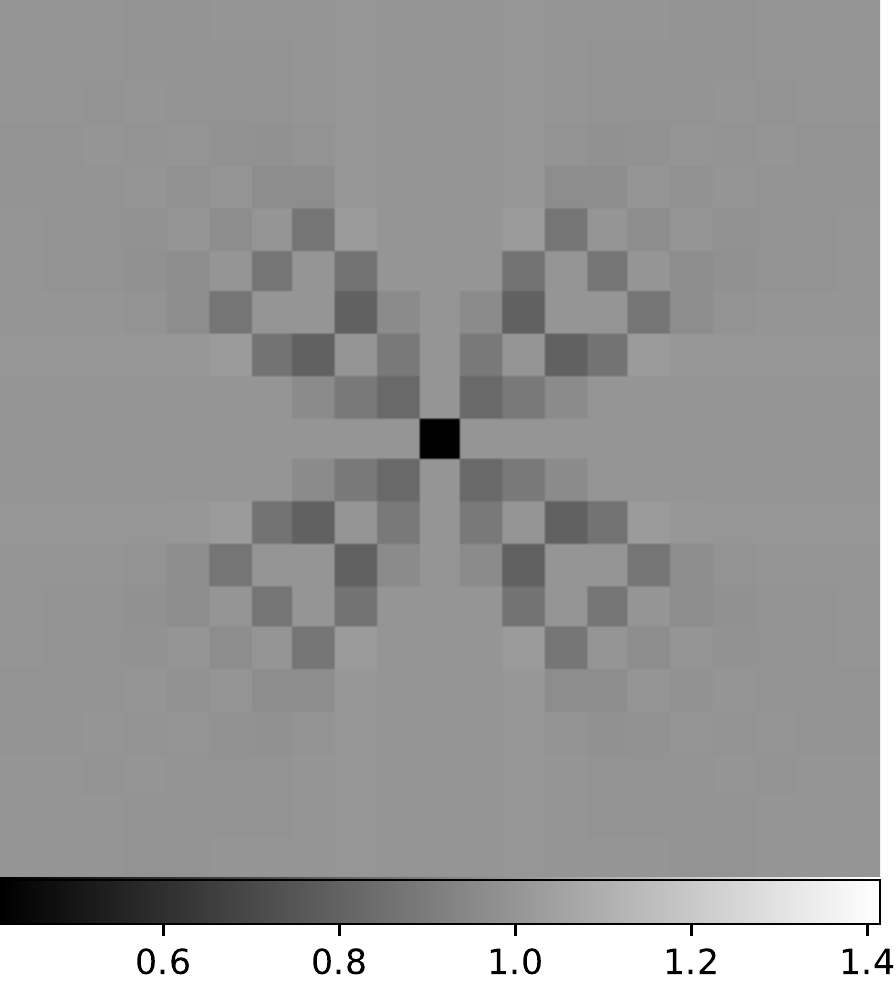}%
      \label{fig:gradients:mag_dfc}
    } \\
    \subfloat[{$\nabla_x \phi(x, 1.5)$}]{
      \includegraphics[width=0.2\linewidth]{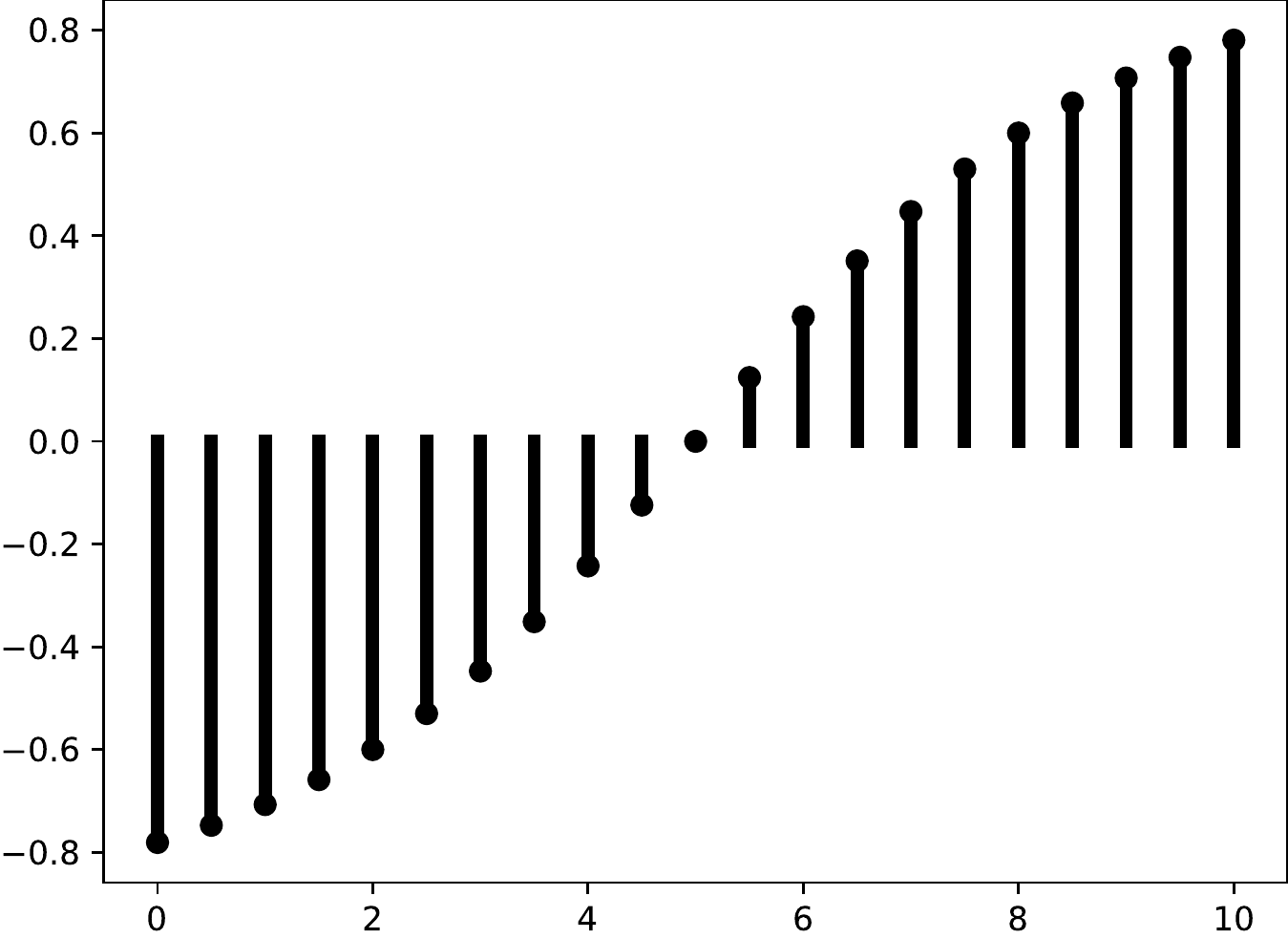}%
      \label{fig:grad-flat:del_phi}
    } &
    \subfloat[{$D^{0x} \tilde{\phi}(x, 1.5)$}]{
      \includegraphics[width=0.2\linewidth]{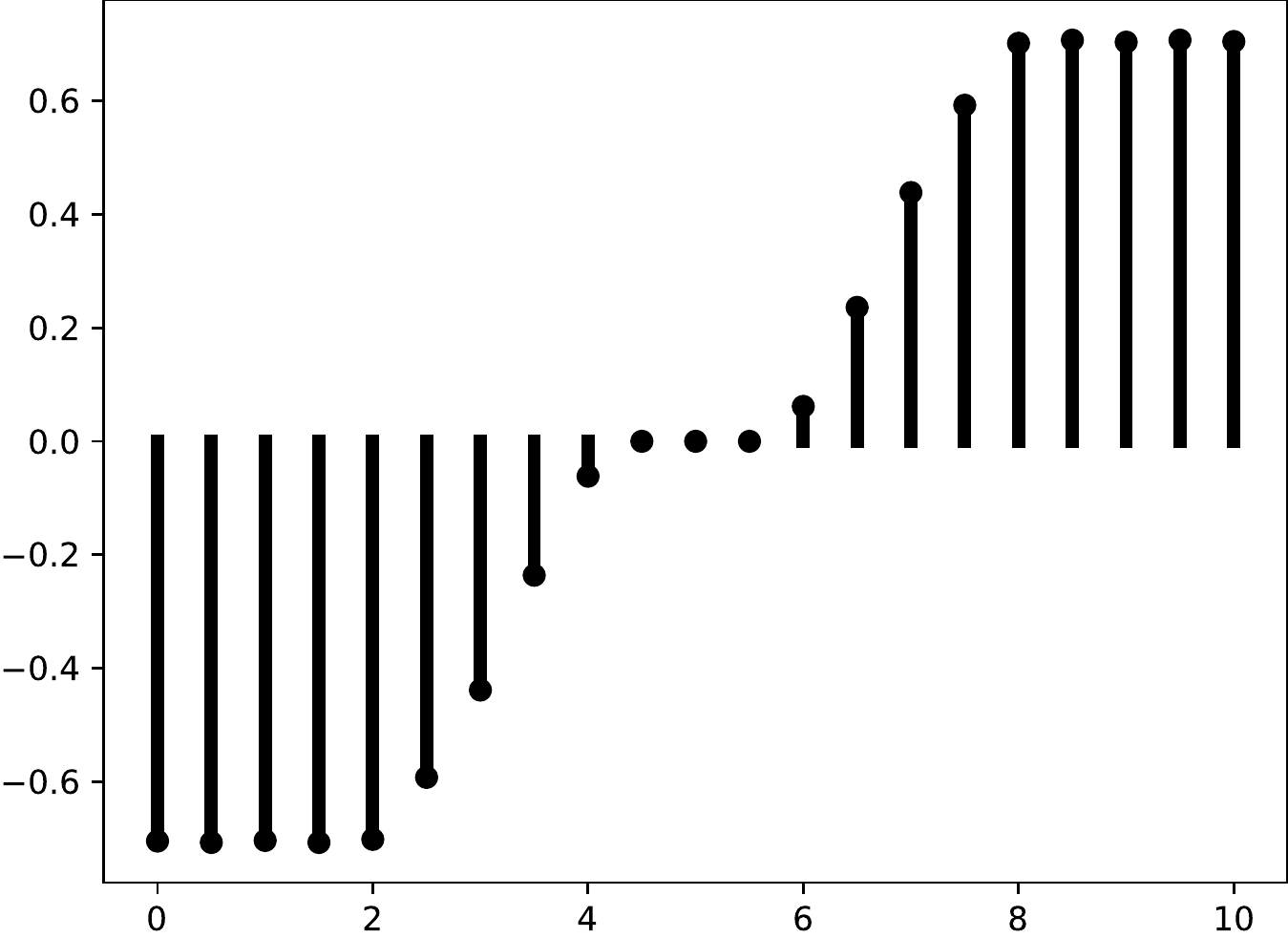}%
      \label{fig:grad-flat:del_d}
    } &
    \subfloat[{$\lVert \nabla \phi(x, 1.5) \rVert$}]{
      \includegraphics[width=0.2\linewidth]{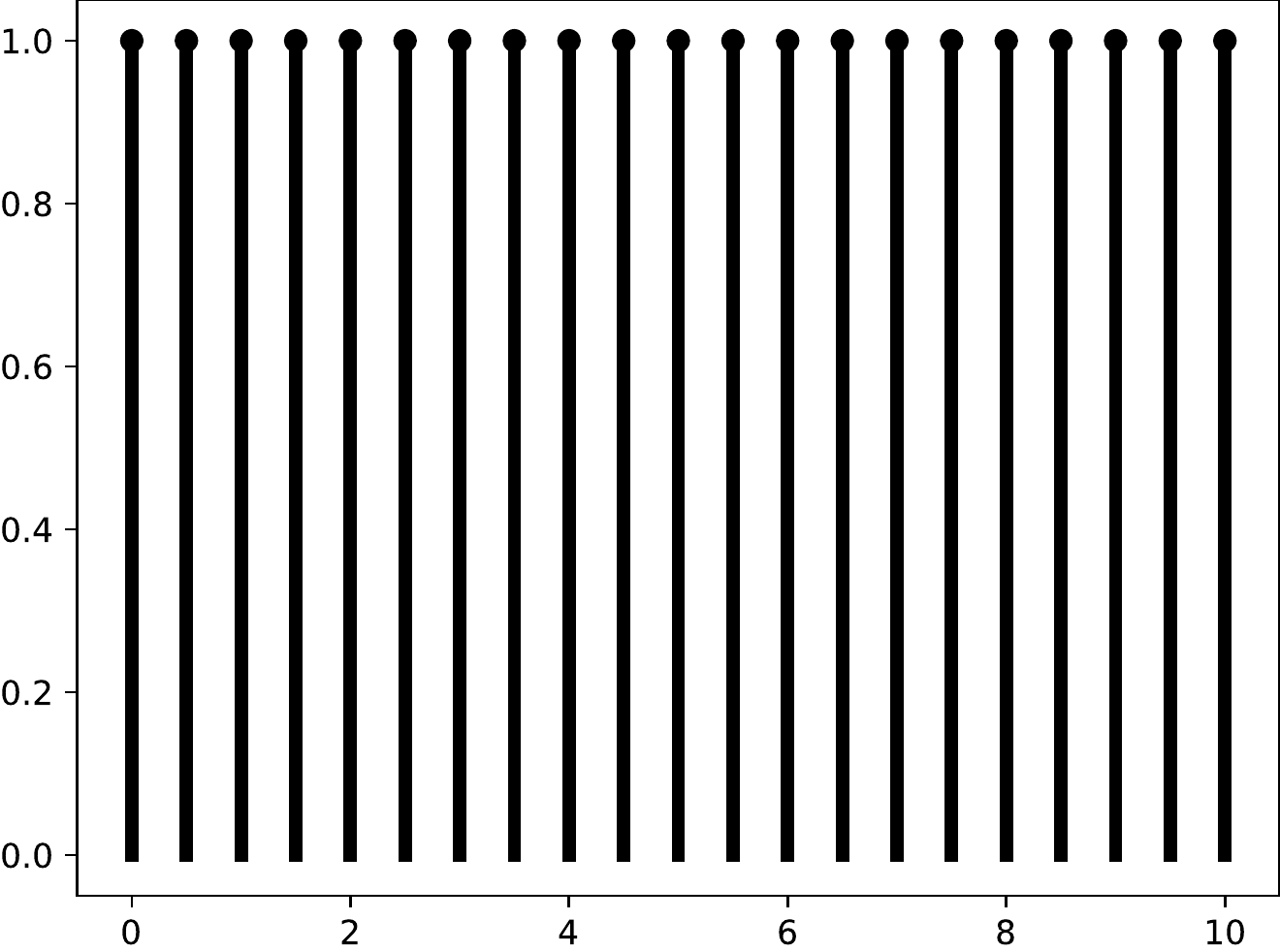}%
      \label{fig:grad-flat:mag_grad_phi}
    } &
    \subfloat[{$\lVert D^{0} \tilde{\phi}(x, 1.5) \rVert$}]{
      \includegraphics[width=0.2\linewidth]{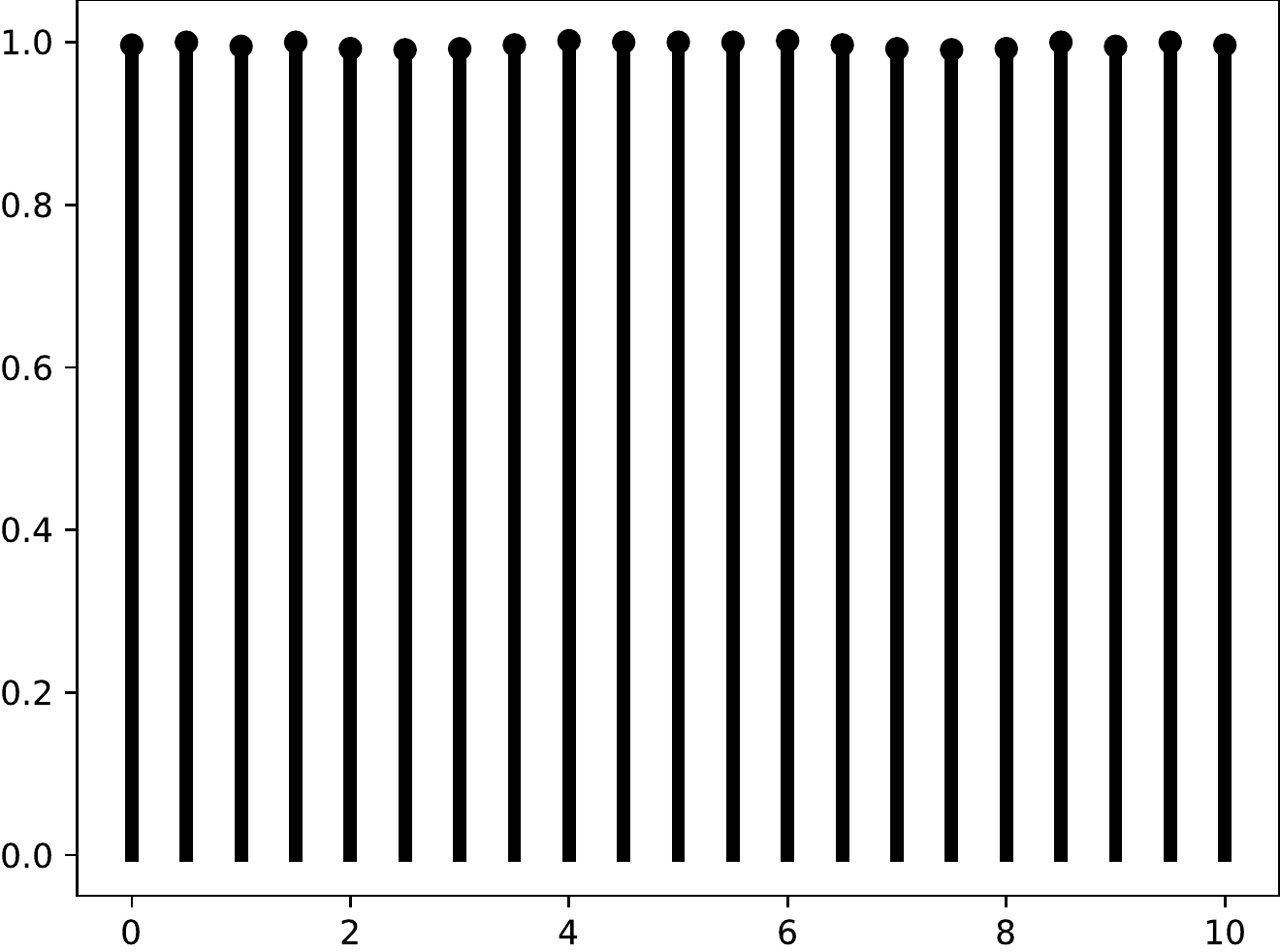}%
      \label{fig:grad-flat:mag_grad_d}
    }
  \end{tabular}
  \caption{Accuracy of numerical gradients of quantized distance transforms to represent true gradients of distance transforms using a 2D sphere ($x_0=\left(5.0, 5.0\right)^T$, $r=2.5$, $h=0.5$). (a-c) demonstrate the quantized signed distance transform. (d-e, h-i) demonstrate banding and flat \textit{x} gradients. (f-g, j-k) demonstrate a near ideal magnitude gradient.}
  \label{fig:gradients}
\end{figure*}

Numerical gradients of the quantized signal are compared to the ideal gradients in Figure~\ref{fig:gradients}.
At the \textit{y} axis, the derivative in the \textit{x} direction is zero as expected (Figure~\ref{fig:gradients:dc}).
However, just forward or backwards from the \textit{x} axis the derivative remains exactly zero.
For the same point in the magnitude gradient image (Figure~\ref{fig:gradients:mag_dfc}), the magnitude gradient is positive one.
The fact that the gradient for some points is flat, while the magnitude of the gradient is one, is important for reinitialization algorithms~\cite{sussman1994level,peng1999pde} where removal of the artifact will converge very slowly.

\subsection{Interpretation from Voronoi Diagrams}
\begin{figure*}[h]
  \centering
  \begin{tabular}{cccc}
    \subfloat[Voronoi]{
      \includegraphics[width=0.2\linewidth]{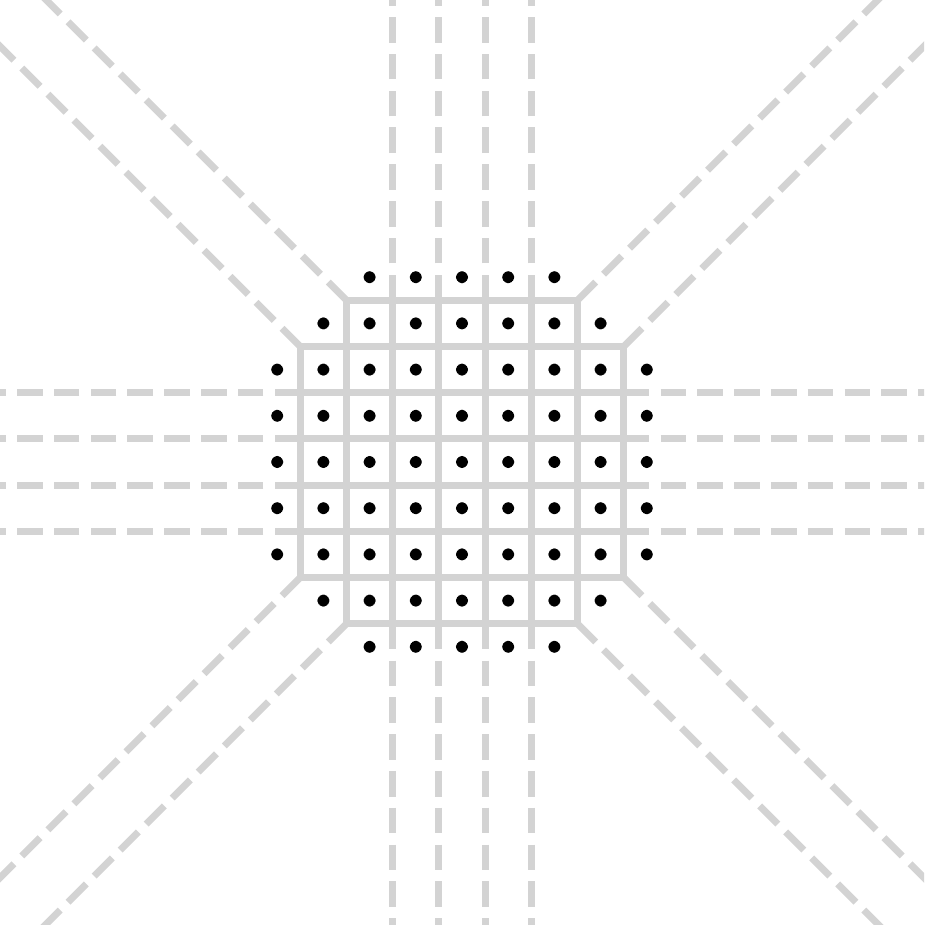}%
      \label{fig:voronoi:voronoi}
    } &
    \subfloat[{$\tilde{\phi}$}]{
      \includegraphics[width=0.2\linewidth]{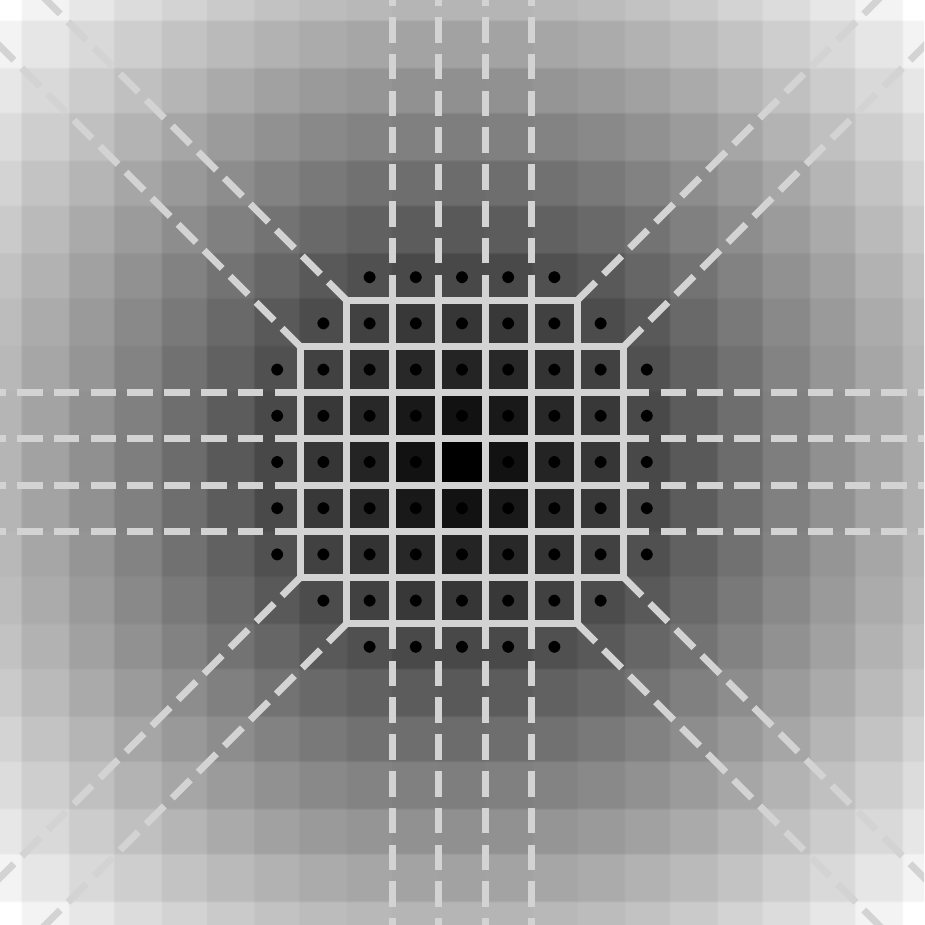}%
      \label{fig:voronoi:phi}
    } &
    \subfloat[{$D^{0x} \tilde{\phi}$}]{
      \includegraphics[width=0.2\linewidth]{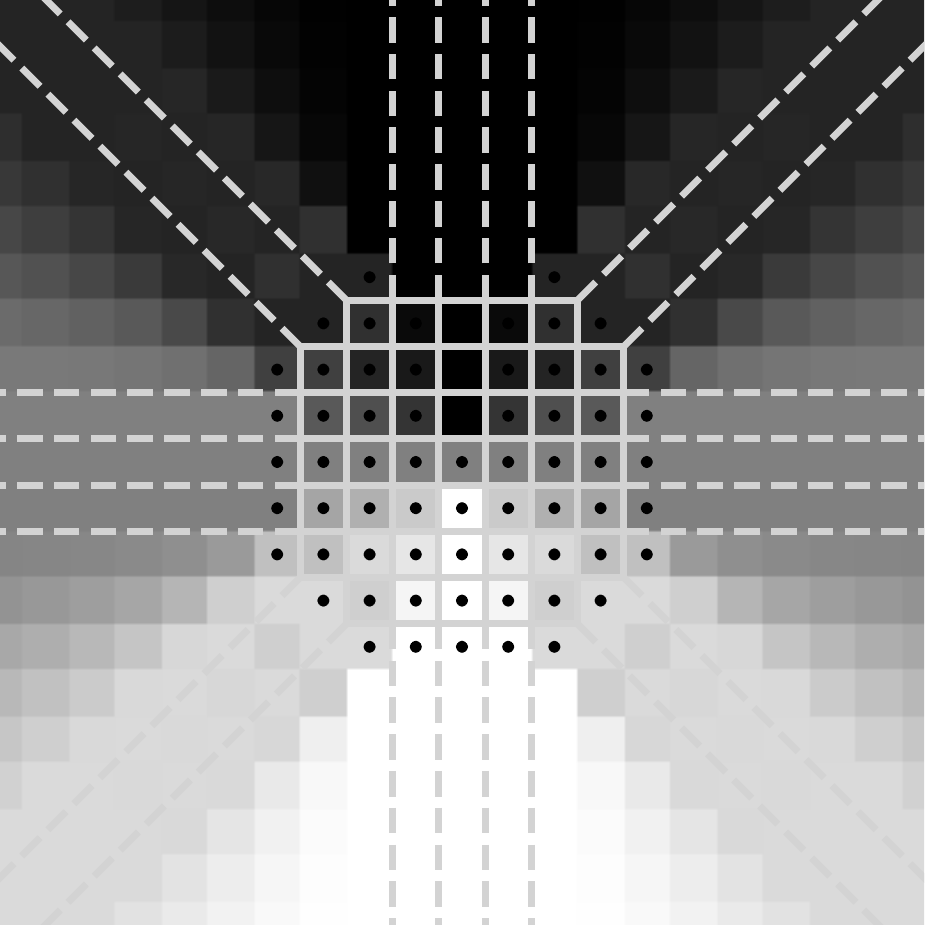}%
      \label{fig:voronoi:d_x}
    } &
    \subfloat[{$\lVert D^{0} \tilde{\phi} \rVert$}]{
      \includegraphics[width=0.2\linewidth]{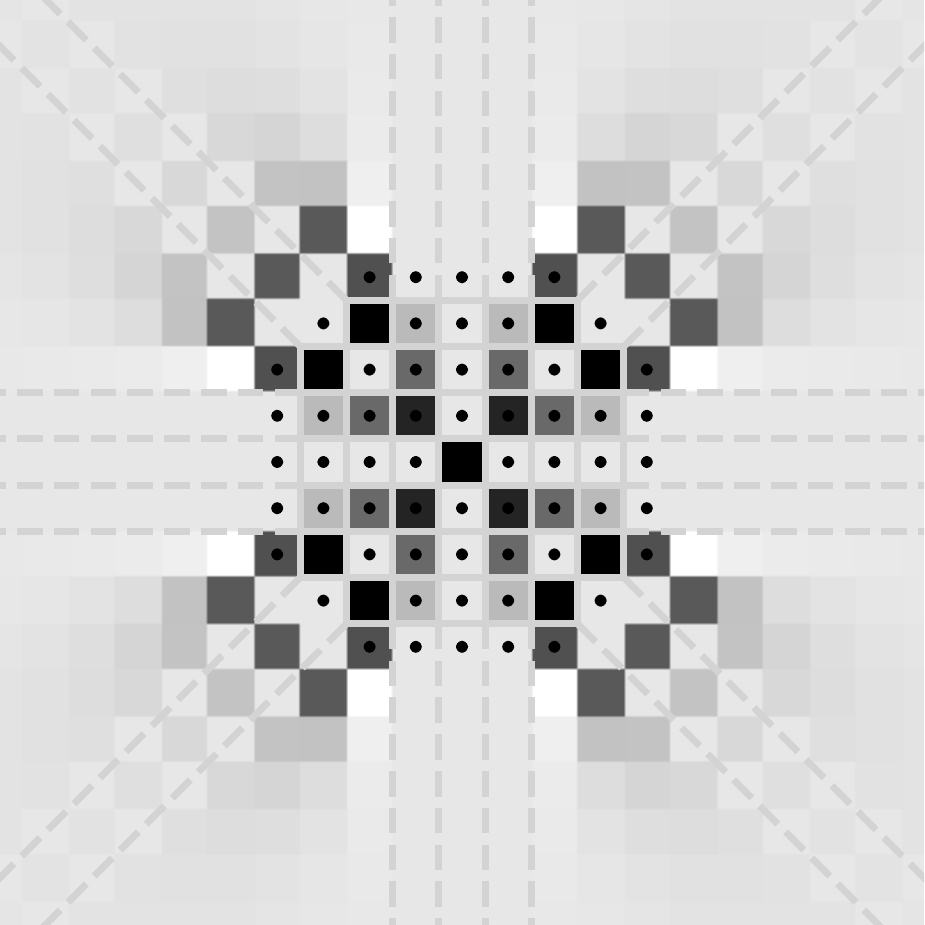}%
      \label{fig:voronoi:mag_grad}
    }
  \end{tabular}
  \caption{Voronoi diagram (\ref{fig:voronoi:voronoi}) overlaid on the signed distance transform (\ref{fig:voronoi:phi}), finite difference in the x-direction (\ref{fig:voronoi:d_x}), and magnitude gradient (\ref{fig:voronoi:mag_grad}).}
  \label{fig:voronoi}
\end{figure*}

The Voronoi diagram ~\cite{voronoi1908nouvelles} of the binary image is overlaid on the gradient and gradient magnitude in Figure~\ref{fig:voronoi}.
Parallel Voronoi lines correspond to bands of constant quantization in the distance transform.
Placing a finite difference stencil across the parallel Voronoi edges leads to gradients forced flat, visualized as banding in the gradient image.
However, the gradient magnitude remains close to one.
As the finite difference stencil increases, this error will not change until the stencil extends beyond parallel Voronoi edges.

\subsection{Higher Order Gradients}
Finally, higher order derivatives are explored.
The finite difference approximation to the second derivative is straight forward.
\begin{equation}
  D^{xx} \phi = \frac{\phi(x+h) - 2\phi(x) + \phi(x-h)}{h^2}
\end{equation}
The mixed derivative $D^{yx}$ has a similar form.
From the first and second derivatives, the Laplacian can be computed.
\begin{equation}
  \Delta \phi = \phi_{xx} + \phi_{yy}
\end{equation}
Similarily, the curvature can also be computed.
\begin{equation}
  \kappa = \frac{\phi_y^2\phi_{xx} - 2 \phi_x \phi_y \phi_{xy} + \phi_x^2 \phi_{yy}}{\left(\phi_x^2 + \phi_y^2\right)^{3/2}}
\end{equation}
Again, infinitesimal differences are replaced by their respective finite difference approximations.
Since an analytic function exists for the distance transform, all these functions are known exactly.
These ideal images are now compared to the finite difference approximation of the quantized signal.

\begin{figure*}[h]
  \centering
  \begin{tabular}{ccccc}
    \subfloat[$\phi$]{
      \includegraphics[width=0.15\linewidth]{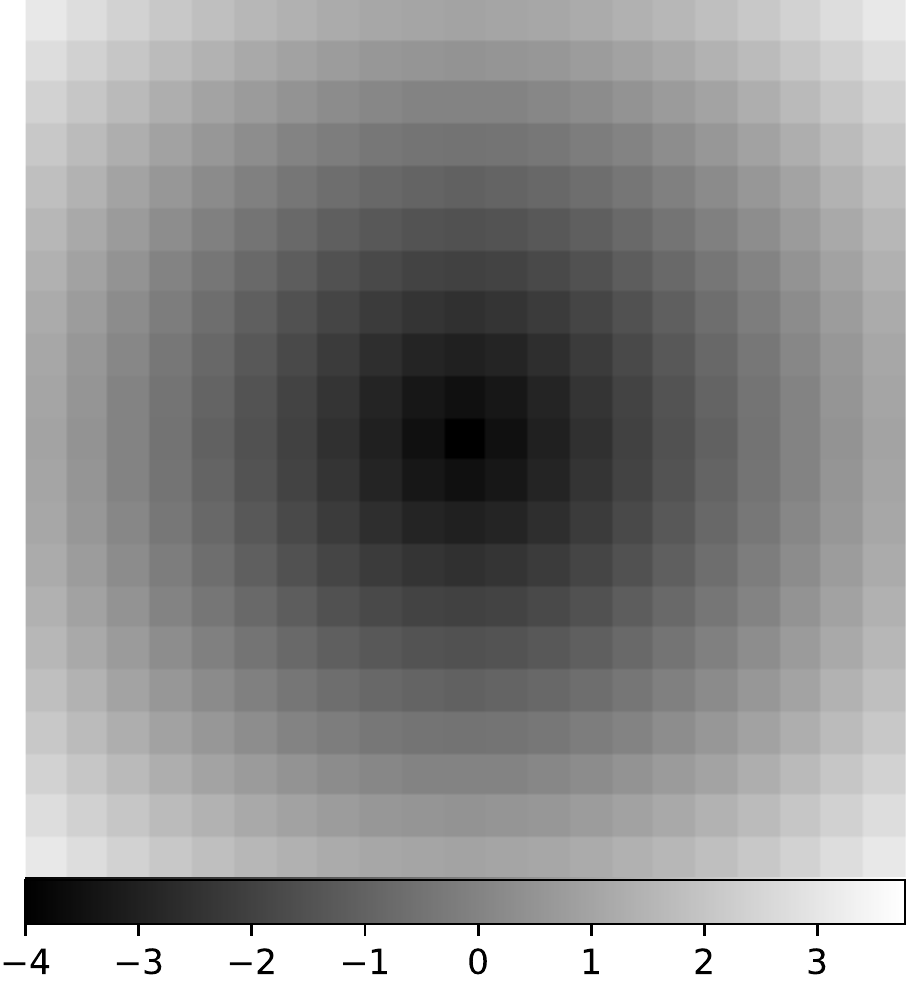}%
      \label{fig:higher_grads:phi}
    } &
    \subfloat[{$d^2\phi/dx^2$}]{
      \includegraphics[width=0.15\linewidth]{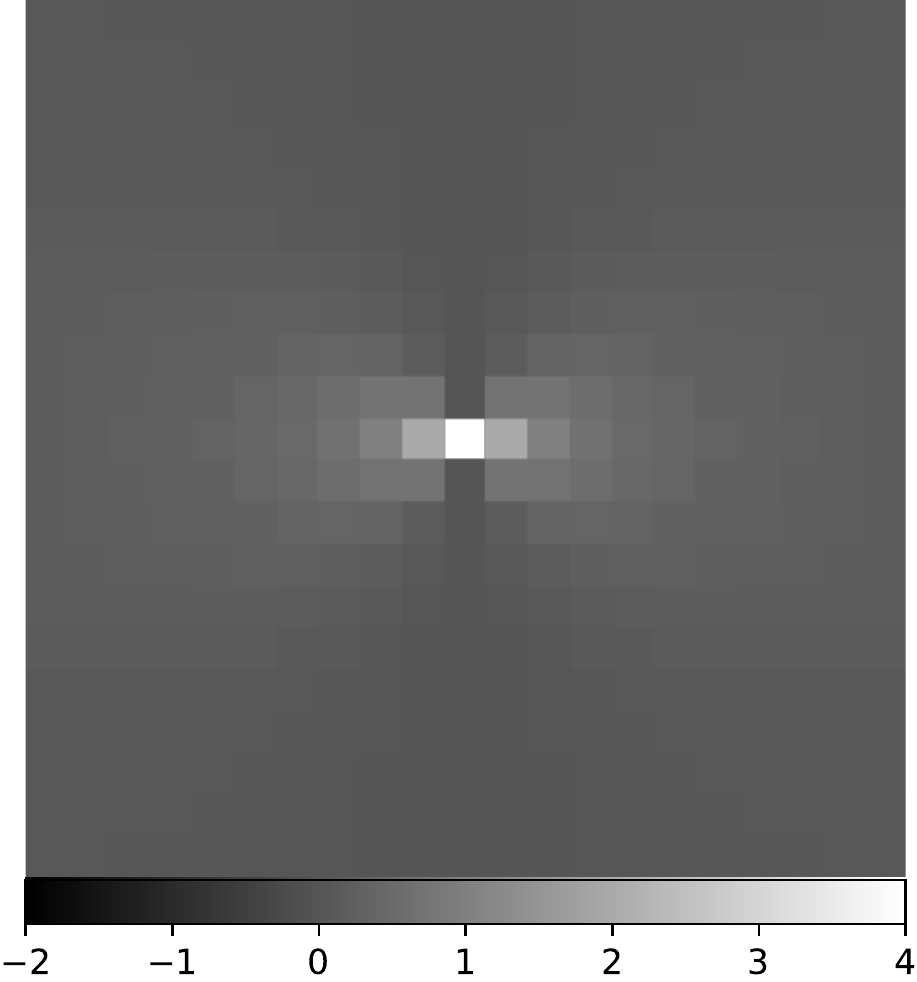}%
      \label{fig:higher_grads:phi_xx}
    } &
    \subfloat[{$d^2\phi/dxdy$}]{
      \includegraphics[width=0.15\linewidth]{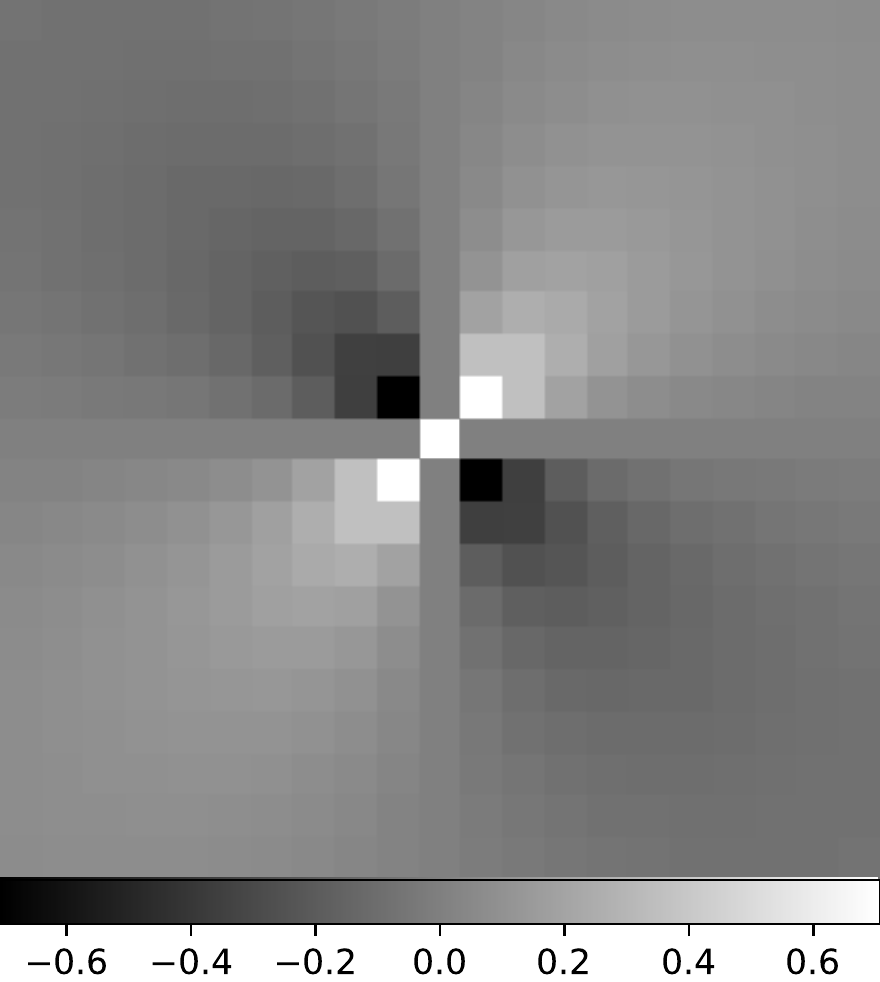}%
      \label{fig:higher_grads:phi_xy}
    } &
    \subfloat[{$\Delta \phi$}]{
      \includegraphics[width=0.15\linewidth]{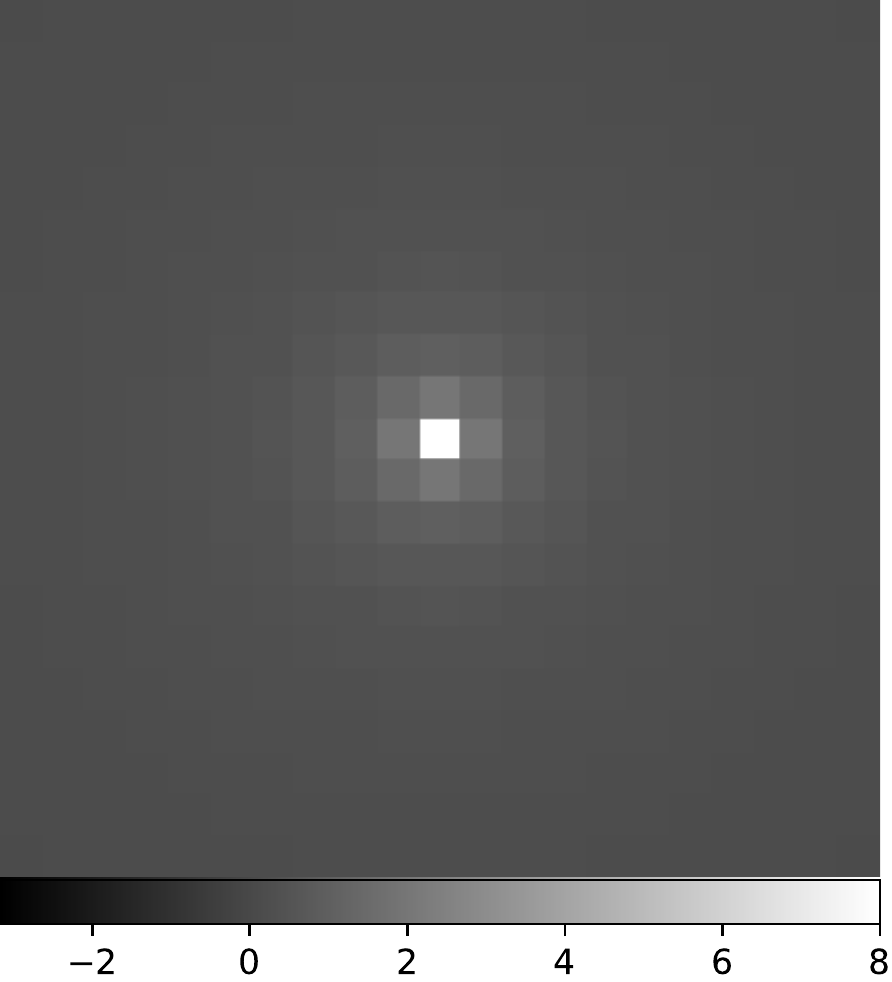}%
      \label{fig:higher_grads:phi_lap}
    } &
    \subfloat[{$\kappa$}]{
      \includegraphics[width=0.15\linewidth]{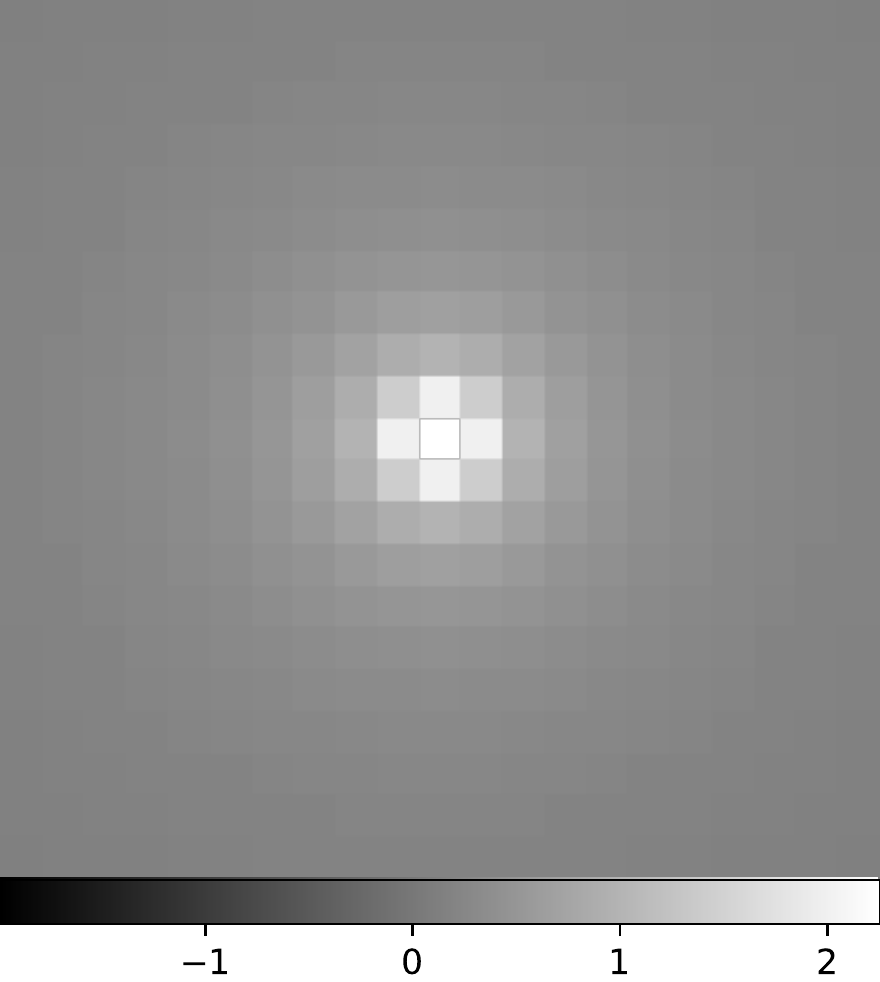}%
      \label{fig:higher_grads:phi_curve}
    } \\
    \subfloat[{$\tilde{\phi}$}]{
      \includegraphics[width=0.15\linewidth]{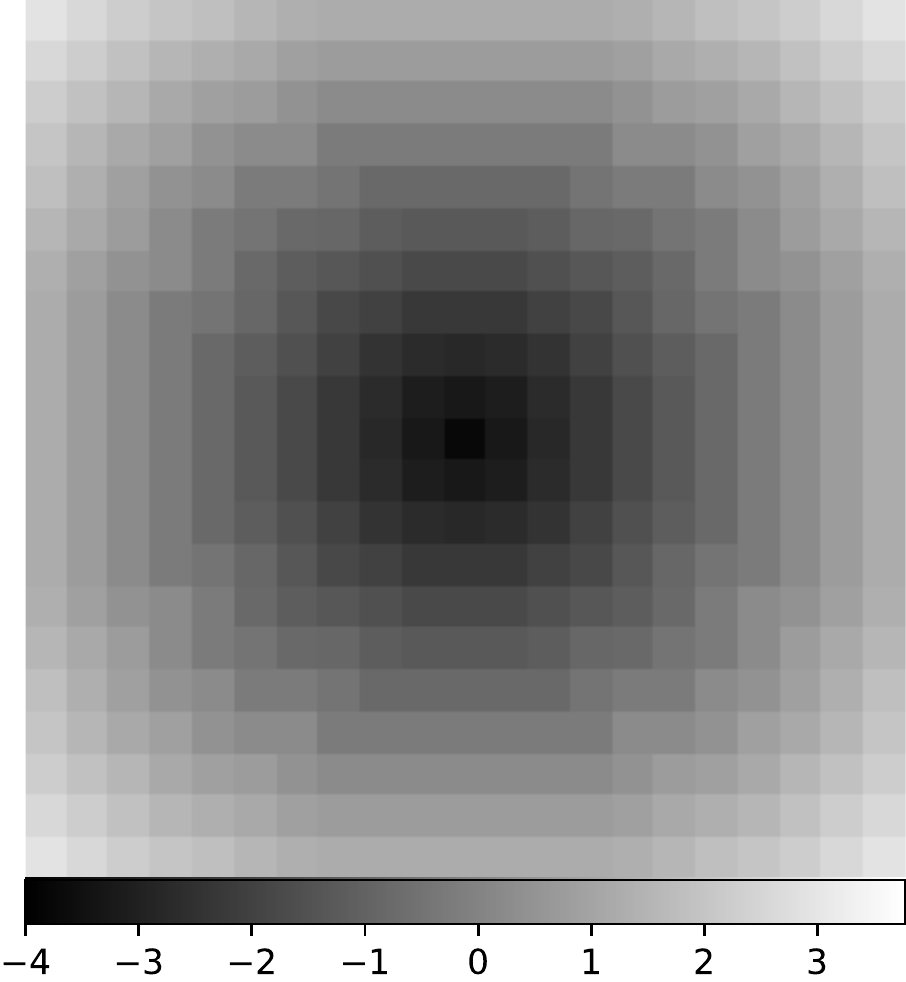}%
      \label{fig:higher_grads:d}
    } &
    \subfloat[{$D^{xx}\tilde{\phi}$}]{
      \includegraphics[width=0.15\linewidth]{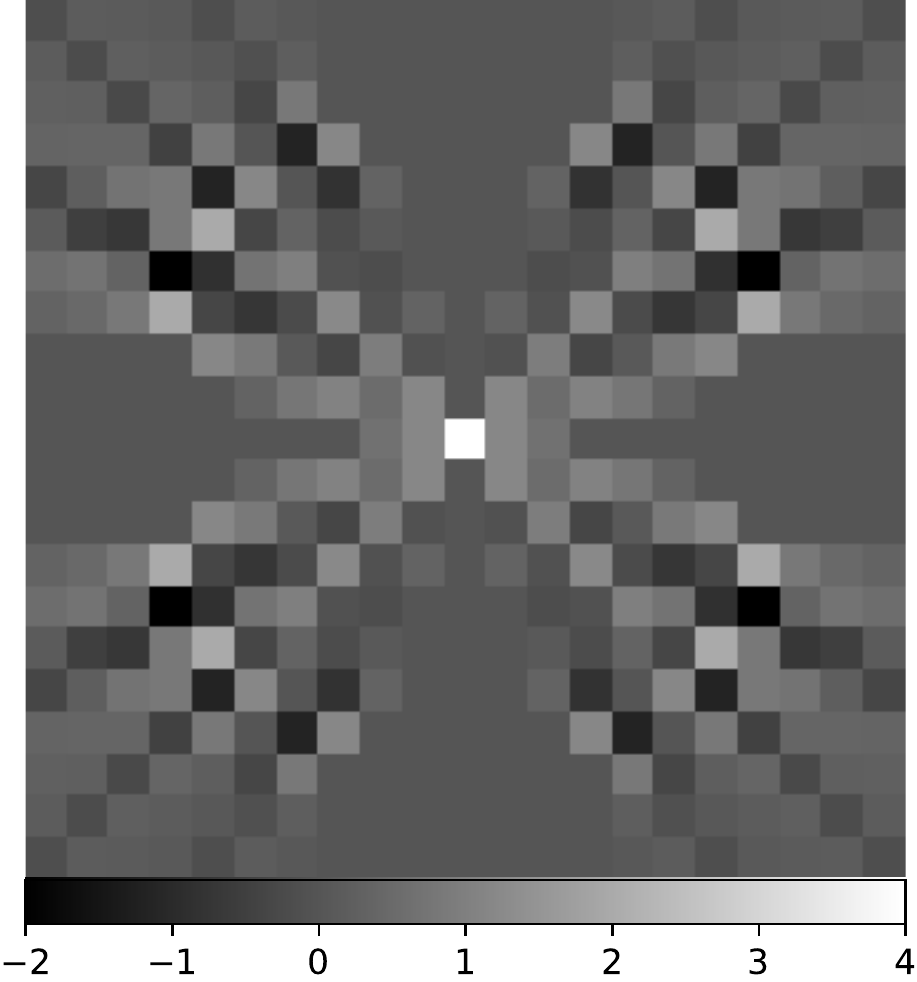}%
      \label{fig:higher_grads:d_xx}
    } &
    \subfloat[{$D^{xy}\tilde{\phi}$}]{
      \includegraphics[width=0.15\linewidth]{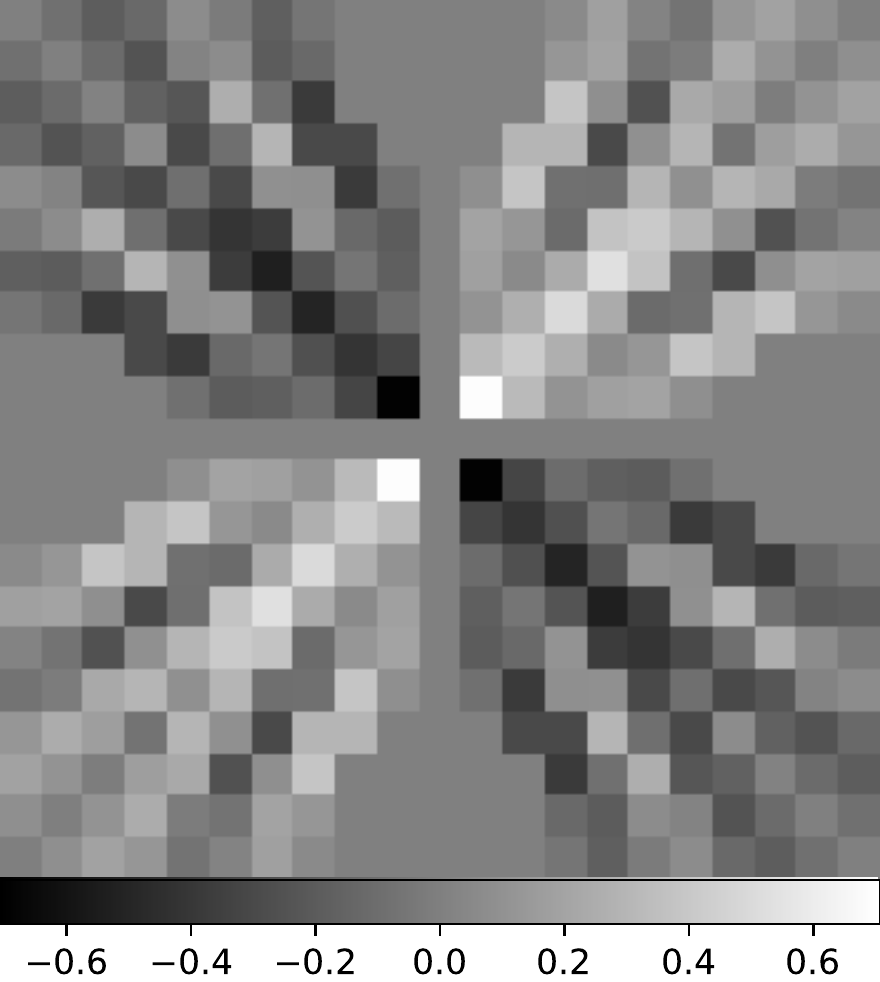}%
      \label{fig:higher_grads:d_xy}
    } &
    \subfloat[{$\Delta \tilde{\phi}$}]{
      \includegraphics[width=0.15\linewidth]{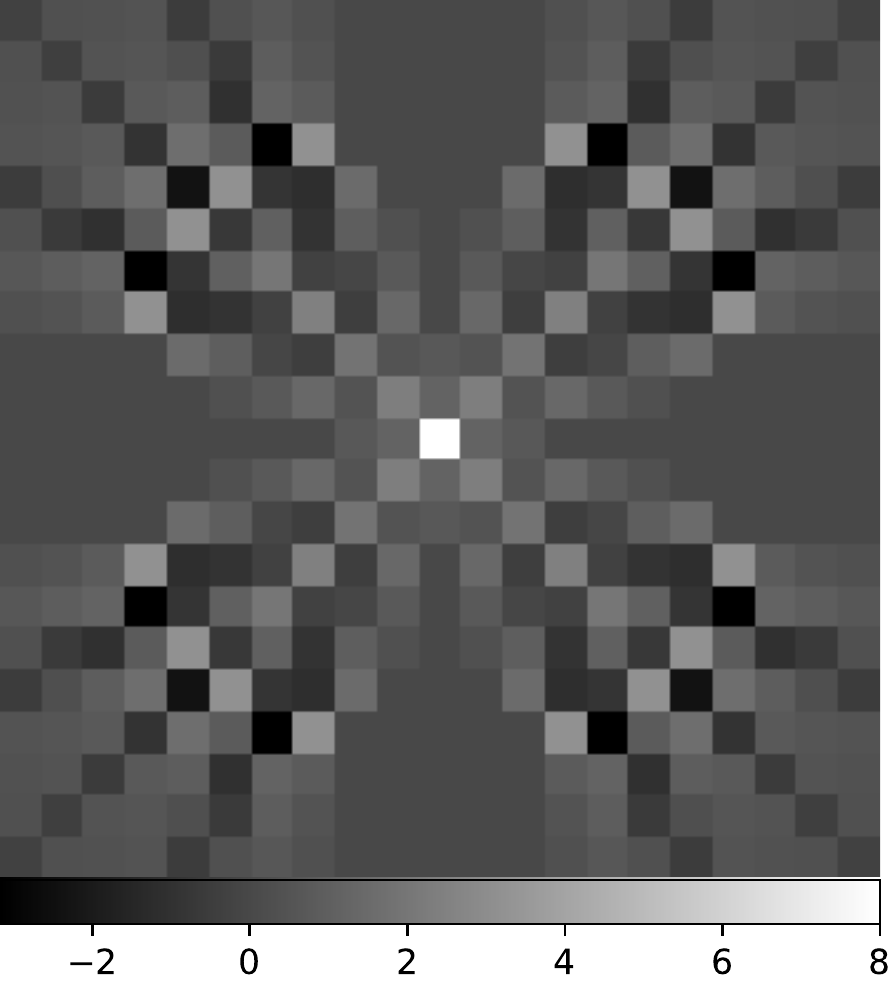}%
      \label{fig:higher_grads:d_lap}
    } &
    \subfloat[{$\tilde{\kappa}$}]{
      \includegraphics[width=0.15\linewidth]{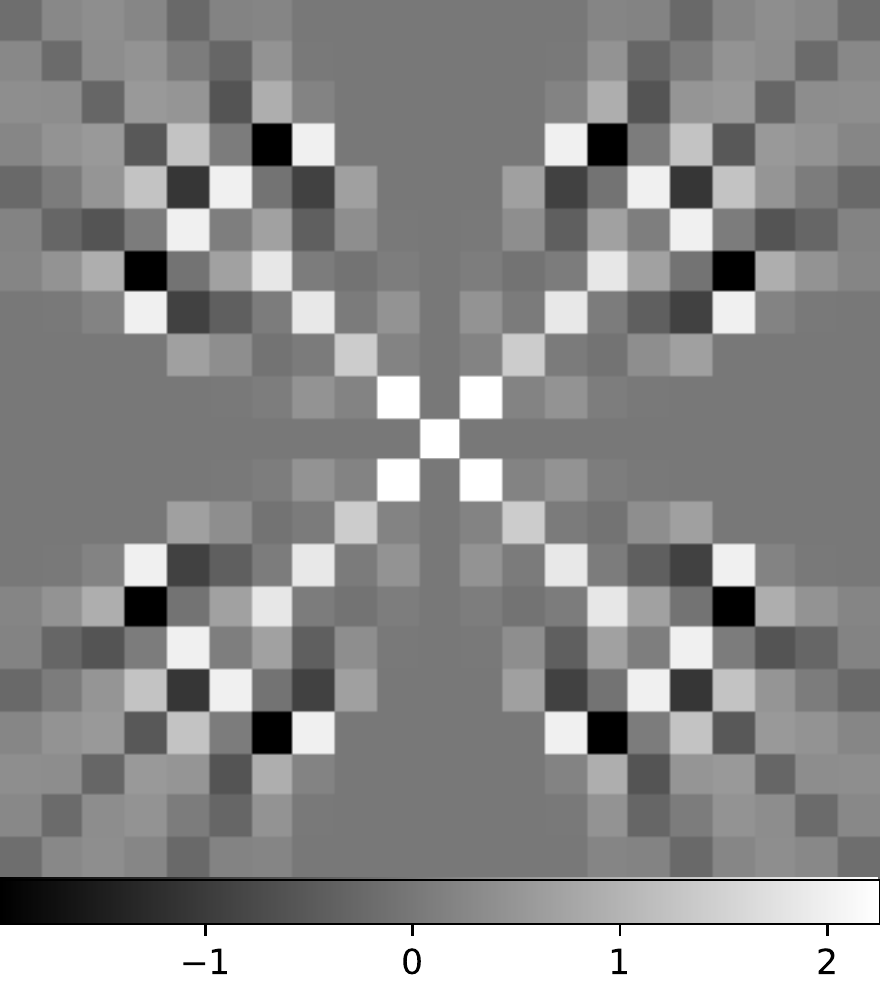}%
      \label{fig:higher_grads:d_curve}
    } \\
    &
    \subfloat[Error]{
      \includegraphics[width=0.15\linewidth]{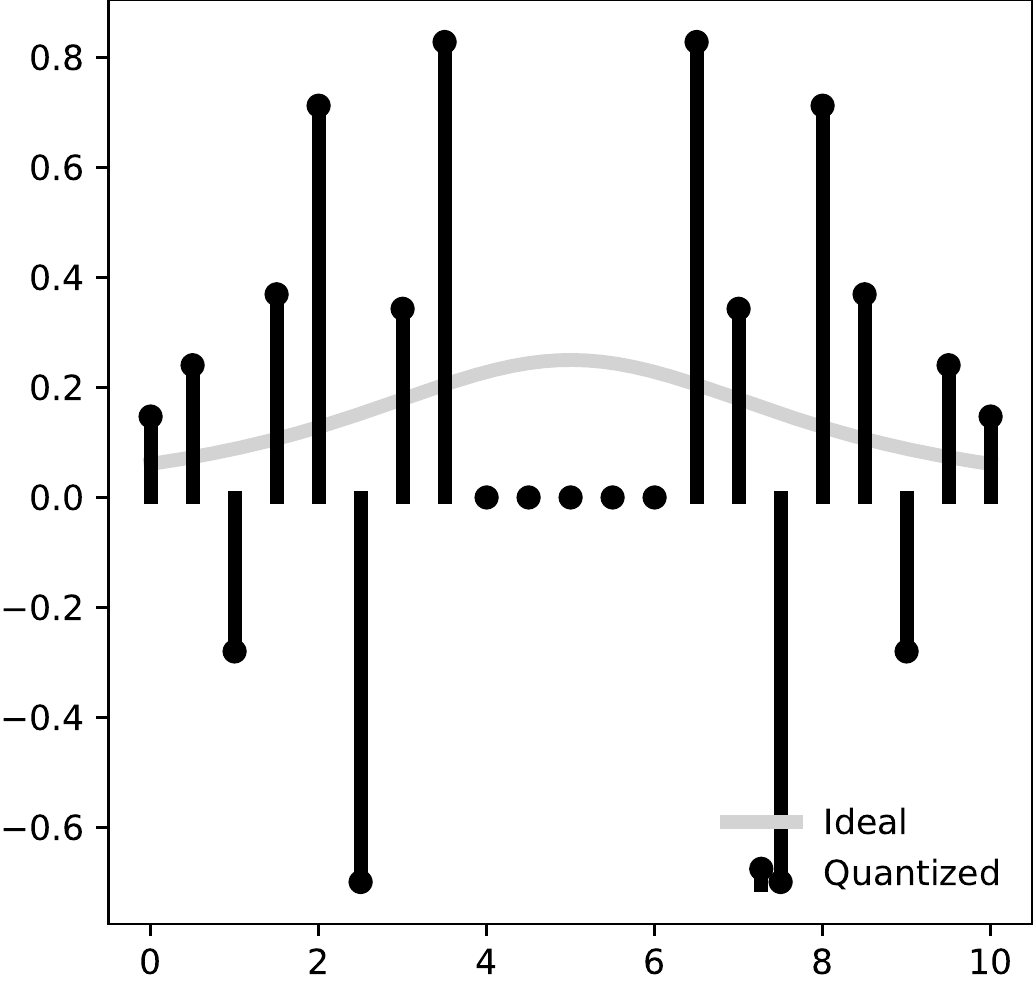}%
      \label{fig:higher_grads:error_xx}
    } &
    \subfloat[Error]{
      \includegraphics[width=0.15\linewidth]{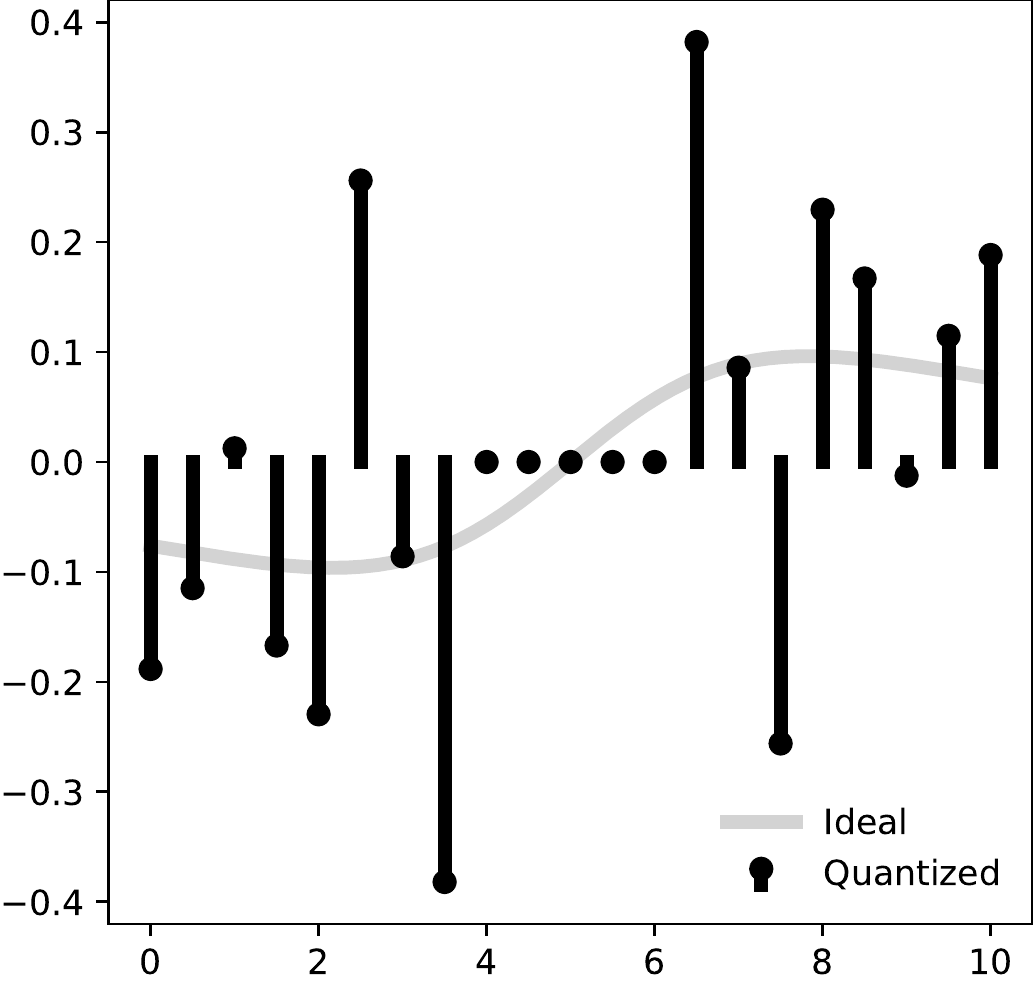}%
      \label{fig:higher_grads:error_xy}
    } &
    \subfloat[Error]{
      \includegraphics[width=0.15\linewidth]{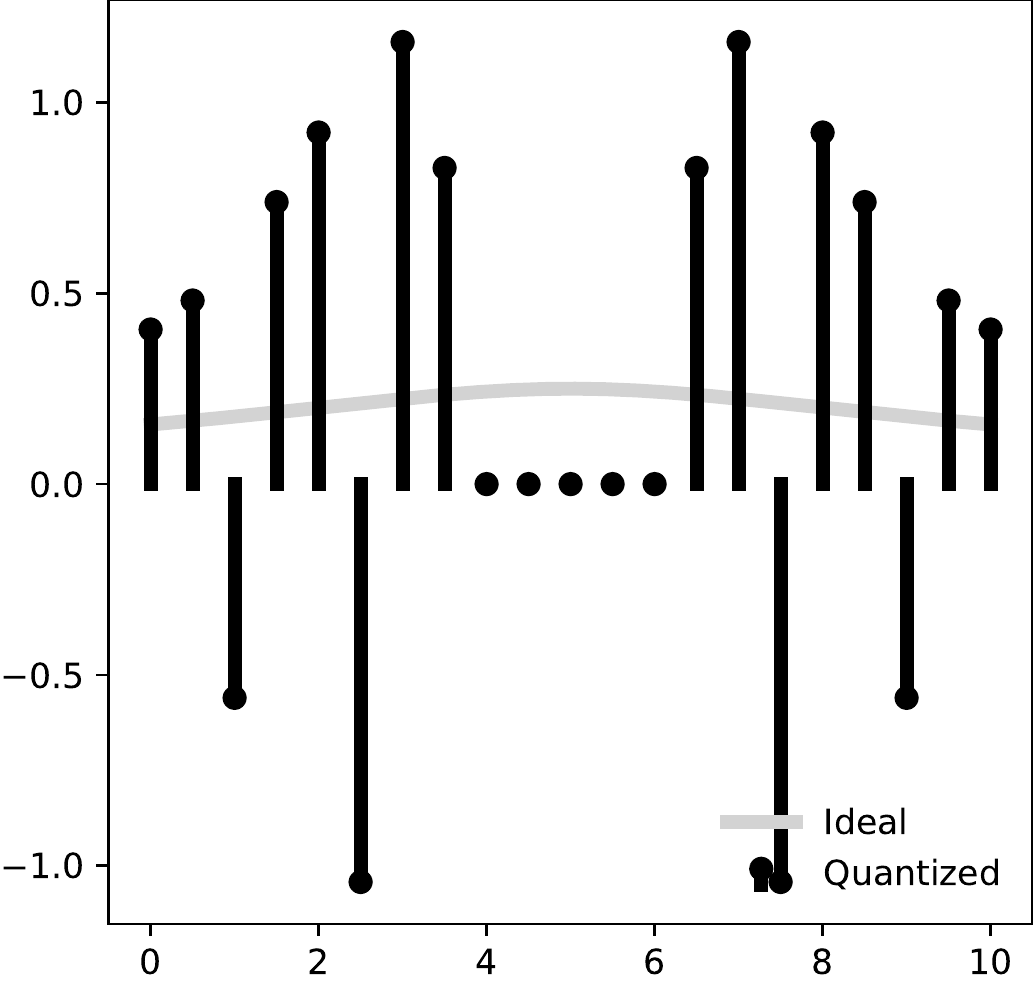}%
      \label{fig:higher_grads:error_lap}
    } &
    \subfloat[Error]{
      \includegraphics[width=0.15\linewidth]{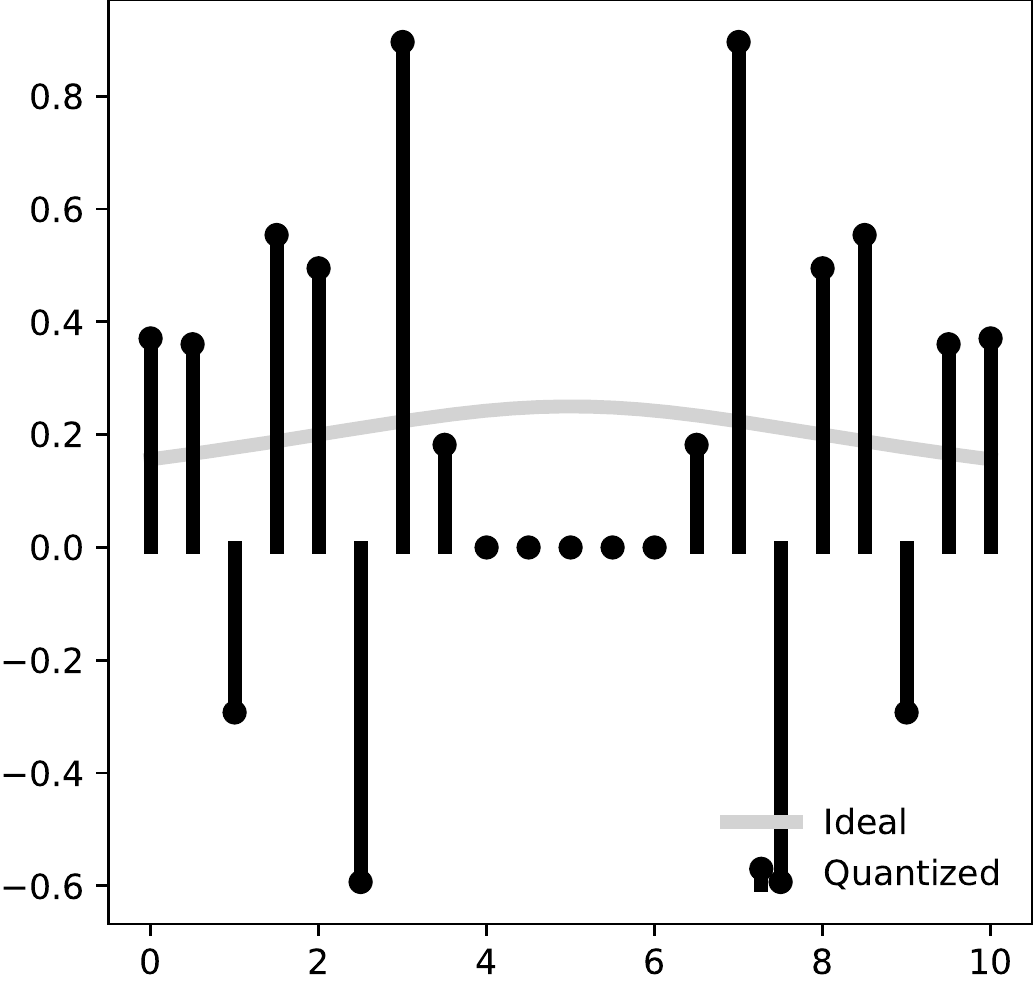}%
      \label{fig:higher_grads:error_lap}
    }
  \end{tabular}
  \caption{Errors in the finite difference approximation of second derivatives and functions thereof from a quantized signed distance signal. Error signal plotted along the path $y=1.0$. Column (b,g,k) demonstrate second derivatives along \textit{x}. Column (c,h,l) demonstrate the mixed derivative. Column (d, i, m) demonstrate the laplacian. Column (e, j, n) demonstrate the curvature.}
  \label{fig:higher_grads}
\end{figure*}

Figure~\ref{fig:higher_grads} plots errors in the second derivative, mixed derivative, Laplacian, and curvature of the signed distance signal.
Large errors are seen in all values, sometimes amplified and sometimes flat.
Structure is seen in the gradient along the diagonals.
Higher order derivatives experience banding and extremely large errors from quantization.

\subsection{Real Data}
\begin{figure*}[h]
  \centering
  \begin{tabular}{ccccc}
    \subfloat[$I$]{
      \includegraphics[width=0.17\linewidth]{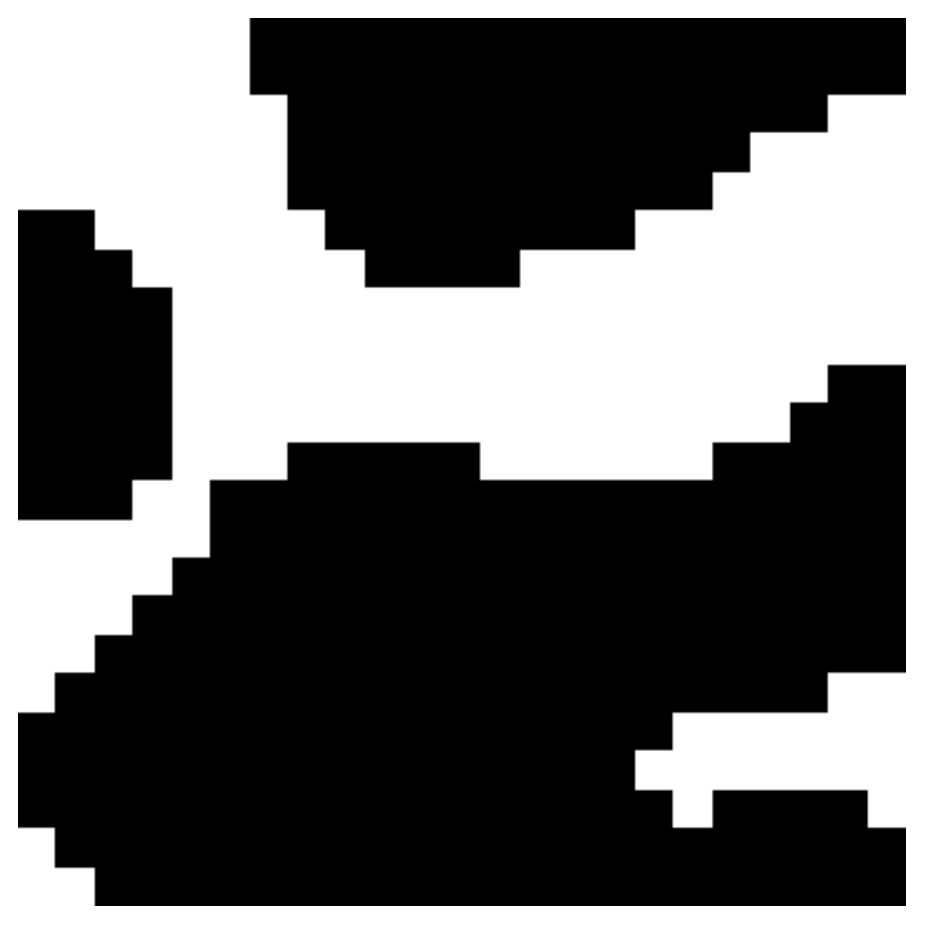}%
      \label{fig:real-data:voronoi}
    } &
    \subfloat[Voronoi]{
      \includegraphics[width=0.17\linewidth]{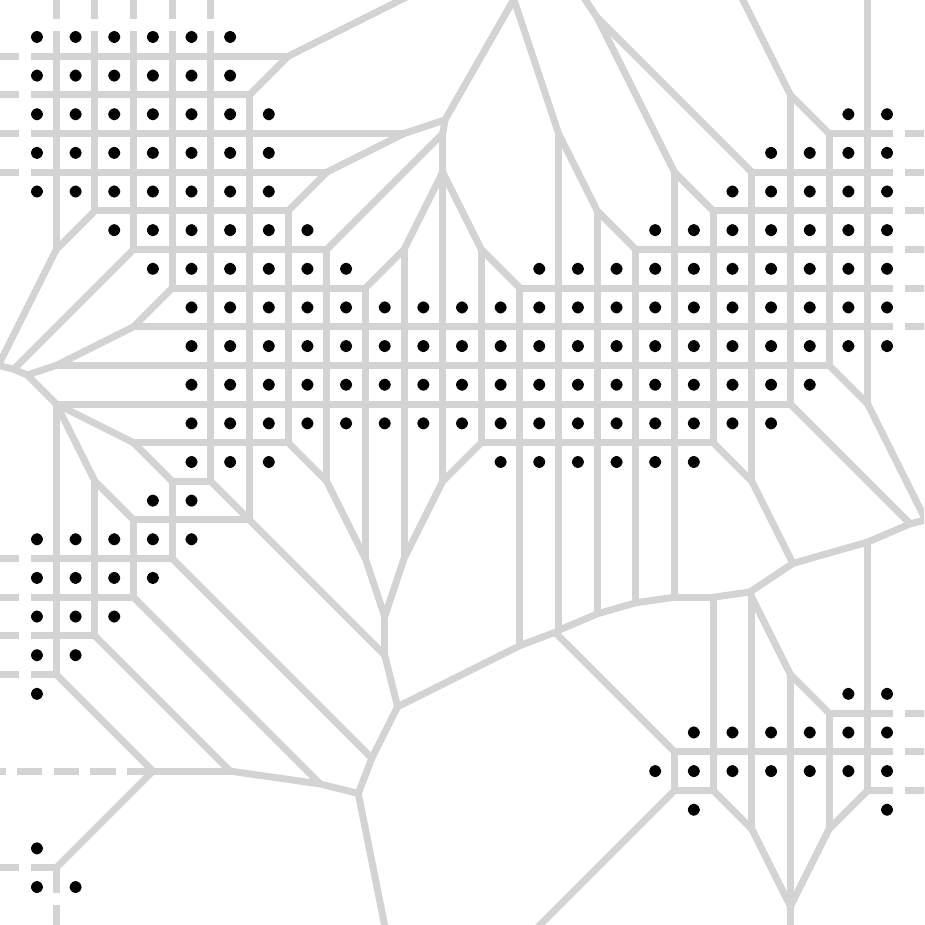}%
      \label{fig:real-data:phi}
    } &
    \subfloat[{$\tilde{\phi}$}]{
      \includegraphics[width=0.17\linewidth]{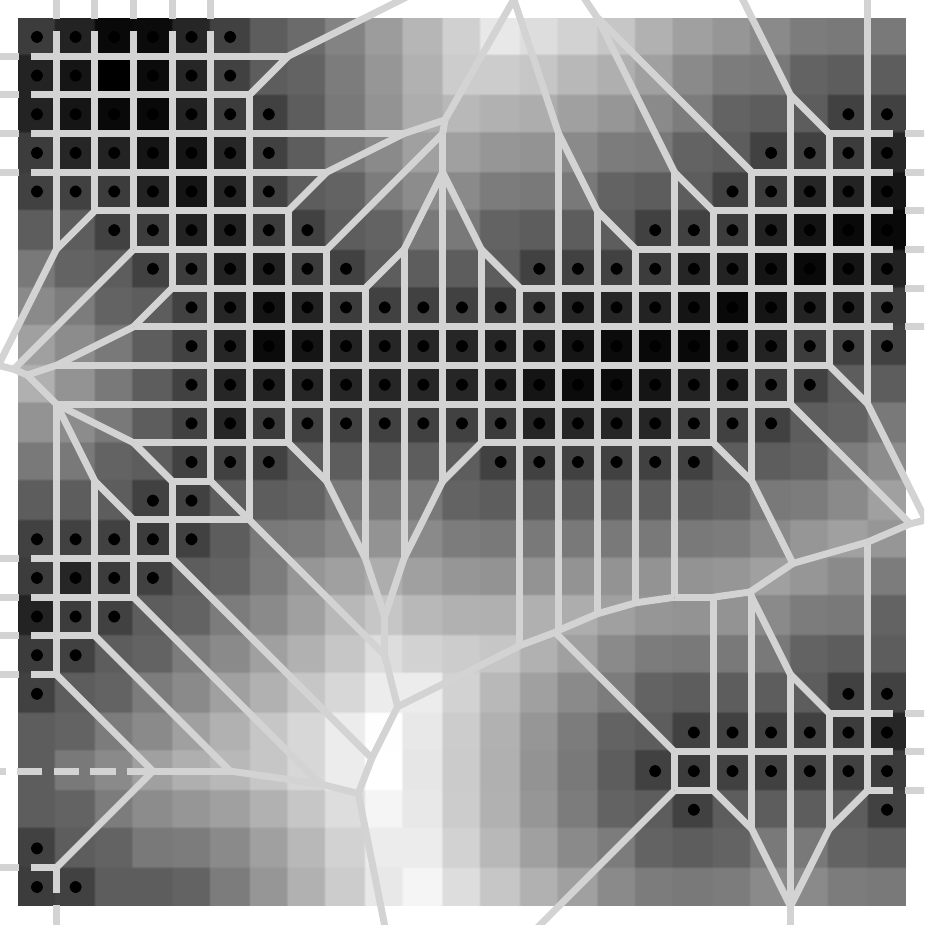}%
      \label{fig:real-data:phi}
    } &
    \subfloat[{$D^{0y} \tilde{\phi}$}]{
      \includegraphics[width=0.17\linewidth]{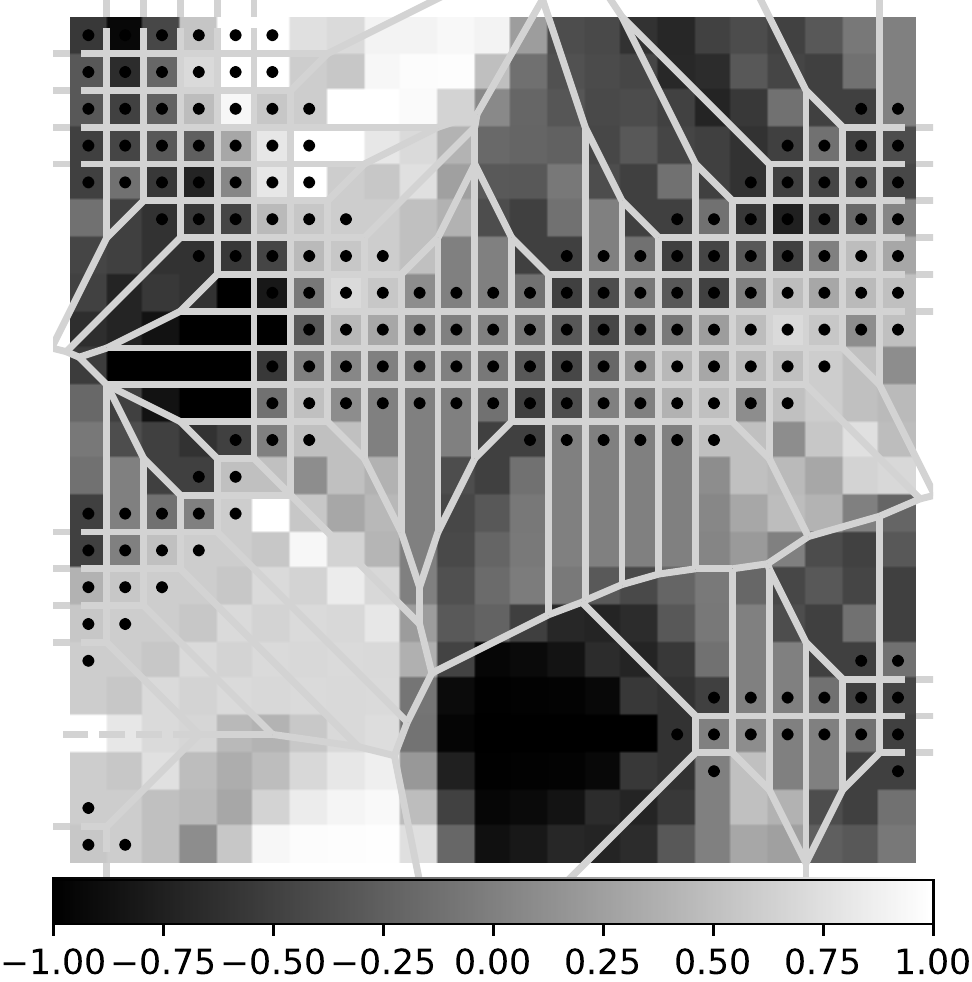}%
      \label{fig:real-data:d_x}
    } &
    \subfloat[{$\lVert D^{0} \tilde{\phi} \rVert$}]{
      \includegraphics[width=0.17\linewidth]{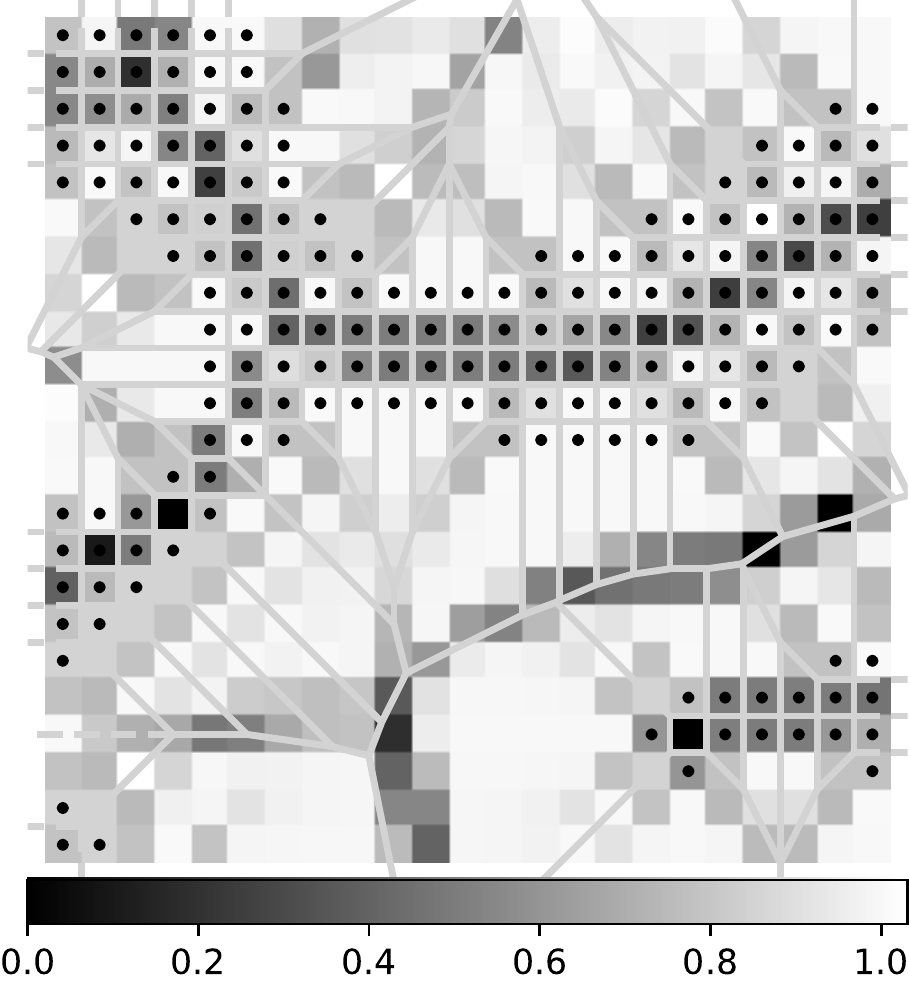}%
      \label{fig:real-data:mag_grad}
    }
  \end{tabular}
  \caption{Voronoi diagram and gradients for a real trabecular bone image.}
  \label{fig:real-data}
\end{figure*}

Finally, this result is experimentally investigated using real data.
A $24$ x $24$ 2D slice of canine trabecular bone is used (isotropic nominal resolution of $34 \: \mu m$).
The image, Voronoi diagram, signed distance transform, and gradient computations are visualized in Figure~\ref{fig:real-data}.
Banding artifacts are seen in the y-gradient while the magnitude gradient is near $+1$.
Note that the magnitude gradient approaches zero near the medial axis of the object, consistent with the central difference operator.

\subsection{Analysis and Synthesis}
The problem is now summarized.
Based on Equation~\ref{eqn:quant-metric}, the computed finite difference can be computed.
\begin{eqnarray}
  D^{+} \phi &=& \frac{h g(p,0) - h g(q,0)}{h} \\
  \label{eqn:finite_diff_quant}
  &=& g(p,0) - g(q,0)
\end{eqnarray}
This can be visualized by laying the finite difference stencil onto the origin of Figure~\ref{fig:discretized_metric}.
This implies that the finite differences of the quantized signed distance transform takes on a finite set of values given by the difference of the metrics between the points.
\begin{equation}
  D \phi \in \left\{g(p, 0) - g(q, 0) \given p,q \in Z^n\right\}
\end{equation}
Of importance is that the finite difference approximation of the first order derivative (Equation~\ref{eqn:finite_diff_quant}) does not depend on the sample period, $h$, because the underlying signal is quantized to multiples of $h$.
This result is independent of the order of the finite difference approximation.
That is, only once the stencil extends beyond parallel Voronoi edges will the derivative not be flat.
The required size of the stencil could be arbitrarily large.
The origin of the error is that the binary image is flat compared to the unsampled signal.

A similar analysis can be applied to higher derivatives, which are seen to vary with some power of $h$.
However, if the first derivative is flat, so will be the second derivative.
Where the higher derivatives are not flat, quantization errors amplify into higher derivative errors.
As a result, the error cannot be changed by modifying the sampling period.
Whether the structure has 2 or 2000 samples across its width, the error will persist.

Finally, the gradient magnitude is also discretized.
\begin{equation}
  \lVert D \phi \rVert  \in \left\{\sqrt{\sum_i^n \left(g(p_i, 0) - g(q_i, 0)\right)^2} \:\middle|\: p_i, q_i \in Z^n\right\}
\end{equation}
However, the quantization error is small due to two factors.
First, the operator is the square-root of the sum-of-squares, taking on finite but many values.
Second, while the gradient in one direction may be flat, the gradient in another is very close to one.
The result is a good estimation of gradient magnitude.

In summary, quantization leads to artifacts in finite difference calculations. 
Gradients are flat at points where the Voronoi cell is aligned with the finite difference stencil.
Higher derivatives are either flat as well or end up amplifying the quantization noise.
This error is independent of sample period.

\section{Handling Quantization }
Attention is now placed on handling the structured artifacts introduced by the quantization of the distance transform.
Prior analysis made no assumptions on the metric $g$ and thus is a problem for any distance transform.
For example, the Chamfer approximation to a Euclidian metric~\cite{montanari1968method} will also suffer artifacts due to quantization.
An important feature of the artifact is that obvious attempts to correct it are challenging.
For instance, any function of the embedding $f(\phi)$ will also be quantized.
Furthermore, by the chain rule, the spatial gradients will be flat.

The most obvious solution to quantization is to perform distance transforms based on a different representation.
Examples include distances from meshes~\cite{baerentzen2005signed}, parametric curves~\cite{pottmann2003geometry}, or other vector objects.
However, this is not possible in many applications where only the binary image is given.

Many possible methods could be designed for handling quantization.
Examples include weighted distance transforms such as replacing the infimum with the equation of the $\ell^p$ norm for negative values of $p$~\cite{brunet2016generalized}\footnote{This is not a valid norm but is a valid smooth approximation to minimum} or computing distance maps from greyscale signals~\cite{kimmel1996sub}.
In cases were some error in representation is allowed, these are exceptionally fast and robust algorithms.
However, it is not guaranteed that given a binary image, the properties of the embedding are preserved.

\subsection{Required Properties}
The first is that the Heaviside function should recover the true signal.
\begin{equation}
  H(-\phi) = I
\end{equation}
This can be equivalently stated as not allowing $\phi$ to change sign.
The second is that the magnitude gradient is $+1$ everywhere except at the medial axis.
\begin{equation}
  \lVert \nabla \phi \rVert = +1
\end{equation}

\subsection{Proposed Methods}
As is commonly done to handle quantization errors, dithering is applied to remove dependence between the quantized samples~\cite{lipshitz1992quantization}.
First, noise will be added such that the sign of the distance transform does not change.
Second, a reinitialization algorithm is performed to smooth the noise.
Importantly, the reinitialization algorithm preserves the sign of the distance transform while gradually shifting the magnitude gradient to $+1$.
The limitations of the method are that the procedure is extremely computationally intensive and adds a stochastic process to analysis.
However, it guarantees that the embedding does not change sign and can guarantee the magnitude gradient is within some convergence criterion of $+1$.

\subsubsection{Model of Noise}
\begin{figure*}[h]
  \centering
  \begin{tabular}{cccccc}
    \subfloat[$\phi$]{
      \includegraphics[width=0.12\linewidth]{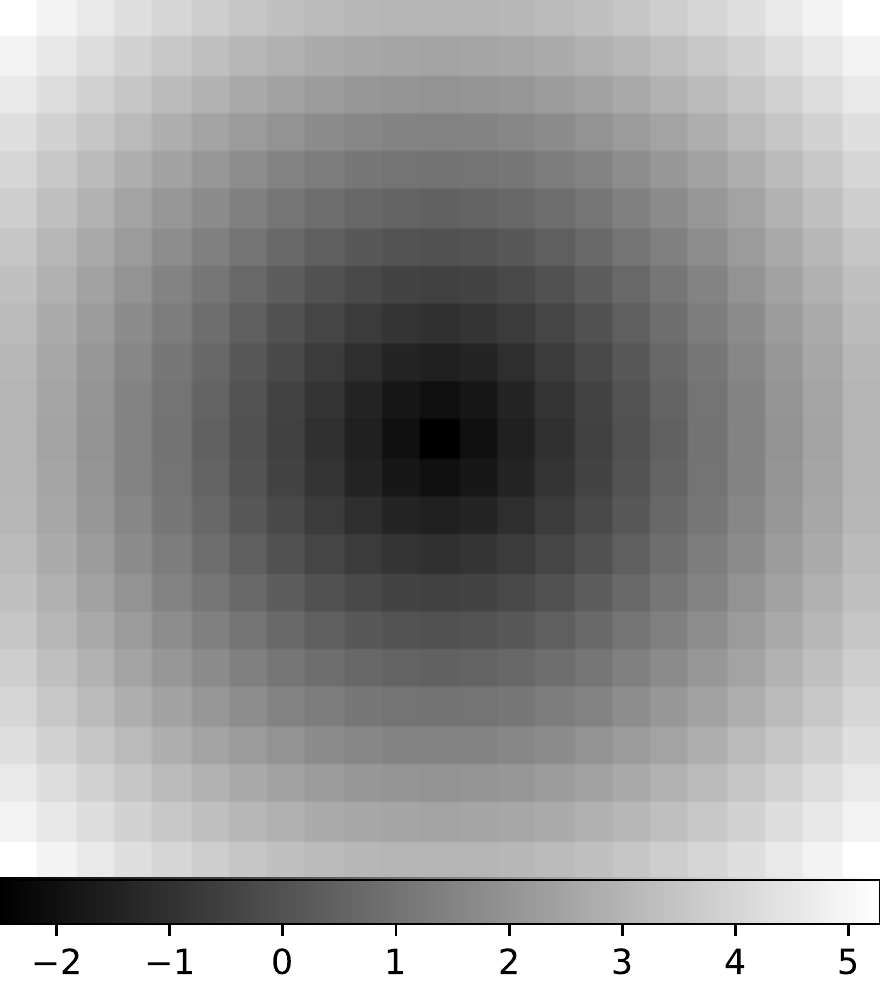}%
      \label{fig:dither:phi}
    } &
    \subfloat[{$I$}]{
      \includegraphics[width=0.13\linewidth]{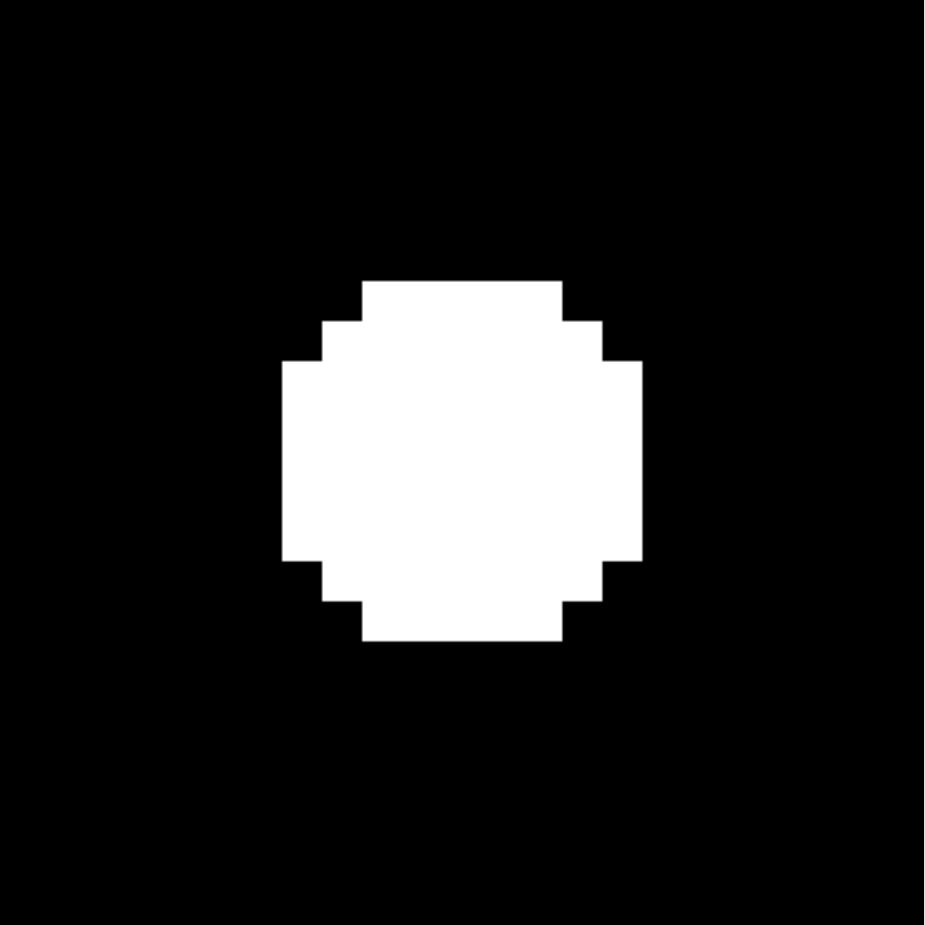}%
      \label{fig:dither:I}
    } &
    \subfloat[$\tilde{\phi}$]{
      \includegraphics[width=0.12\linewidth]{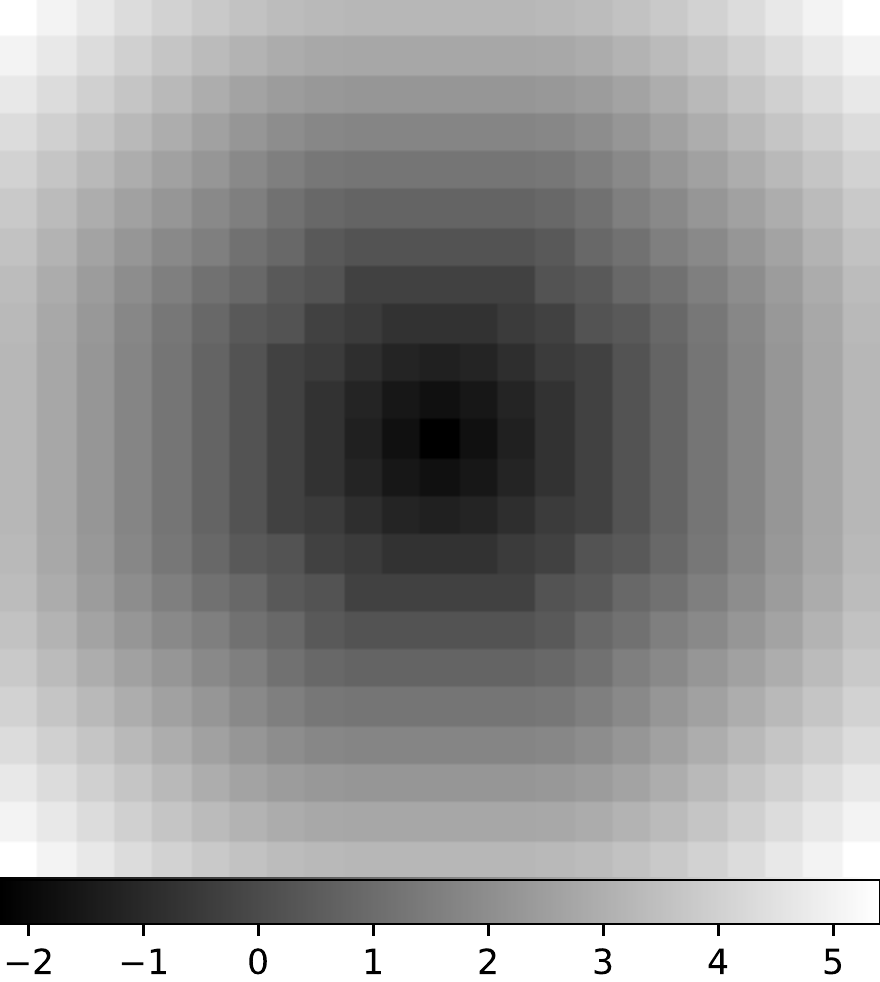}%
      \label{fig:dither:phi_quant}
    } &
    \subfloat[{$\hat{\phi}$}]{
      \includegraphics[width=0.12\linewidth]{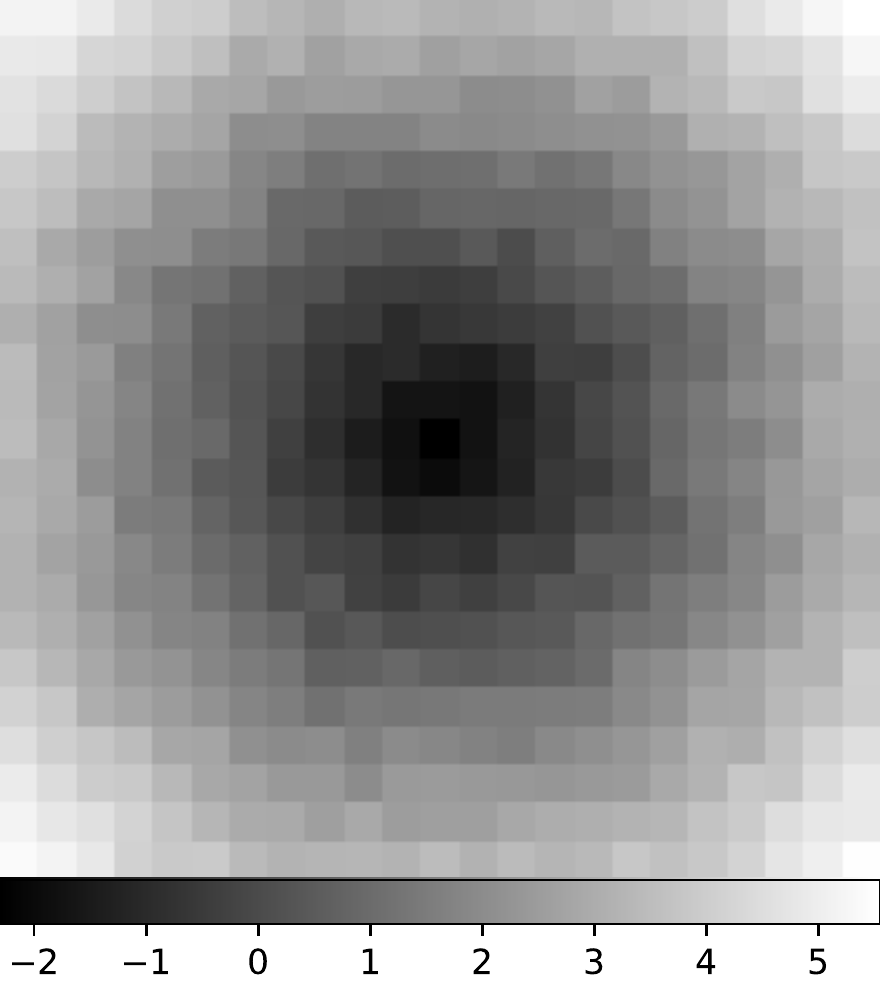}%
      \label{fig:dither:dither}
    } &
    \subfloat[$A\cdot\mathcal{U}$]{
      \includegraphics[width=0.12\linewidth]{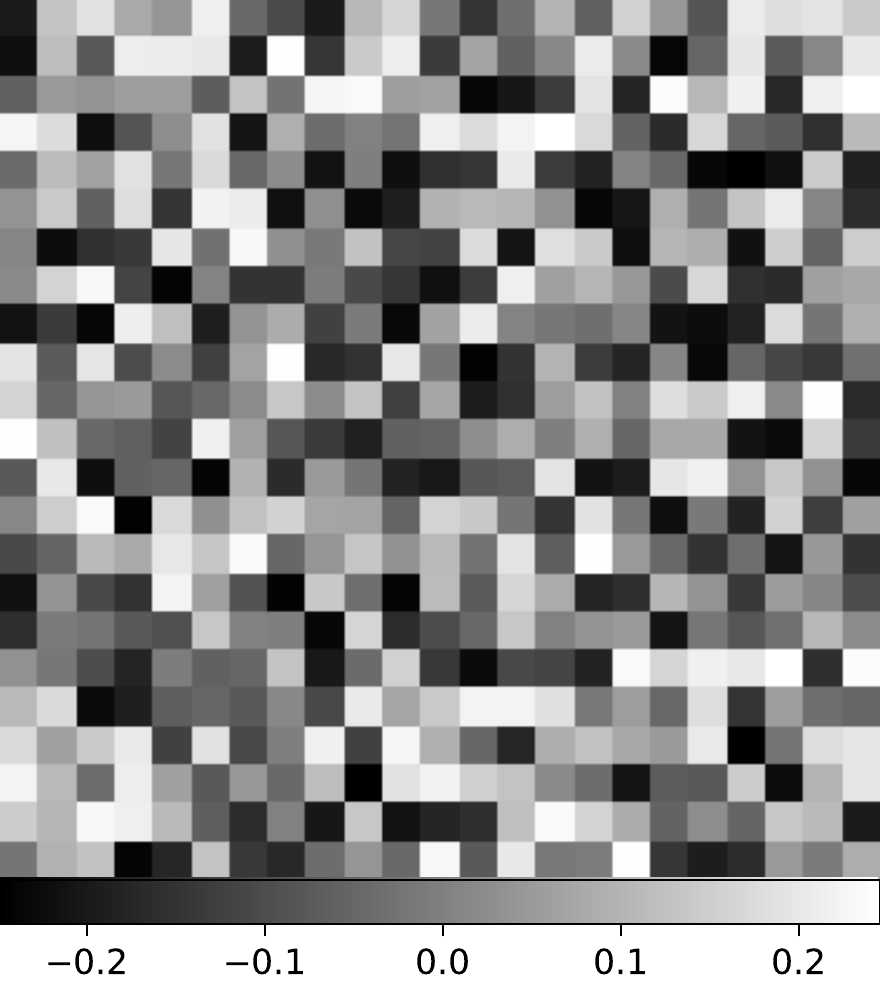}%
      \label{fig:dither:noise}
    } &
    \subfloat[{$H(-\hat{\phi})$}]{
      \includegraphics[width=0.13\linewidth]{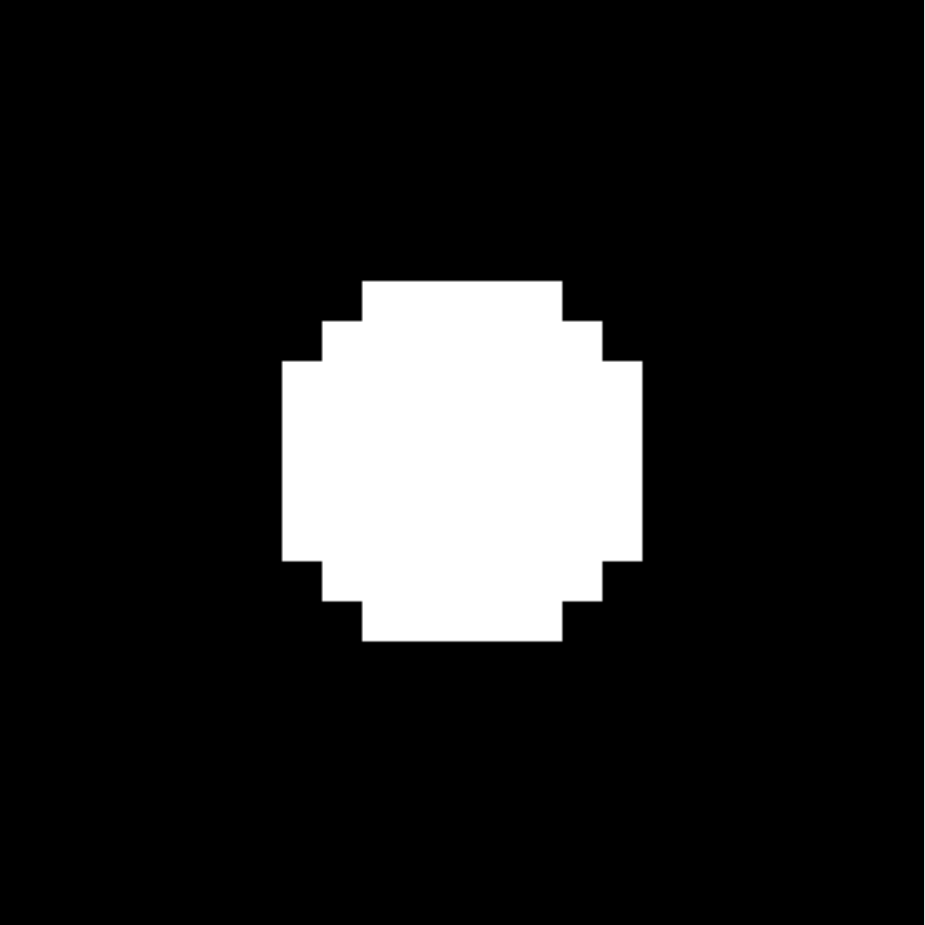}%
      \label{fig:dither:h_dither}
    }
  \end{tabular}
  \caption{Dithering of the quantized signal ($h = 0.5$, $\alpha = 2.0$). (\ref{fig:dither:h_dither}) The binary image does not change.}
  \label{fig:dither}
\end{figure*}

First, noise is added to the signed distance map.
Noise is drawn from a uniform distribution between $-1$ and $+1$ and added to the image.
The dithered signal is represented by a hat, $\hat{\phi}$.
However, the amplitude is modified such that the sign of the distance transform does not change
\begin{equation}
  \hat{\phi}(x) = \tilde\phi(x) + A(x) \cdot \mathcal{U}(-1, 1)
\end{equation}
where $\mathcal{U}$ is the uniform distribution and $A$ the amplitude given by
\begin{equation}
  A(x) = \min\left(\frac{h}{\alpha}, \lvert\tilde{\phi}(x)\rvert\right)
\end{equation}
where $\alpha > 1$ controls the amplitude relative to the sampling period.
Since the signal is quantized to values of $h$, selecting $\alpha$ smaller than one adds noise larger than the quantization error.
On the other hand, selecting $\alpha$ too large will not add enough noise to sufficiently dither the image.
Given that the noise amplitude is clipped to the magnitude of the signal, it is impossible for the dithered signal to change sign.
An example of a dithered signal is given in Figure~\ref{fig:dither}.

\subsubsection{Reinitialization}
Next, the noise is removed.
Here, a reinitialization algorithm is used for smoothing out the noise~\cite{sussman1994level,peng1999pde}.
This is done by solving a partial differential equation that moves the embedding around until its magnitude gradient is $1$ everywhere:
\begin{equation}
  \label{eqn:reinitialization}
  \phi_t = \text{sgn}(\phi) \left(\lVert \nabla \phi \rVert - 1 \right)
\end{equation}
Here, $\phi_t$ is the partial derivative in time and $\text{sgn}\left(\cdot\right)$ is the sign function.

When implemented appropriately, this algorithm has some exceptional properties.
First, the algorithm works for very sharp surfaces such as would be seen in trabecular bone.
Second, the algorithm has fast convergence far away from the original surface as would be advantageous in transforms of large objects.
Finally, and most importantly, this algorithm does not change the sign of the distance transform during evolution.

Details on the implementation are available elsewhere~\cite{peng1999pde}.
For this work, Equation~\ref{eqn:reinitialization} is solved with a third-order total variation diminishing Runge-Kutta upwind scheme in time and a 5th order accurate weighted essentially non-oscillatory scheme in space~\cite{jiang2000weighted}.
A high order scheme was needed for accurate reinitialization.

\subsection{Quantifying Error}
Three errors are defined for measuring the accuracy of the model.
The first is the error in representation as the maximum difference between the Heaviside function and the binary image.
\begin{equation}
  \label{eqn:rep_error}
  e_{R} = \lVert H(-\phi) - I \rVert_\infty
\end{equation}
The second is the difference in the magnitude gradient from $1$
\begin{equation}
  \label{eqn:mag_grad_error}
  e_{MG} = \frac{1}{N} \lVert \lVert \nabla \phi \rVert - 1 \rVert_2
\end{equation}
where $N$ is the number of voxels in the image.
Lastly, the error between the finite difference gradient and ideal gradient is defined.
\begin{equation}
  \label{eqn:grad_error}
  e_{D} = \frac{1}{N} \lVert \lVert D \phi - \nabla \phi \rVert_2 \rVert_2
\end{equation}
Equation~\ref{eqn:grad_error} should be interpreted as first taking the $\ell^2$ norm at each pixel of the difference in gradient vectors and measuring this error across the image using the $\ell^2$ norm again.
Finite differences are computed using 4th order accurate central differences.
All errors are normalized to the size of the image so that they can be interpreted as if they were errors in a pixel.

\subsection{Experiments}

\subsubsection{Convergence of the Algorithm}
\begin{figure*}[h]
  \centering
  \begin{tabular}{ccc}
    \subfloat[{$\lVert \nabla \phi \rVert - 1$}]{
      \includegraphics[width=0.3\linewidth]{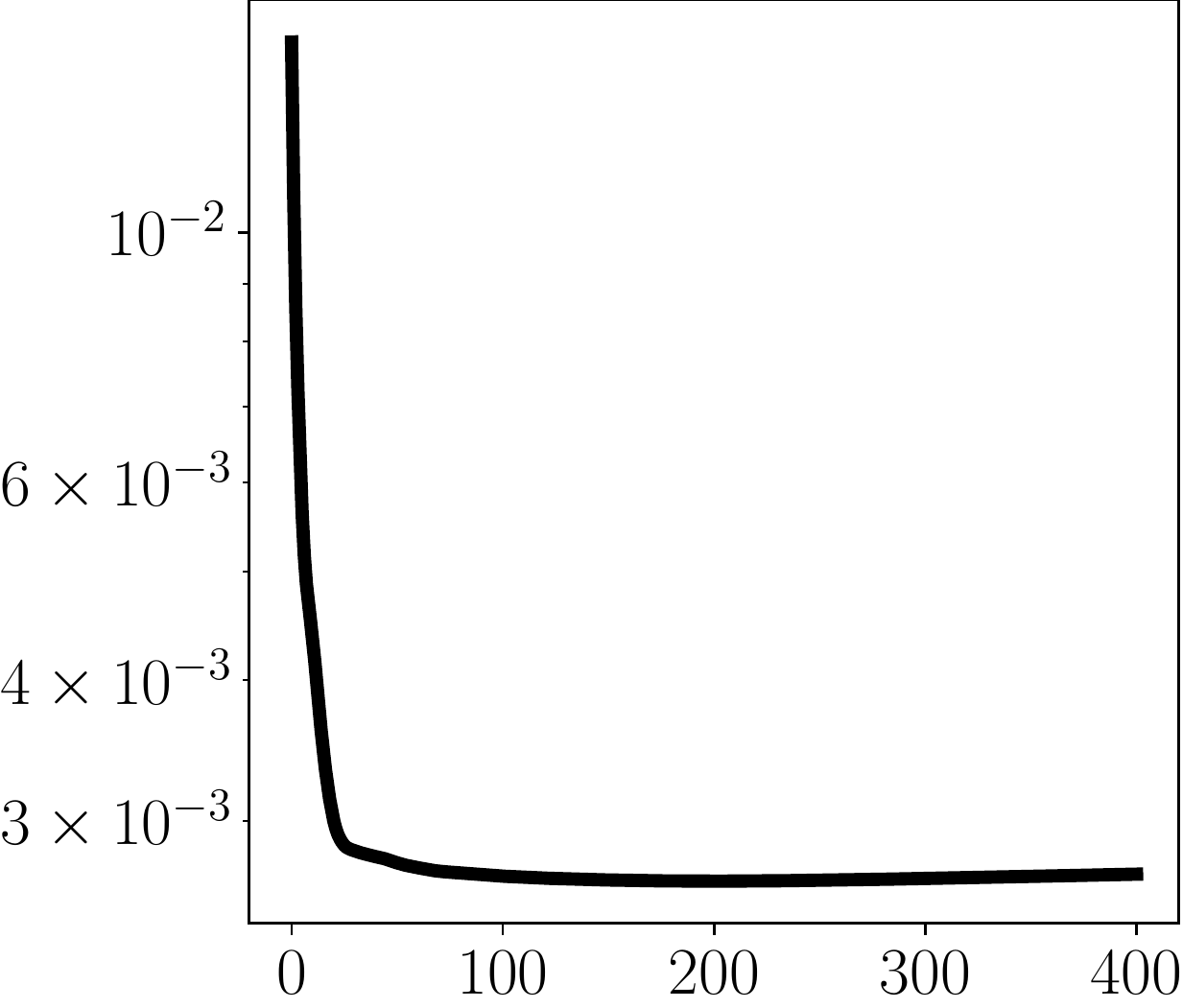}%
      \label{fig:convergence2:maggrad}
    } &
    \subfloat[{$\lVert D \phi - \nabla \phi\rVert$}]{
      \includegraphics[width=0.3\linewidth]{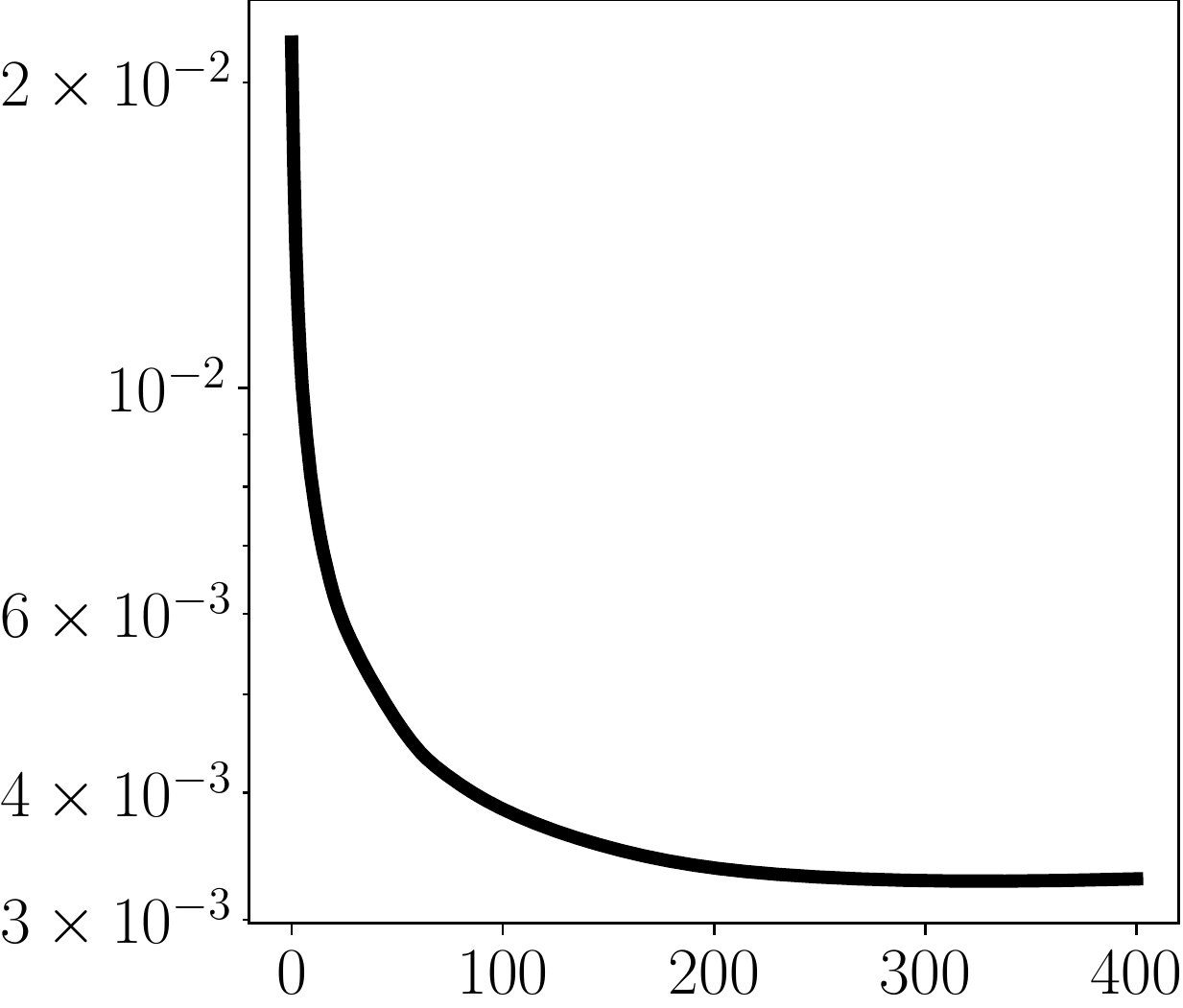}%
      \label{fig:convergence2:grad}
    } &
    \subfloat[{$H(-\phi) - I$}]{
      \includegraphics[width=0.3\linewidth]{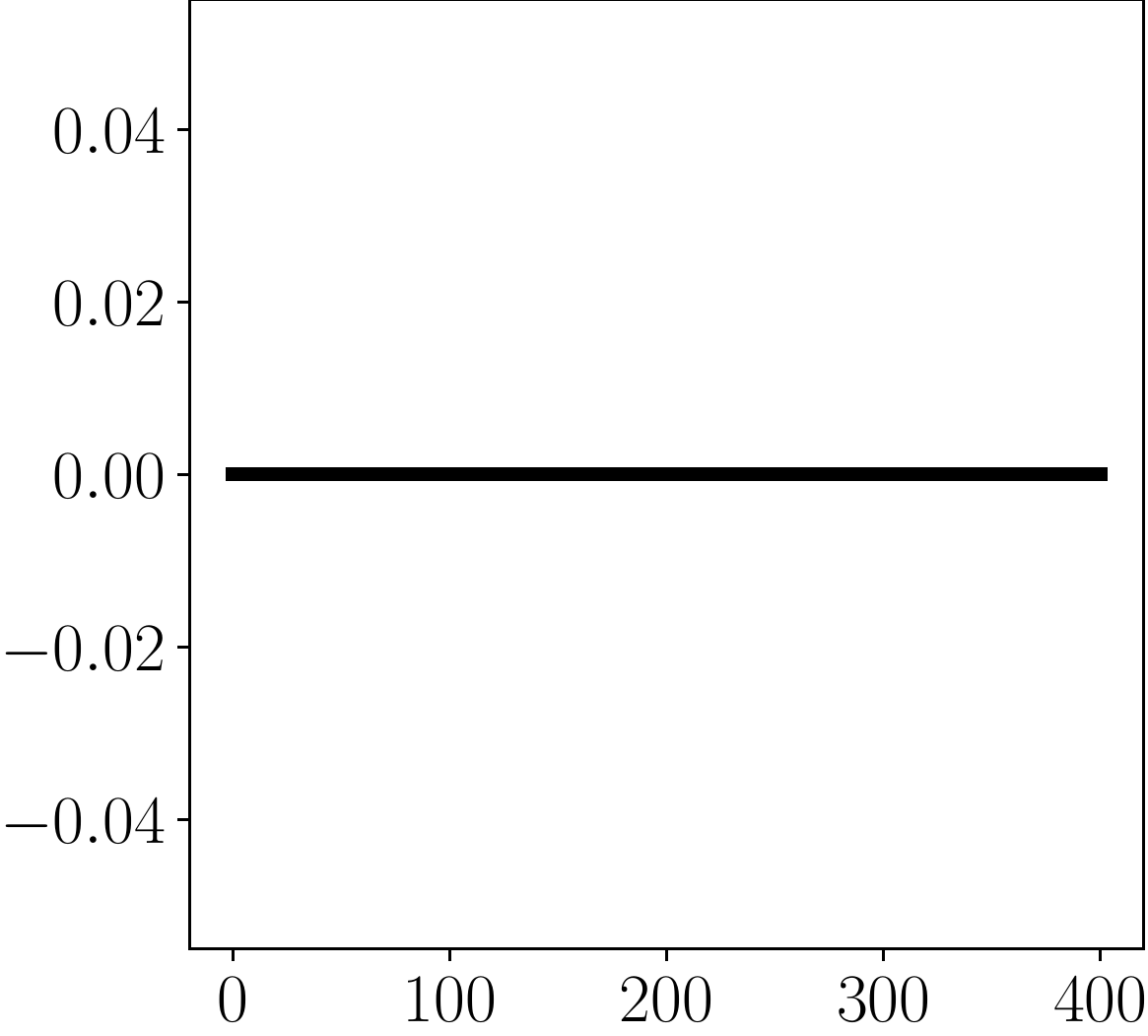}%
      \label{fig:convergence2:rep}
    }
  \end{tabular}
  \caption{Log-plots of convergence measures over 400 iterations ($h=0.5$, $\alpha = 2.0$). The representation error (\ref{fig:convergence2:rep}) is zero.}
  \label{fig:convergence2}
\end{figure*}

\begin{figure}[h]
  \centering
  \begin{tabular}{ccc}
    \subfloat[{$\tilde{\phi}$}]{
      \includegraphics[width=0.25\linewidth]{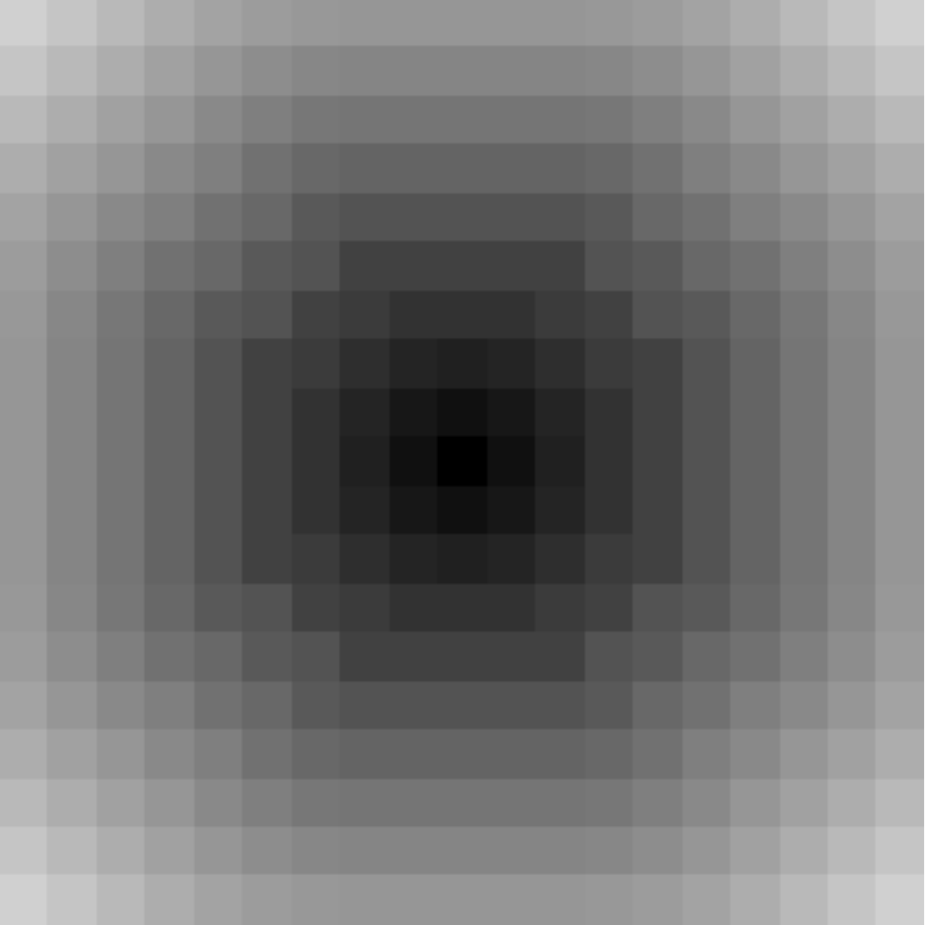}%
      \label{fig:convergence:nonoise:phi}
    } &
    \subfloat[{$|D\tilde{\phi}|$}]{
      \includegraphics[width=0.25\linewidth]{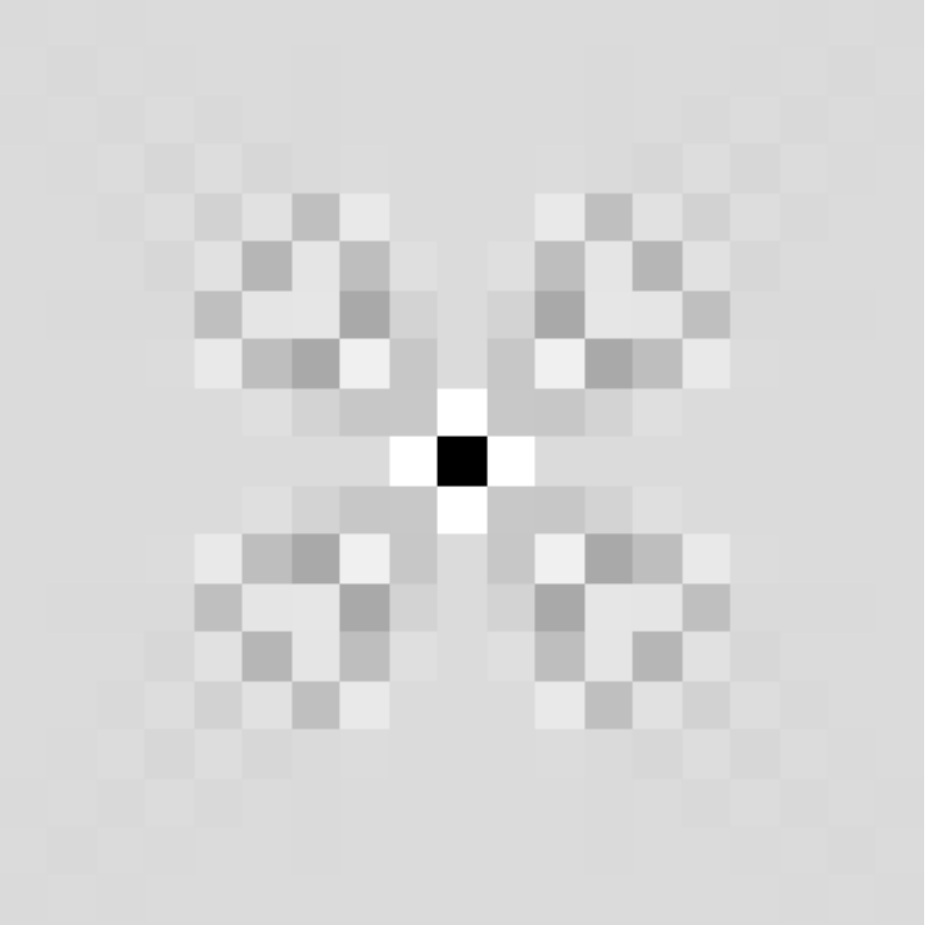}%
      \label{fig:convergence:nonoise:maggrad}
    } &
    \subfloat[{$D^x \tilde{\phi}$}]{
      \includegraphics[width=0.25\linewidth]{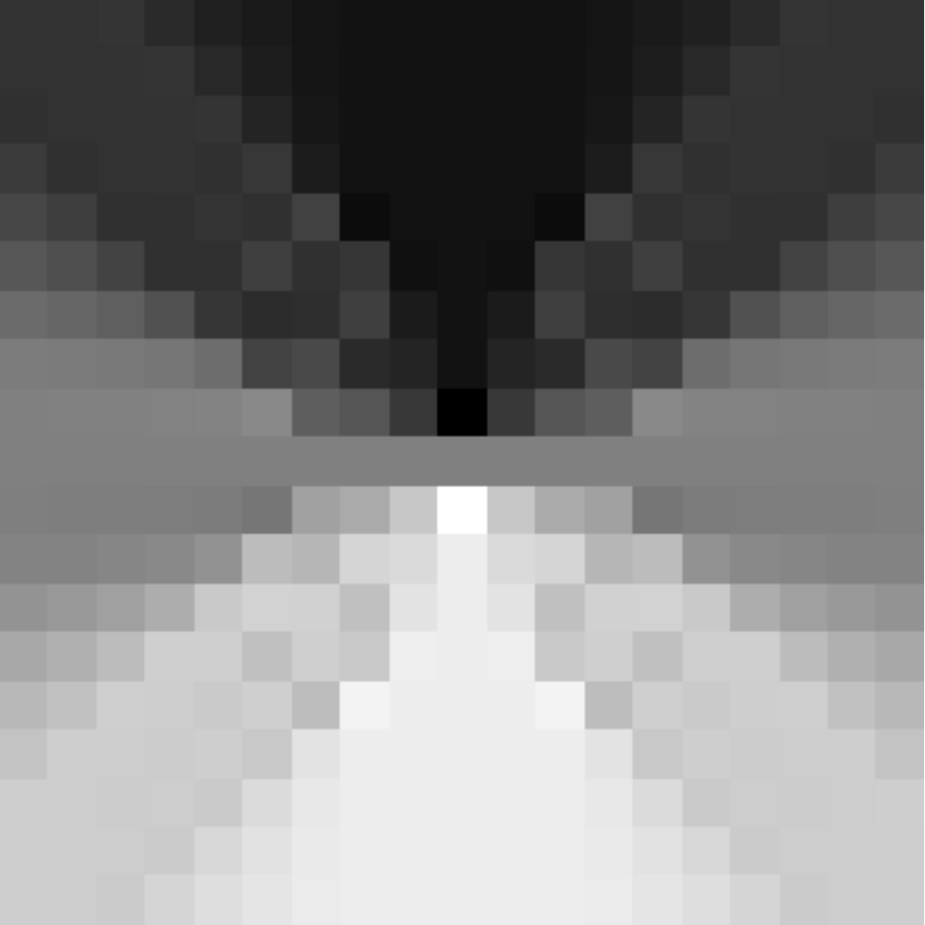}%
      \label{fig:convergence:nonoise:dx}
    } \\
    \subfloat[{$\hat{\phi}$}]{
      \includegraphics[width=0.25\linewidth]{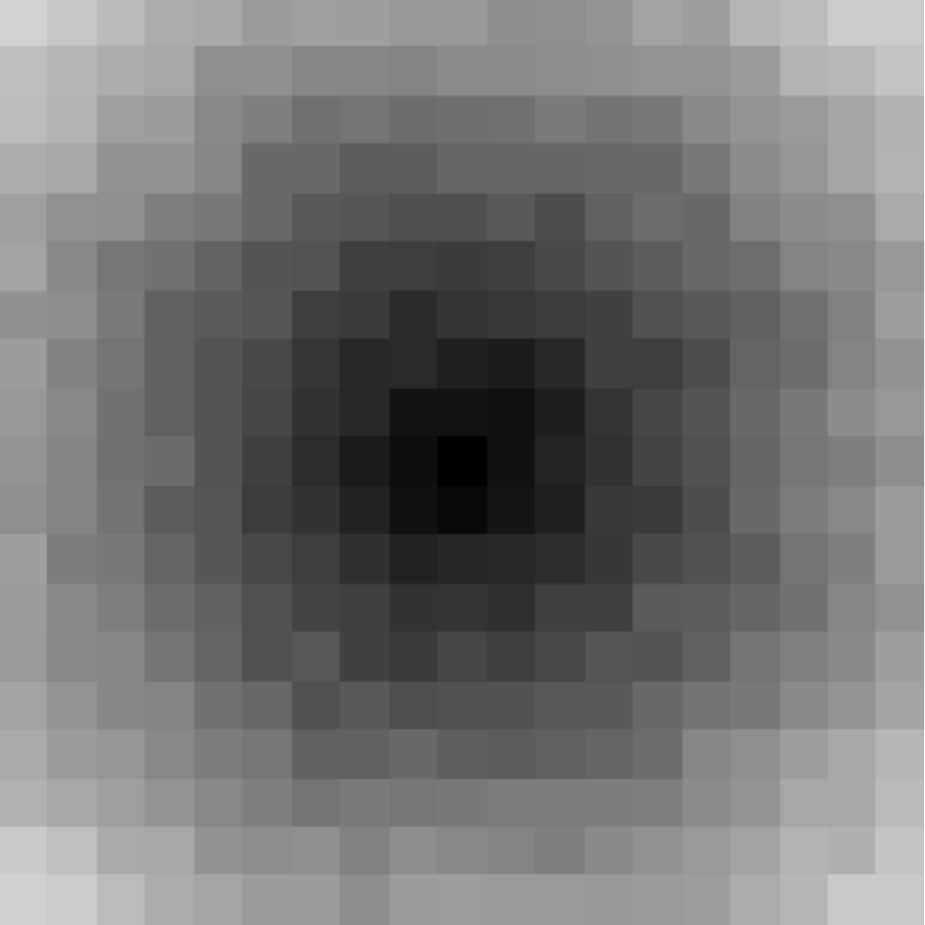}%
      \label{fig:convergence:noise:phi:0}
    } &
    \subfloat[{$|D\hat{\phi}|$}]{
      \includegraphics[width=0.25\linewidth]{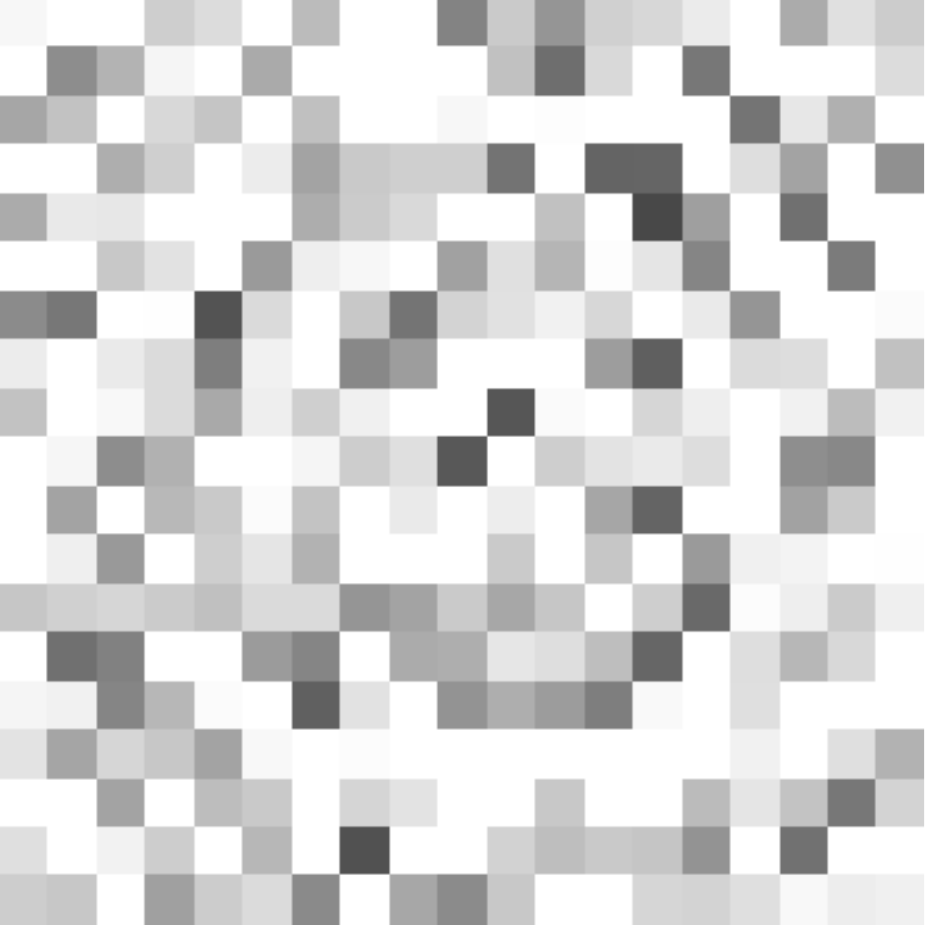}%
      \label{fig:convergence:noise:maggrad:0}
    } &
    \subfloat[{$D^x \hat{\phi}$}]{
      \includegraphics[width=0.25\linewidth]{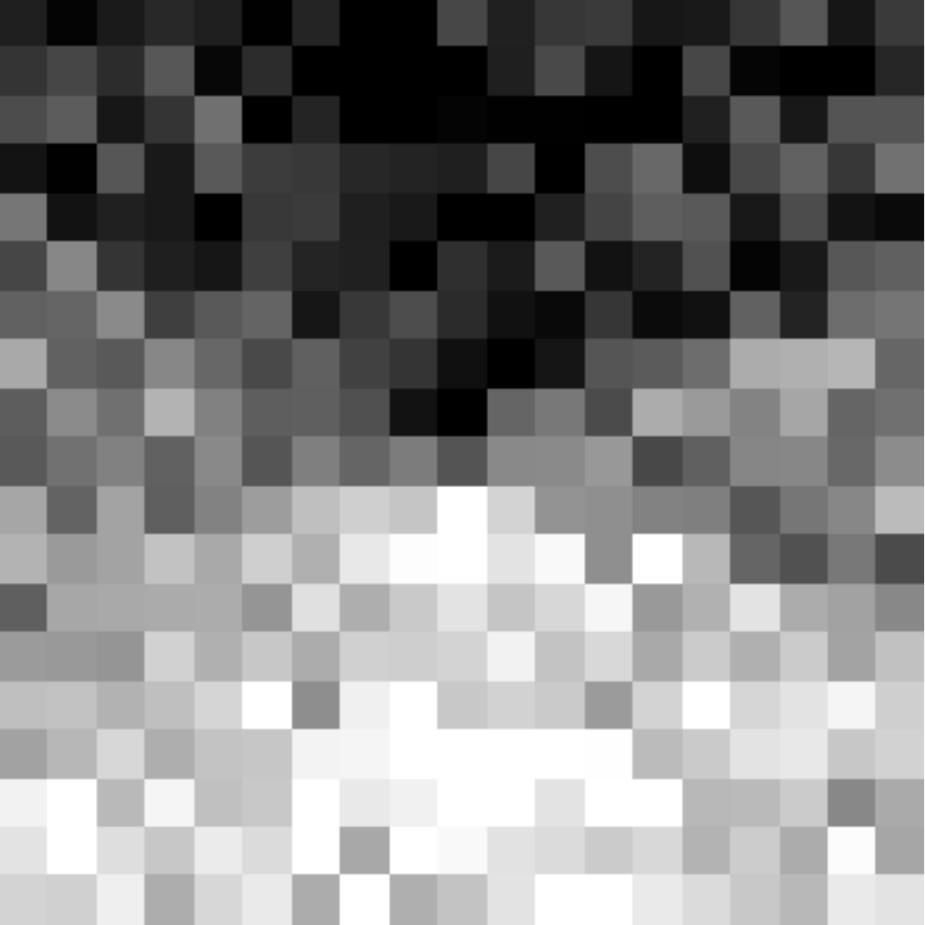}%
      \label{fig:convergence:noise:dx:0}
    } \\
    \subfloat[{$\hat{\phi}_{10}$}]{
      \includegraphics[width=0.25\linewidth]{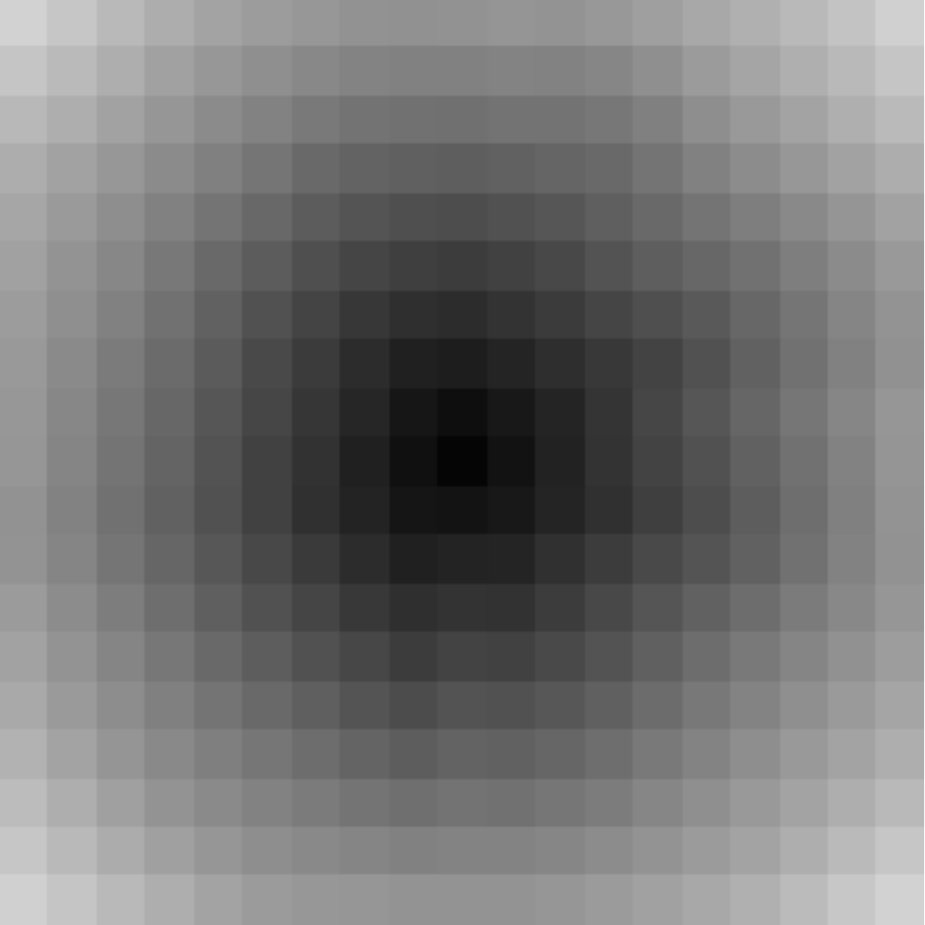}%
      \label{fig:convergence:noise:phi:10}
    } &
    \subfloat[{$|D\hat{\phi}_{10}|$}]{
      \includegraphics[width=0.25\linewidth]{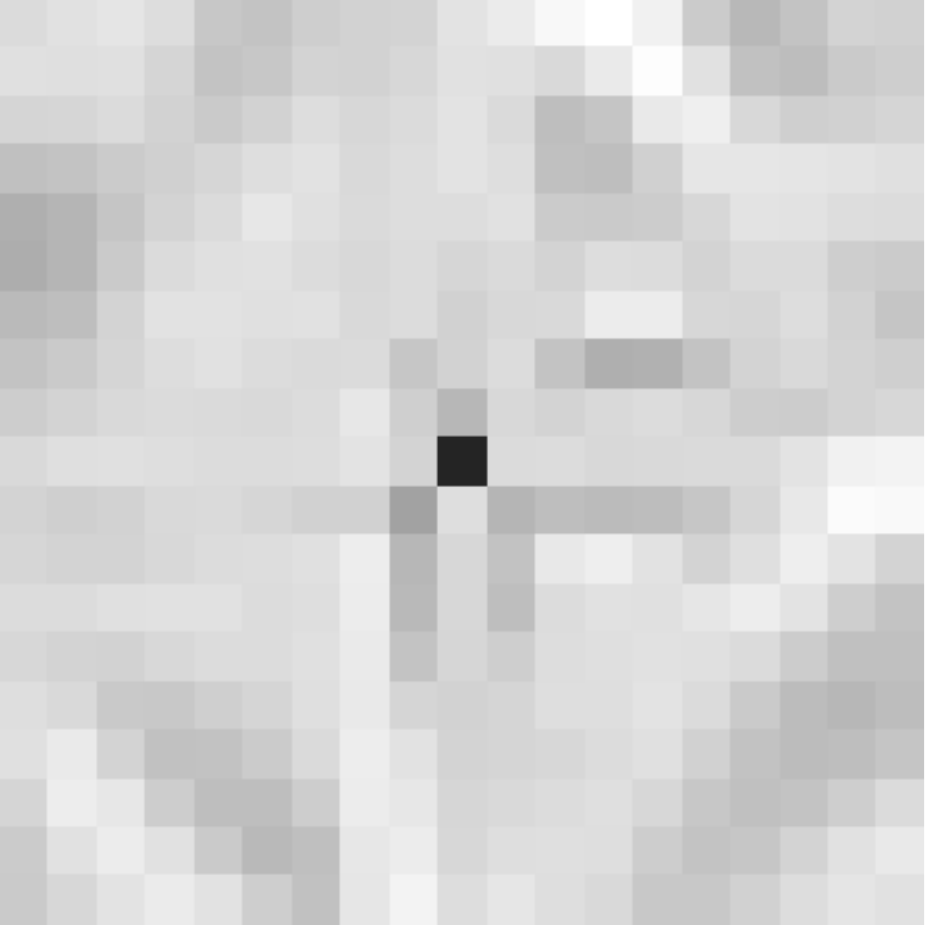}%
      \label{fig:convergence:noise:maggrad:10}
    } &
    \subfloat[{$D^x \hat{\phi}_{10}$}]{
      \includegraphics[width=0.25\linewidth]{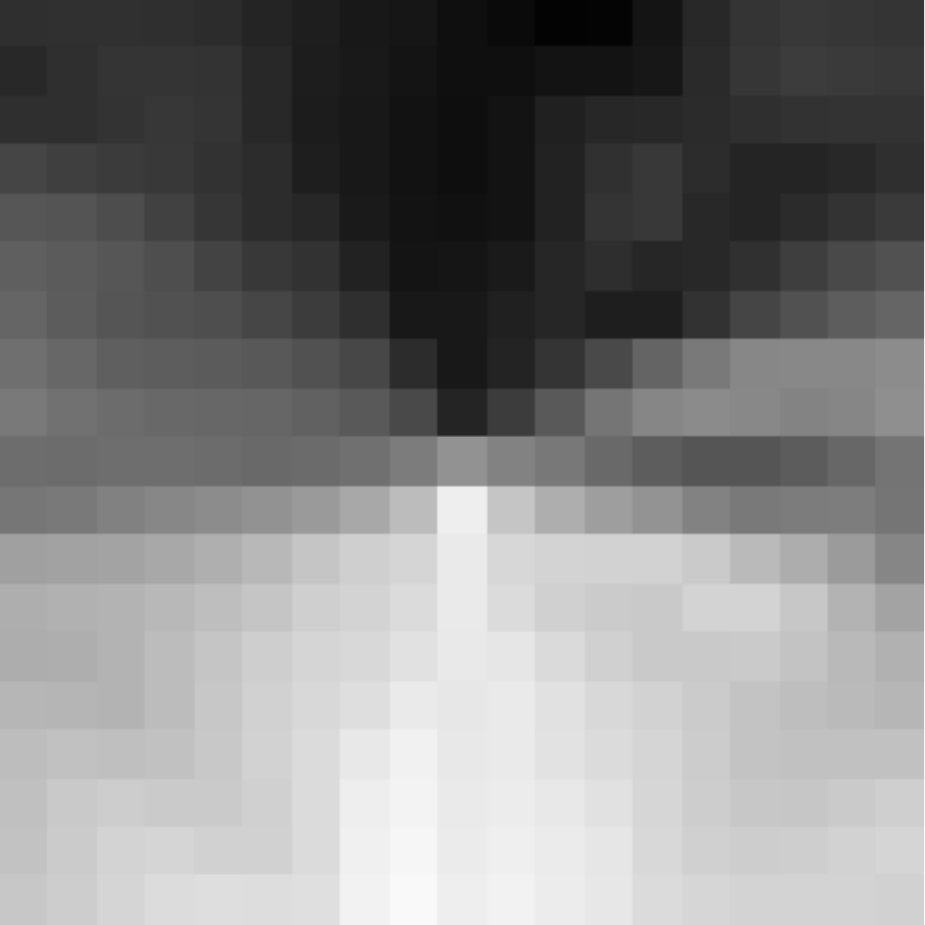}%
      \label{fig:convergence:noise:dx:10}
    } \\
    \subfloat[{$\hat{\phi}_{50}$}]{
      \includegraphics[width=0.25\linewidth]{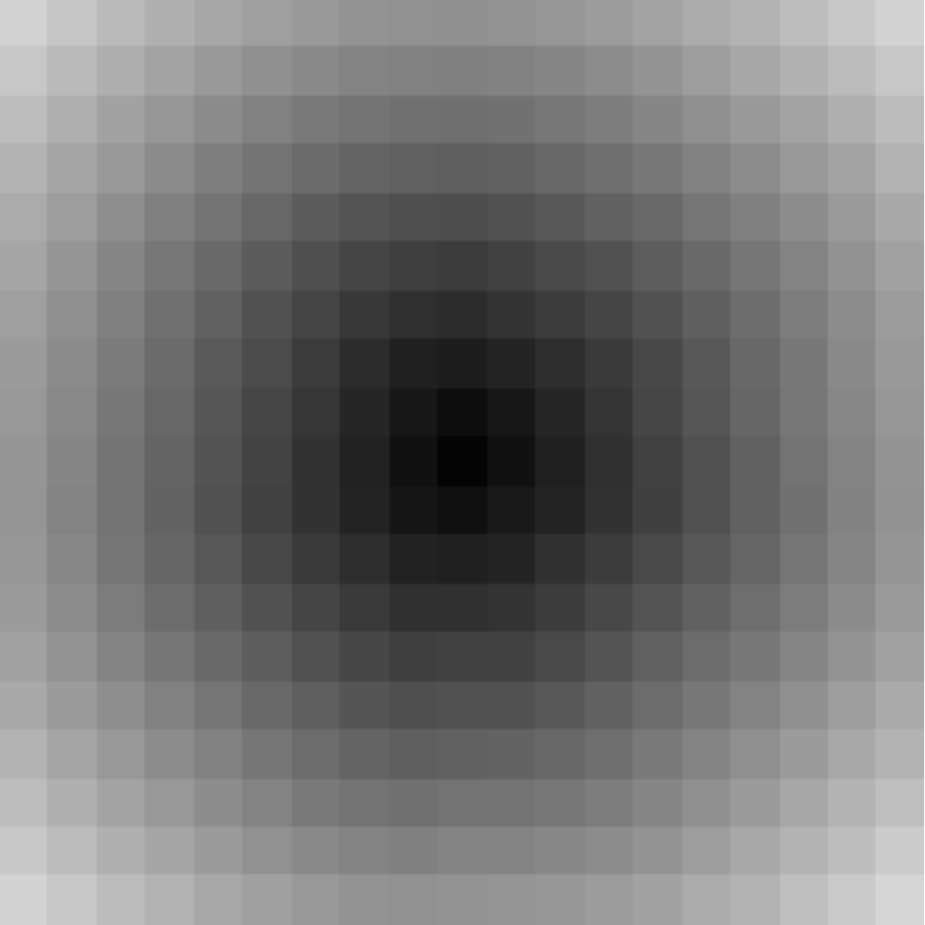}%
      \label{fig:convergence:noise:phi:50}
    } &
    \subfloat[{$|D\hat{\phi}_{50}|$}]{
      \includegraphics[width=0.25\linewidth]{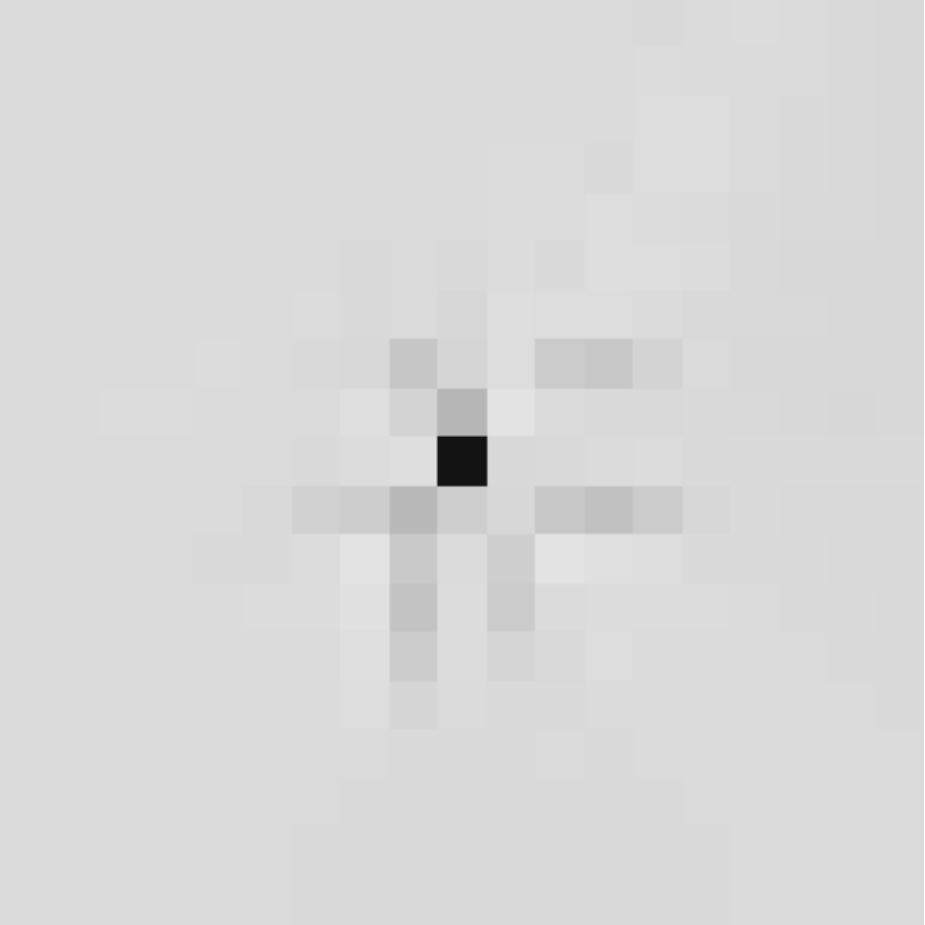}%
      \label{fig:convergence:noise:maggrad:50}
    } &
    \subfloat[{$D^x \hat{\phi}_{50}$}]{
      \includegraphics[width=0.25\linewidth]{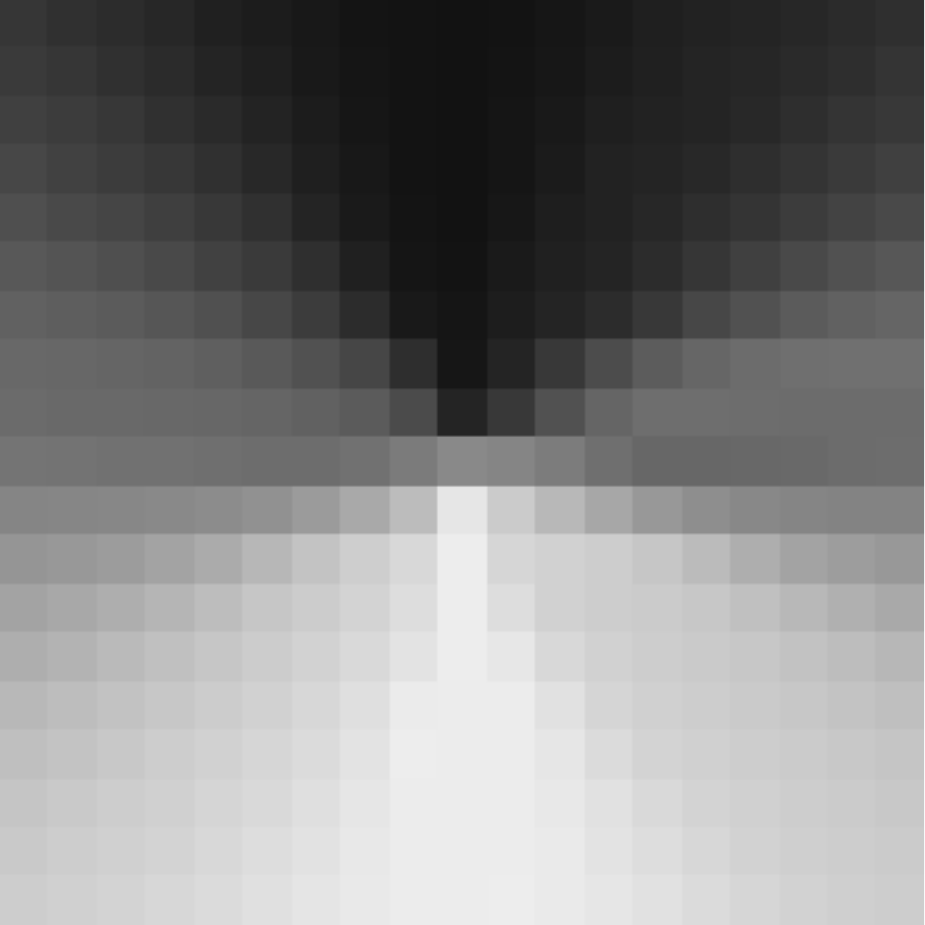}%
      \label{fig:convergence:noise:dx:50}
    } \\
    \subfloat[{$\hat{\phi}_{100}$}]{
      \includegraphics[width=0.25\linewidth]{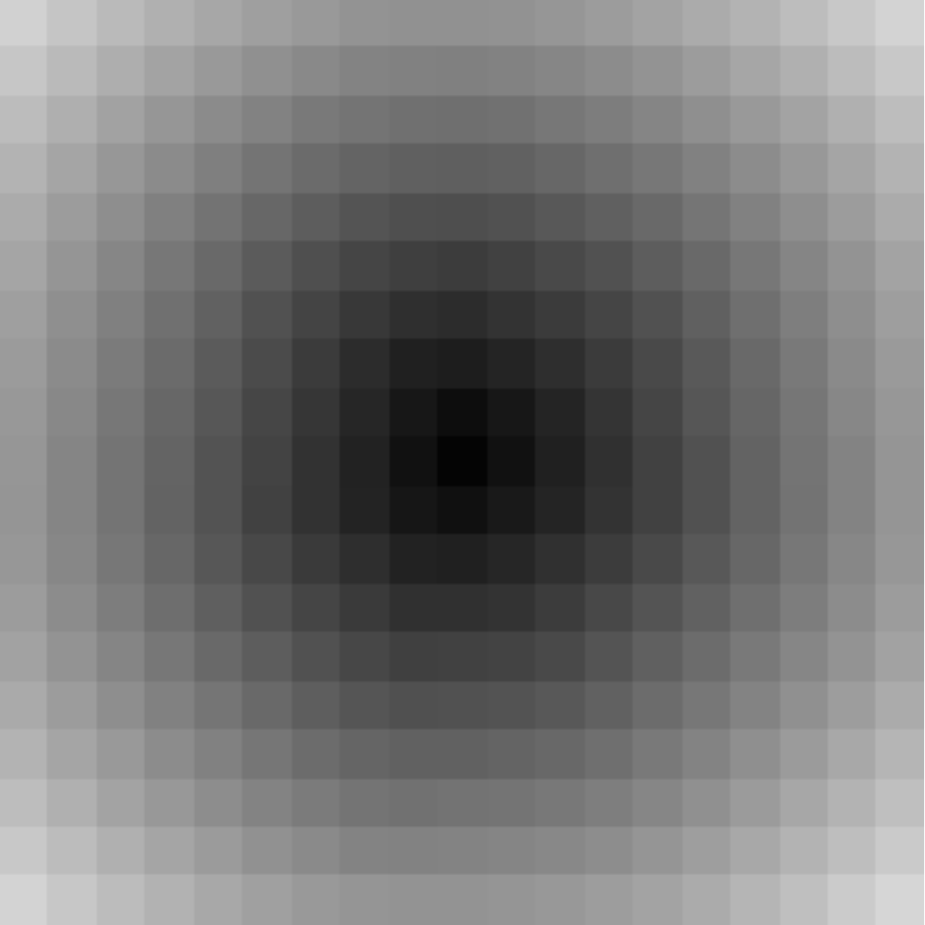}%
      \label{fig:convergence:noise:phi:100}
    } &
    \subfloat[{$|D\hat{\phi}_{100}|$}]{
      \includegraphics[width=0.25\linewidth]{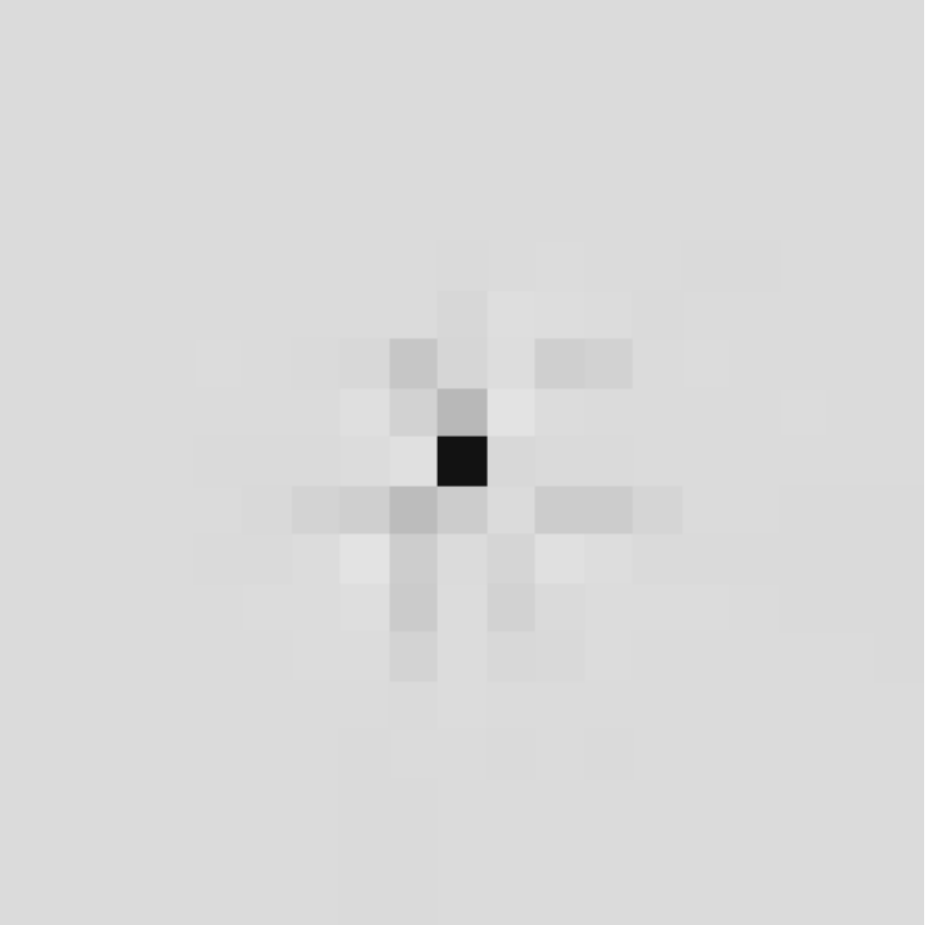}%
      \label{fig:convergence:noise:maggrad:100}
    } &
    \subfloat[{$D^x \hat{\phi}_{100}$}]{
      \includegraphics[width=0.25\linewidth]{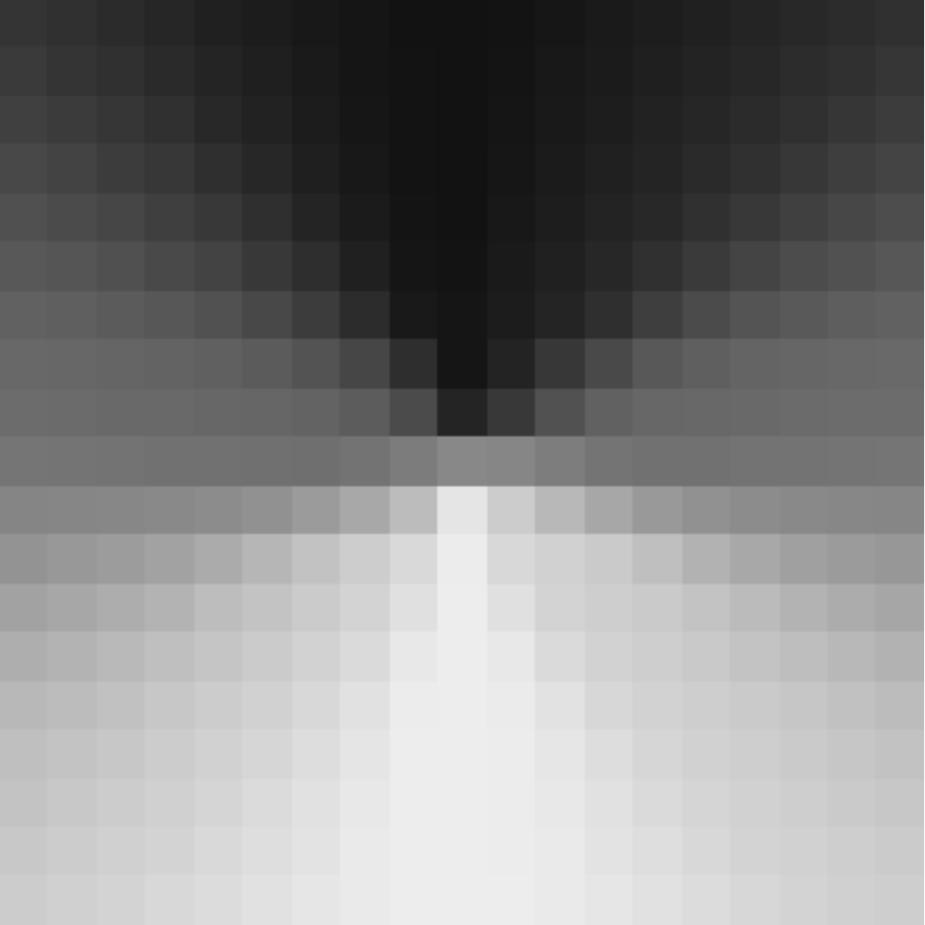}%
      \label{fig:convergence:noise:dx:100}
    } \\
    \subfloat[{$\hat{\phi}_{400}$}]{
      \includegraphics[width=0.25\linewidth]{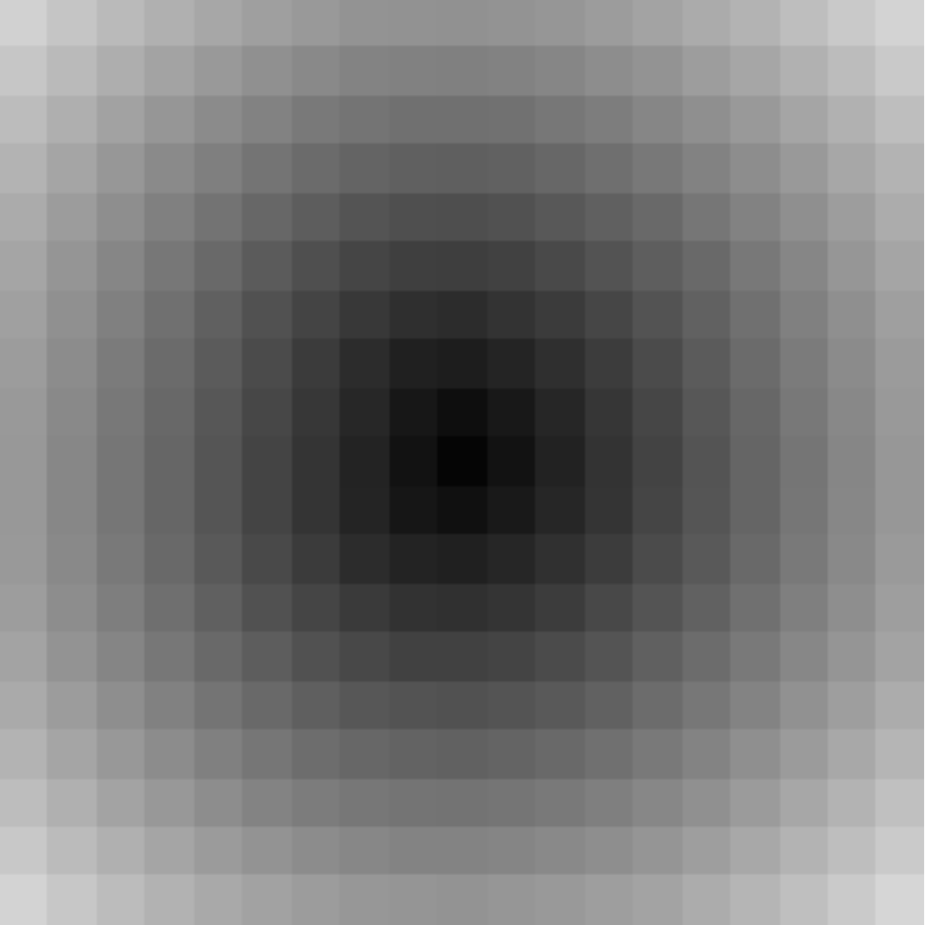}%
      \label{fig:convergence:noise:phi:400}
    } &
    \subfloat[{$|D\hat{\phi}_{400}|$}]{
      \includegraphics[width=0.25\linewidth]{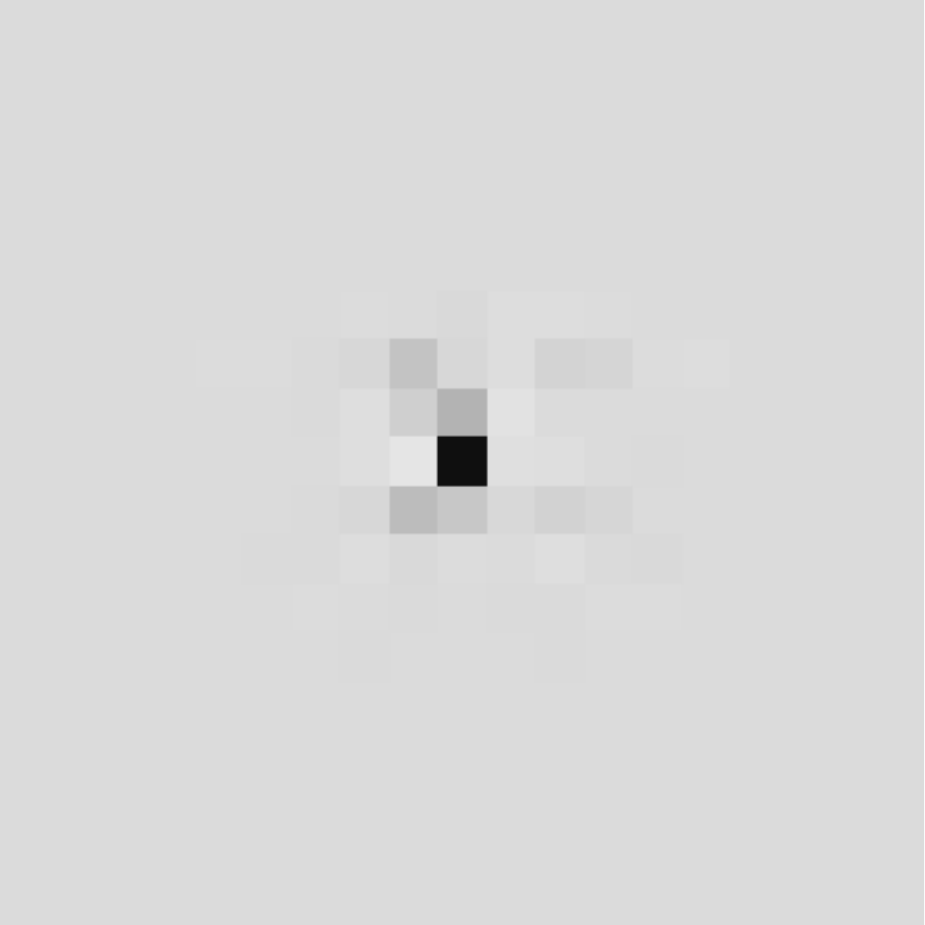}%
      \label{fig:convergence:noise:maggrad:400}
    } &
    \subfloat[{$D^x \hat{\phi}_{400}$}]{
      \includegraphics[width=0.25\linewidth]{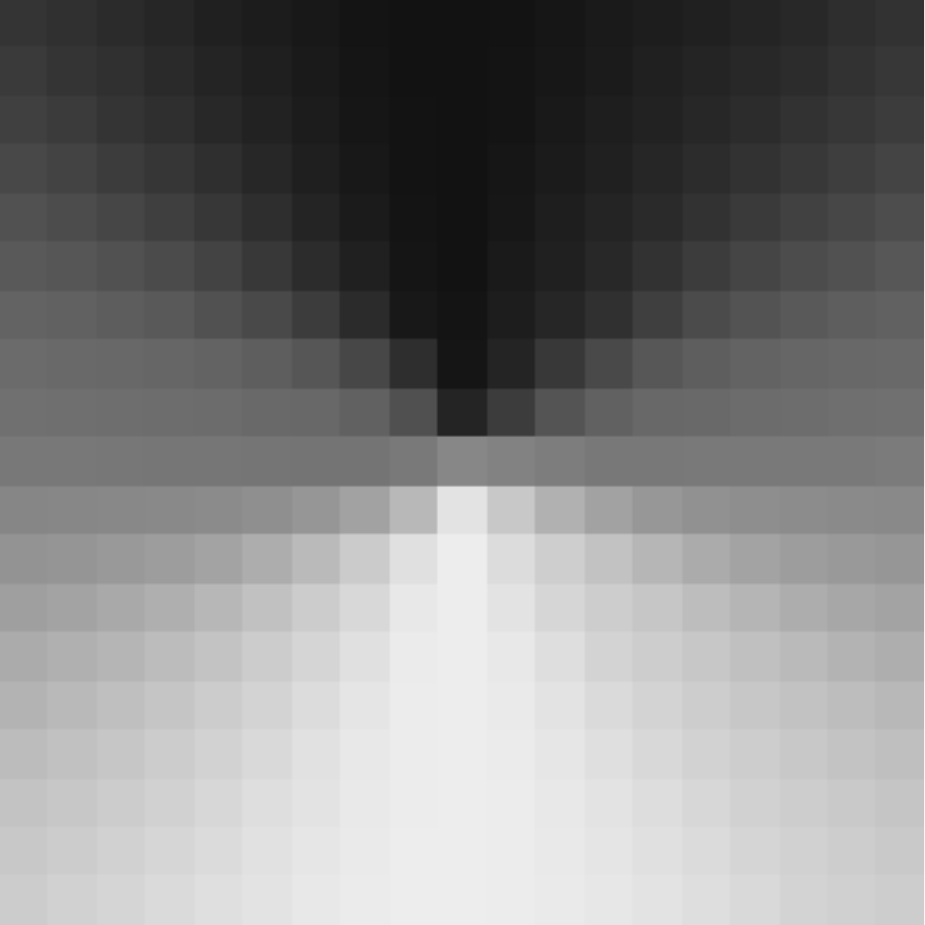}%
      \label{fig:convergence:noise:dx:400}
    } 
  \end{tabular}
  \caption{Visualizing convergence for specific iterations ($h=0.5$, $\alpha = 2.0$). $\phi_n$ indicates iteration $n$.}
  \label{fig:convergence}
\end{figure}

First, the convergence of the algorithm is analyzed.
The three errors are plotted over 400 iterations in Figure~\ref{fig:convergence2}.
Changes in the magnitude gradient error inflect around 20 iterations while gradient error continues to improve until just before 300 iterations
As guaranteed by the algorithm, the representation error is always zero (Figure~\ref{fig:convergence2:rep}).
Select iterations are visualized in Figure~\ref{fig:convergence}.
Banding of the gradient image is immediately removed while noise in the gradient image is visually removed between iteration 50 and 100.

\subsubsection{Is Dithering Needed?}
\begin{figure}[h]
  \centering
  \begin{tabular}{c}
    \subfloat[{$\lVert \nabla \phi \rVert - 1$}]{
      \includegraphics[width=0.75\linewidth]{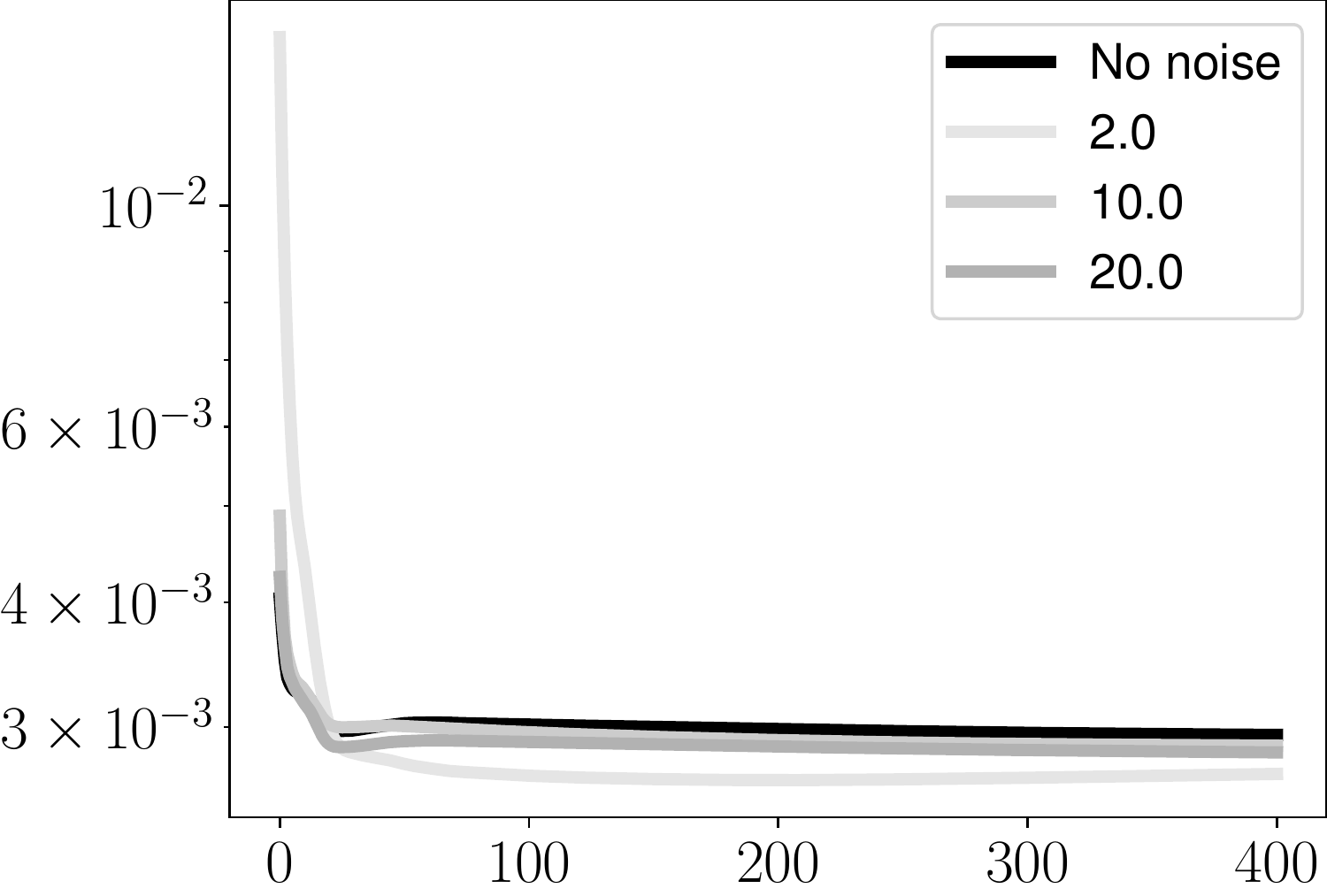}%
      \label{fig:alpha2:maggrad}
    } \\
    \subfloat[{$\lVert D \phi - \nabla \phi \rVert$}]{
      \includegraphics[width=0.75\linewidth]{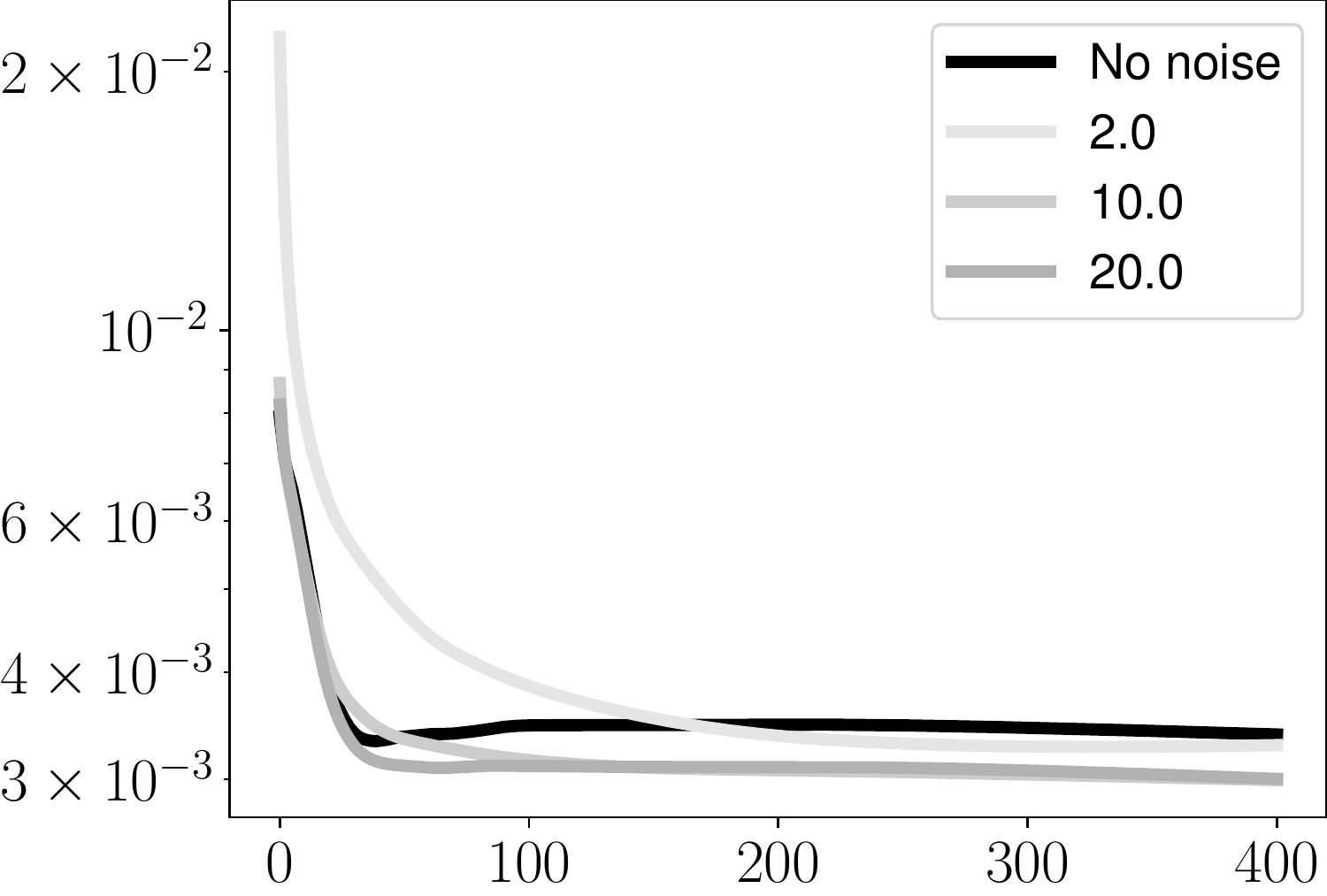}%
      \label{fig:alpha2:grad}
    } 
  \end{tabular}
  \caption{Log-plots of convergence measures over 400 iterations for different $\alpha$ ($h=0.5$).}
  \label{fig:alpha2}
\end{figure}

\begin{figure}[h]
  \centering
  \begin{tabular}{ccc}
    \subfloat[{$D^{x}\tilde{\phi}_0$}]{
      \includegraphics[width=0.25\linewidth]{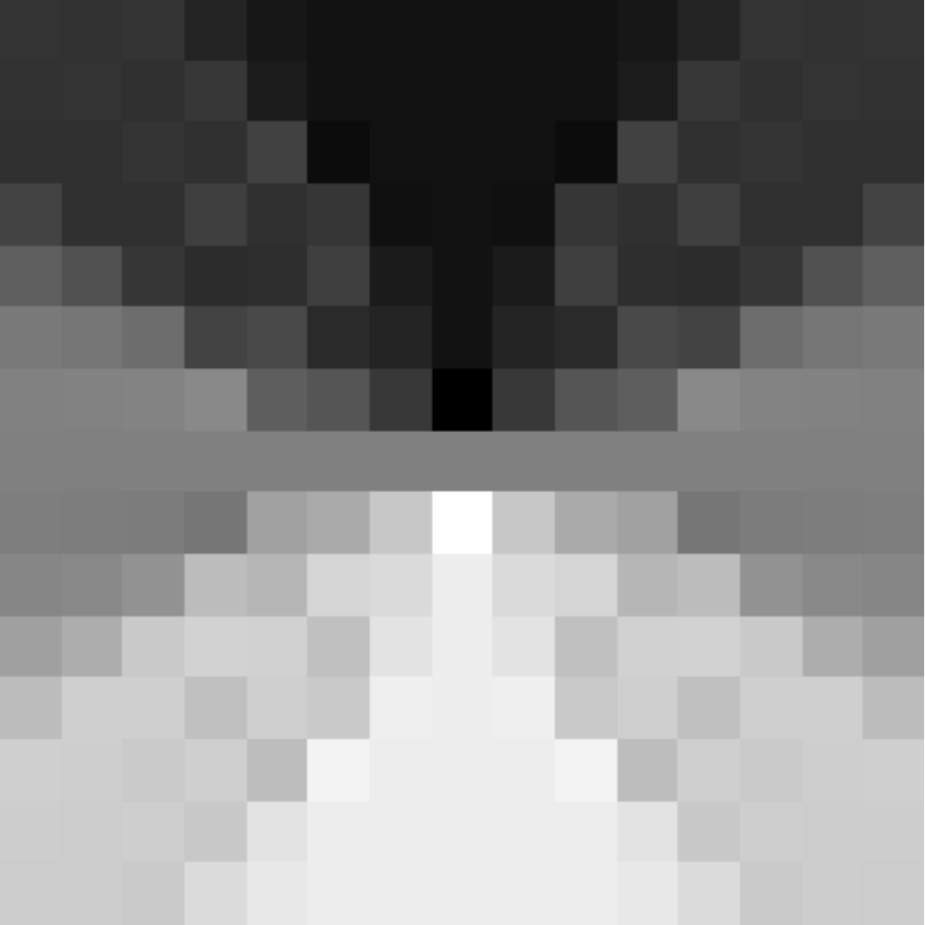}%
      \label{fig:alpha:nonoise:phi}
    } &
    \subfloat[{$D^{x}\hat{\phi}_0^2$}]{
      \includegraphics[width=0.25\linewidth]{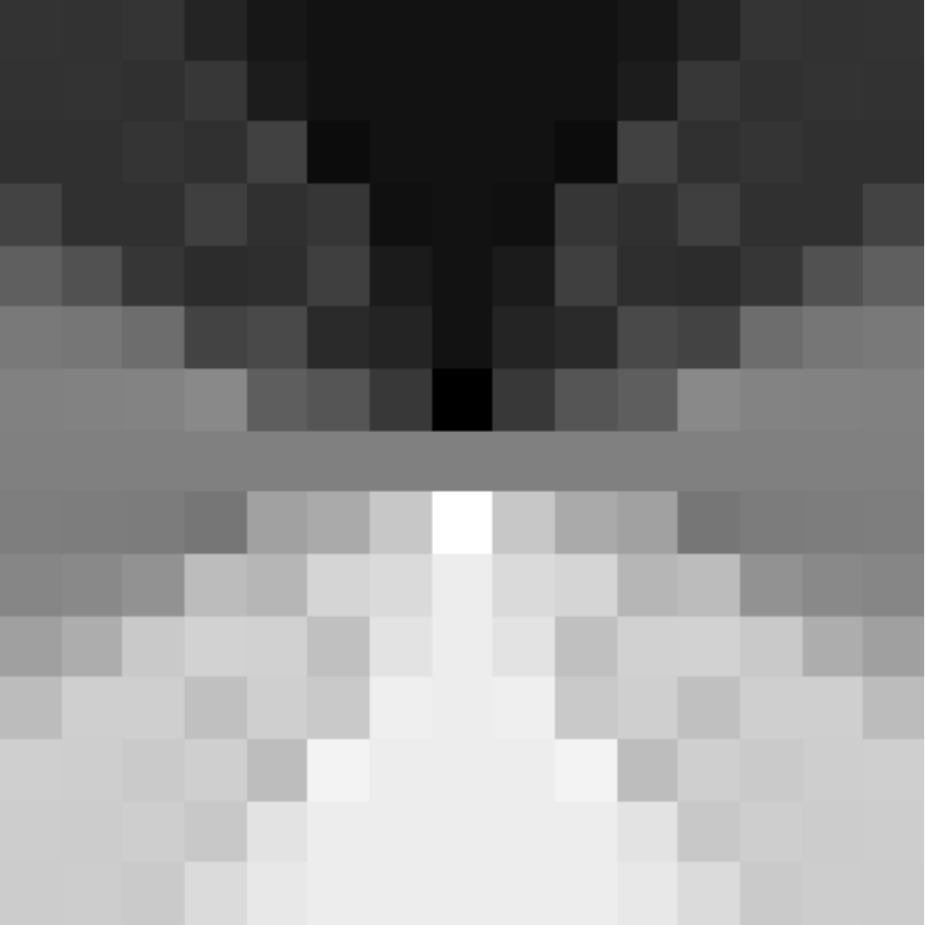}%
      \label{fig:alpha:nonoise:maggrad}
    } &
    \subfloat[{$D^{x}\hat{\phi}_0^{20}$}]{
      \includegraphics[width=0.25\linewidth]{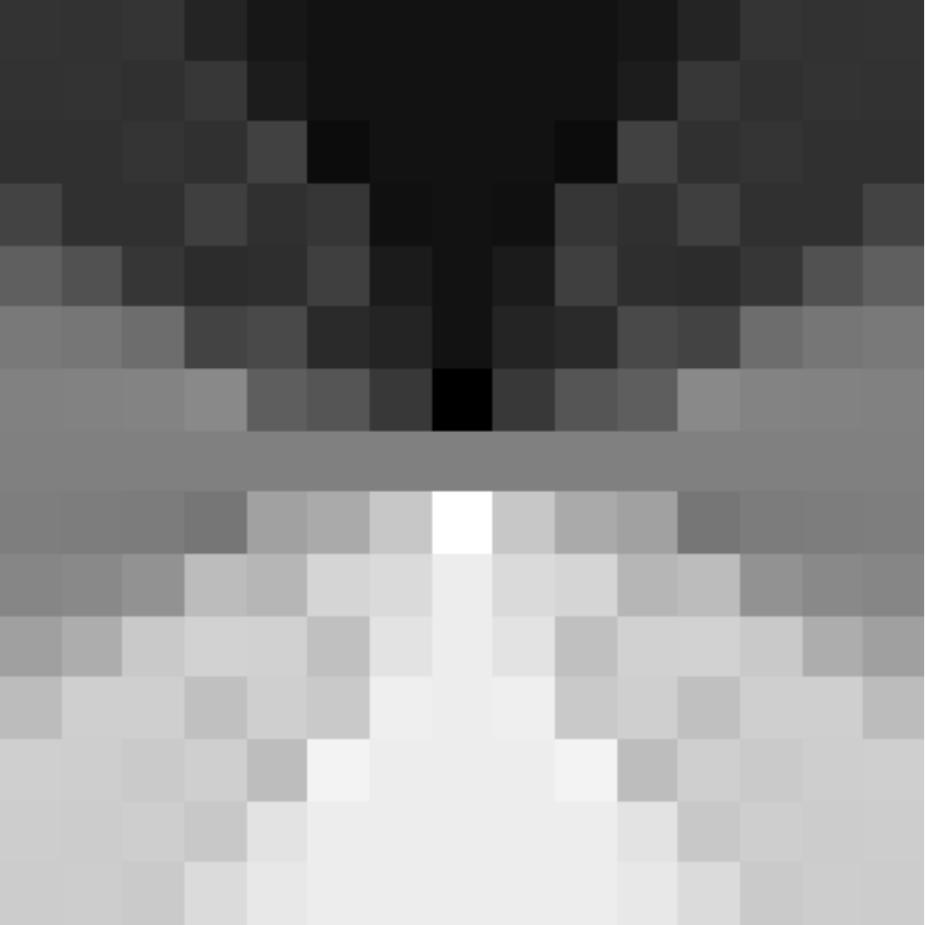}%
      \label{fig:alpha:nonoise:dx}
    } \\
    \subfloat[{$D^{x}\tilde{\phi}_{20}$}]{
      \includegraphics[width=0.25\linewidth]{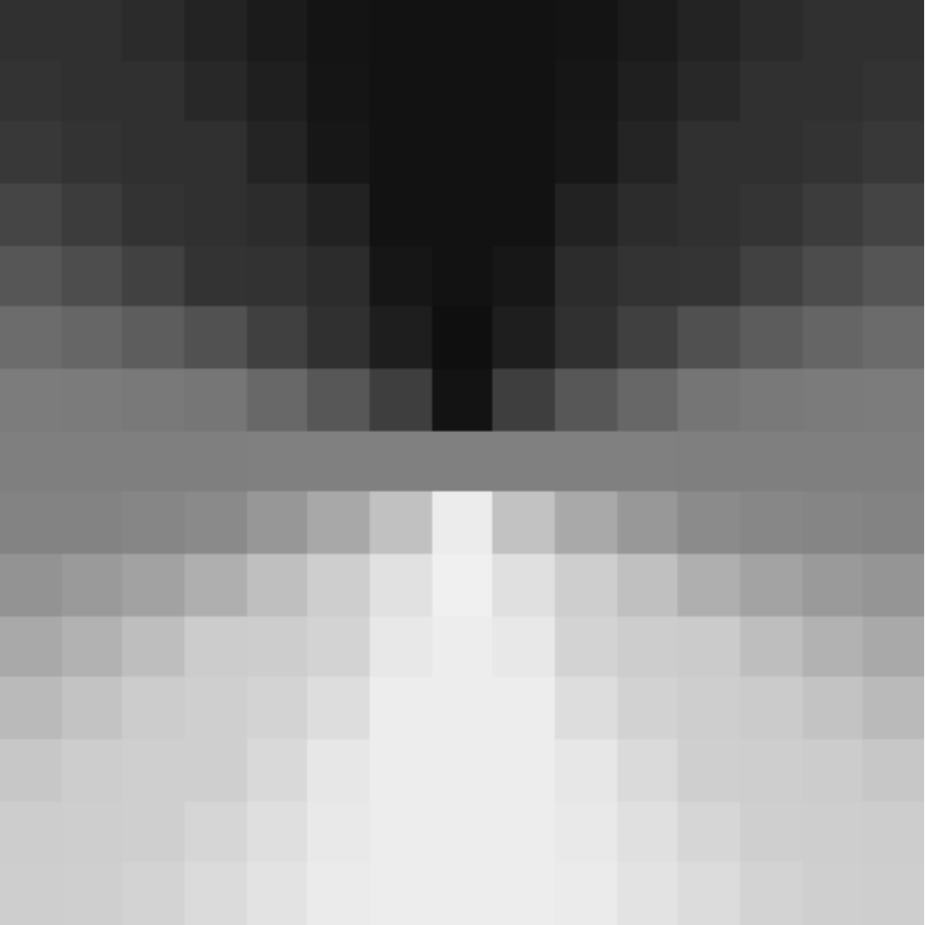}%
      \label{fig:alpha:nonoise:phi}
    } &
    \subfloat[{$D^{x}\hat{\phi}_{20}^2$}]{
      \includegraphics[width=0.25\linewidth]{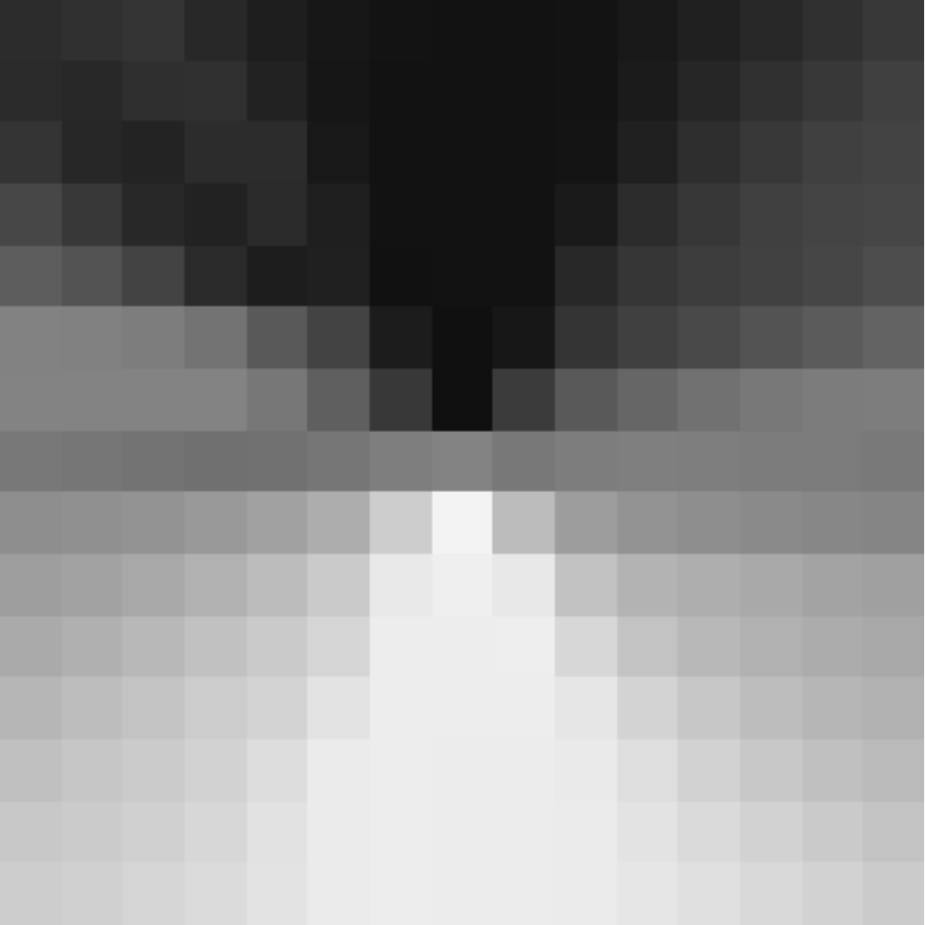}%
      \label{fig:alpha:nonoise:maggrad}
    } &
    \subfloat[{$D^{x}\hat{\phi}_{20}^{20}$}]{
      \includegraphics[width=0.25\linewidth]{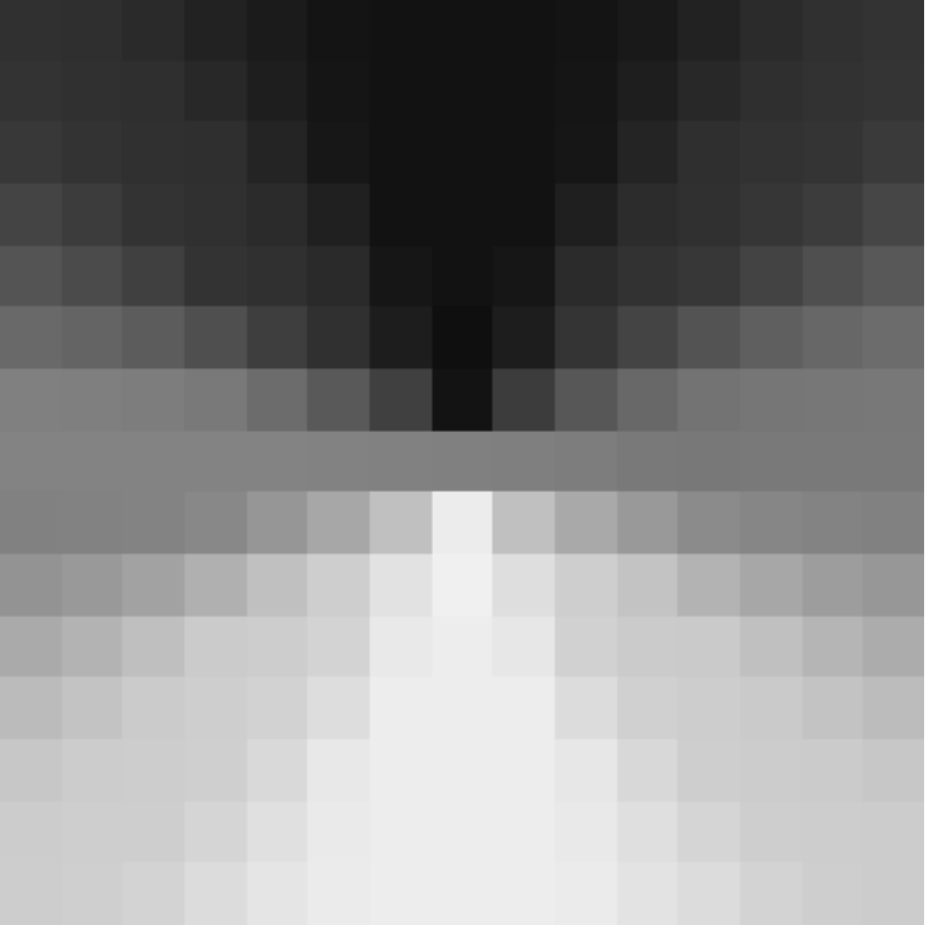}%
      \label{fig:alpha:nonoise:dx}
    } \\
    \subfloat[{$D^{x}\tilde{\phi}_{50}$}]{
      \includegraphics[width=0.25\linewidth]{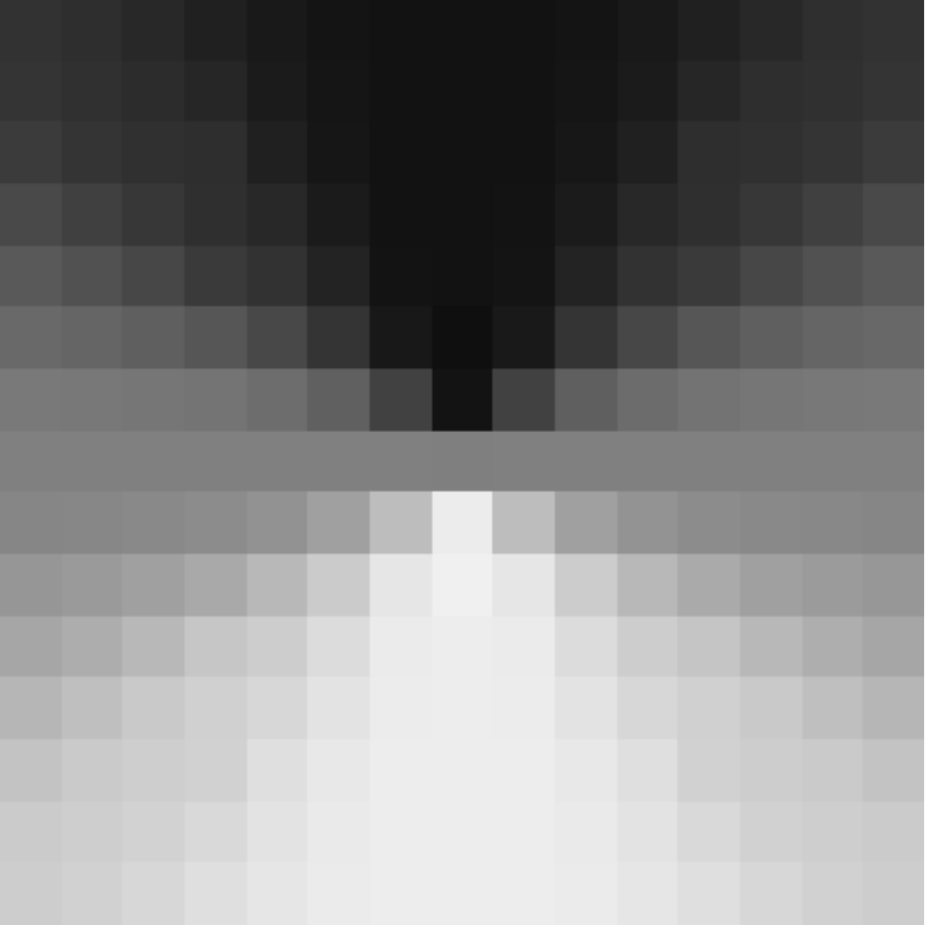}%
      \label{fig:alpha:nonoise:phi}
    } &
    \subfloat[{$D^{x}\hat{\phi}_{50}^{2}$}]{
      \includegraphics[width=0.25\linewidth]{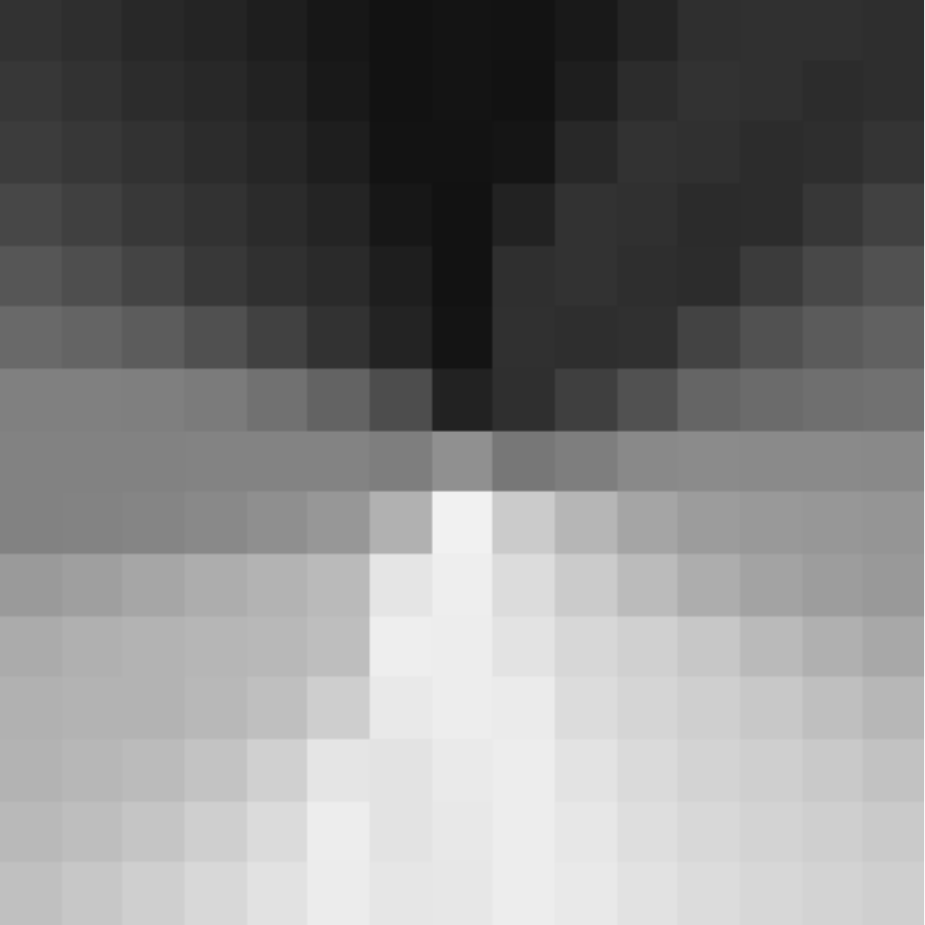}%
      \label{fig:alpha:nonoise:maggrad}
    } &
    \subfloat[{$D^{x}\hat{\phi}_{50}^{20}$}]{
      \includegraphics[width=0.25\linewidth]{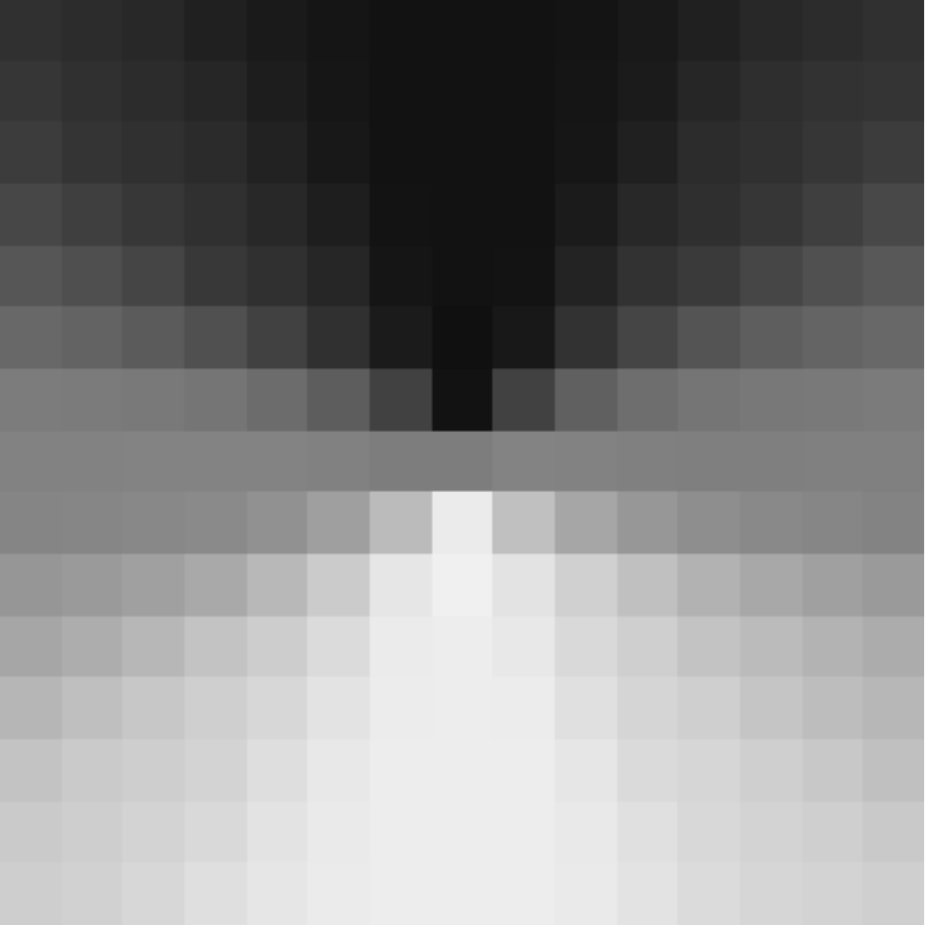}%
      \label{fig:alpha:nonoise:dx}
    } \\
  \end{tabular}
  \caption{Visualizing convergence for specific iterations and $\alpha$ ($h=0.5$). $\phi_n^\alpha$ indicates iteration $n$ for noise parameter $\alpha$.}
  \label{fig:alpha}
\end{figure}

Next, the value of dithering is investigated.
Reinitialization is performed without dithering and for dithering with varying values of $\alpha$.
The convergence of the gradient error and magnitude gradient error are displayed in Figure~\ref{fig:alpha2}.
The representation error was zero for all iterations and is not plotted.
A modest improvement is seen as $\alpha$ increases.
As alpha goes to infinity, the errors will settle at the undithered signal.
Images for select iterations are displayed in Figure~\ref{fig:alpha}.
For small $\alpha$, incorrect gradients last for larger iterations.
For all examples, a large number of iterations ($>20$) is needed to get sufficiently smooth gradients.

In summary, dithering is not imperative, but its use slightly improves accuracy.
Since dithering is essentially computationally free compared to the reinitialization algorithm, it is recommended to add dithering with a large $\alpha$ value.

\subsubsection{Quantization in Mean Curvature}
The ability of the method to remove quantization of the mean curvature is demonstrated.
A single L4 vertebrae was used imaged at an in-plane resolution of 0.7 mm and a slice-thickness of 1.0 mm.
The image was embedded using a signed distance transform ~\cite{danielsson1980euclidean}.
The signed distance map was dithered ($\alpha = 20$) and reinitialized for 100 iterations. 
A surface was extracted from both embeddings using marching cubes~\cite{lorensen1987marching}.
Fourth-order accurate finite difference stencils were placed on the vertices of the mesh and interpolated into the embeddings to compute the mean curvature locally.
This allowed the mean curvature to be visualized on the surface of the mesh and histograms to be computed from the vertices.

\begin{figure}[h]
  \centering
  \begin{tabular}{cc}
    \subfloat[\small sdt]{
      \includegraphics[width=0.44\linewidth]{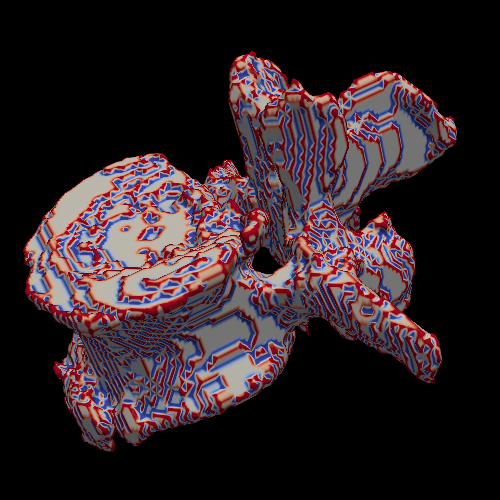}%
      \label{fig:l4:sdt}
    } &
    \subfloat[\small Proposed]{
      \includegraphics[width=0.44\linewidth]{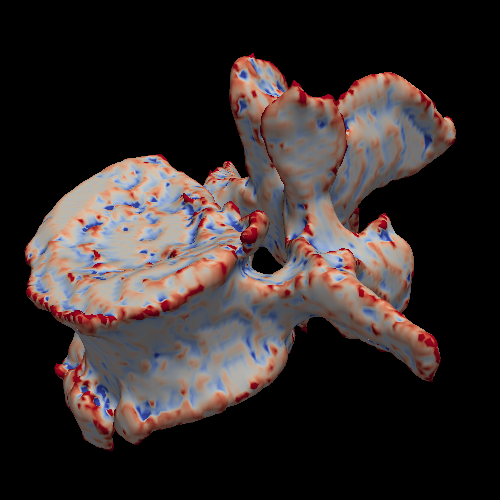}%
      \label{fig:l4:corrected}
    } \\
    \subfloat[\small sdt histogram]{
      \includegraphics[width=0.44\linewidth]{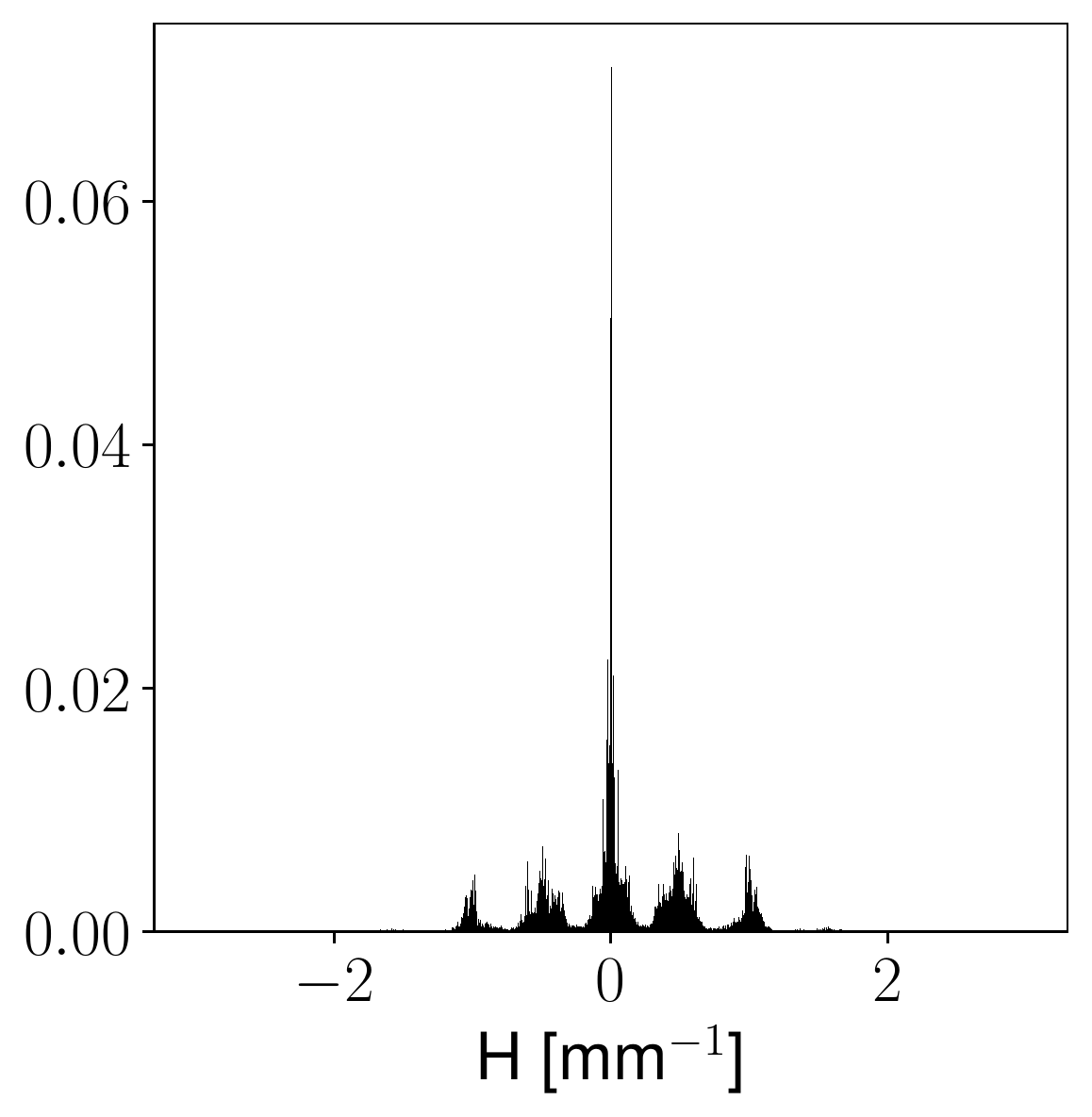}%
      \label{fig:l4:mean}
    } &
    \subfloat[\small Proposed histogram]{
      \includegraphics[width=0.44\linewidth]{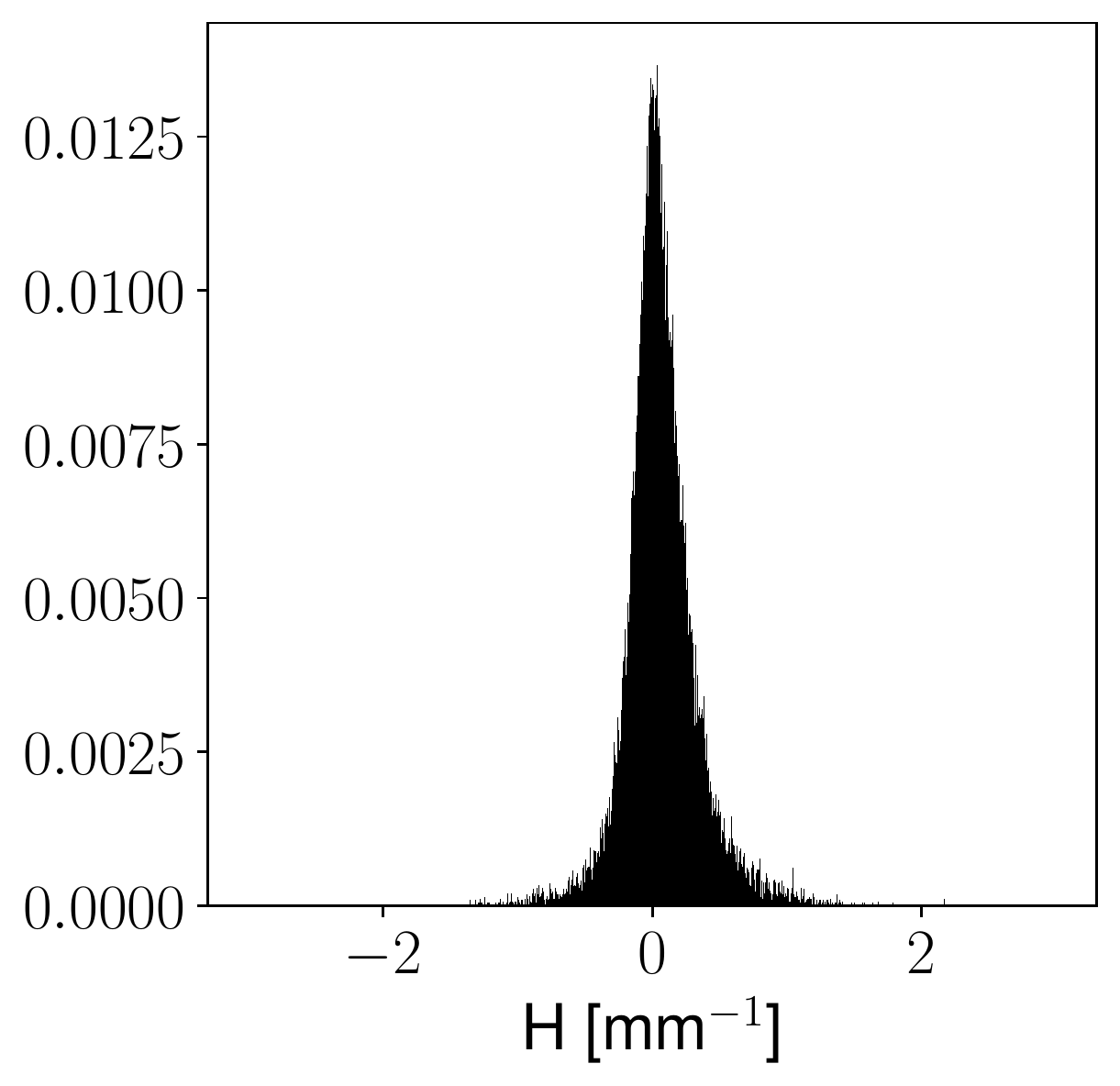}%
      \label{fig:l4:corrected_mean}
    } \\
  \end{tabular}
  \caption{Visualization of mean curvature from an embedding. (\ref{fig:l4:sdt}) Traditional signed distance transform, (\ref{fig:l4:corrected}) dithered and reinitialized, (\ref{fig:l4:mean}) histogram of mean curvature in the traditional embedding, (\ref{fig:l4:corrected_mean}) histogram of mean curvature in the proposed correction.}
  \label{fig:l4}
\end{figure}

Local mean curvature across the trabecular bone is visualized in Figure~\ref{fig:l4}.
Mean curvature appears quantized when computed from the distance map (\ref{fig:l4:sdt}).
After running the proposed algorithm, the mean curvature takes on sensible values and the surface appears more smooth (\ref{fig:l4:corrected}).
We reiterate here that the embedding did not change sign, quantization has just been removed.
Finally, comparing the histogram of the distance map (\ref{fig:l4:mean}) to that of the corrected embedding (\ref{fig:l4:corrected_mean}), quantization of the mean curvature has been removed using the proposed method.
It should be noted that the algorithm showed no further improvements after 100 iterations and the representation cannot be made more accurate. 

\section{Consequences of Quantization}
\subsection{Morphological Image Processing}
Applications of the distance transform in morphological image processing include skeletonization~\cite{blum1967transformation,kimmel1995skeletonization,siddiqi2002hamilton}, shape matching~\cite{barrow1977parametric}, and thickness computation~\cite{hildebrand1997new}.
Many algorithms exist for each task, some making use of gradients and others not.
In gradient-free methods, quantization is unlikely to affect results.
In gradient-based methods~\cite{kimmel1995skeletonization,siddiqi2002hamilton}, these artifacts can affect results.
Most algorithms take explicit steps to improve gradient computation~\cite{kimmel1995skeletonization} or using integral rather than divergence representations~\cite{siddiqi2002hamilton}.
Quantization is unlikely to have noticeable affects in morphology tasks.

\subsection{Curve Evolution}
Many curve evolution problems such as active contours use signed distance signals as an implicit representation of the curve~\cite{osher1988fronts,caselles1993geometric,malladi1995shape}.
These algorithms evolve the embedding according to gradients of the image, knowing the curve can be recovered as the zero level set of the embedding.
The findings of this work demonstrate that that the initialization of an embedding from the signed distance transform of binary image data is no better than first order accurate, $O(h)$, due to quantization.
This is important because curve evolution based on mean curvature flow requires at least second order accuracy initialization~\cite{coquerelle2016fourth}.
Since most curve evolution problems are implemented with total variation diminishing numerical methods, the noise will not amplifying during evolution.
However, the error in the solution will be independent of sampling period.
Solving this issue with reinitialization is impractical because the reinitialization converges slowly, stopping before the representation is second order accurate.
This strongly motivates alternative embedding procedures.

\section{Conclusion}
Distance transforms of sampled signals produce a quantized approximation to the true distance signal.
Quantization leads to an artifact where numerical gradients are made flat.
This artifact is independent of sample period and manifests as banding in the gradient image.
If needed and where possible, the initial distance signal should be constructed from a representation other than a binary image.
However, given no other options, a dithering and reinitialization algorithm is proposed that removes artifacts in the gradients while preserving the representation of the binary signal and keeping the gradient magnitude nearly equal to $+1$.

\appendices

\section{Implementing the Signed Distance Transform}
\label{app:sdt}
\begin{figure}[h]
  \centering
  \begin{tabular}{cc}
    \multicolumn{2}{c}{
    \subfloat[{$I[nh]$}]{
      \includegraphics[width=0.45\linewidth]{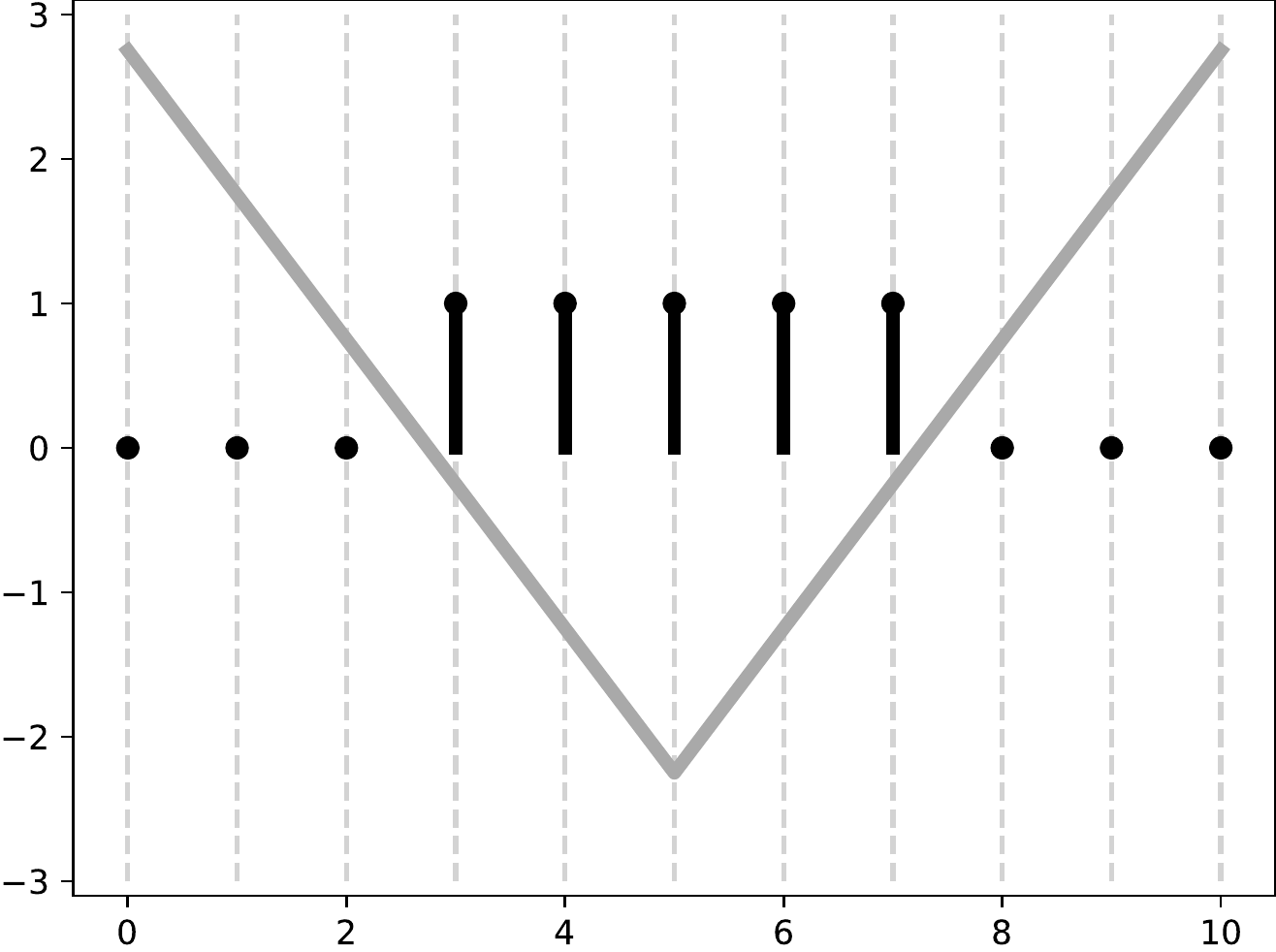}%
      \label{fig:sdt:binary}
    }} \\
    \subfloat[{$d(nh, A^C)$}]{
      \includegraphics[width=0.45\linewidth]{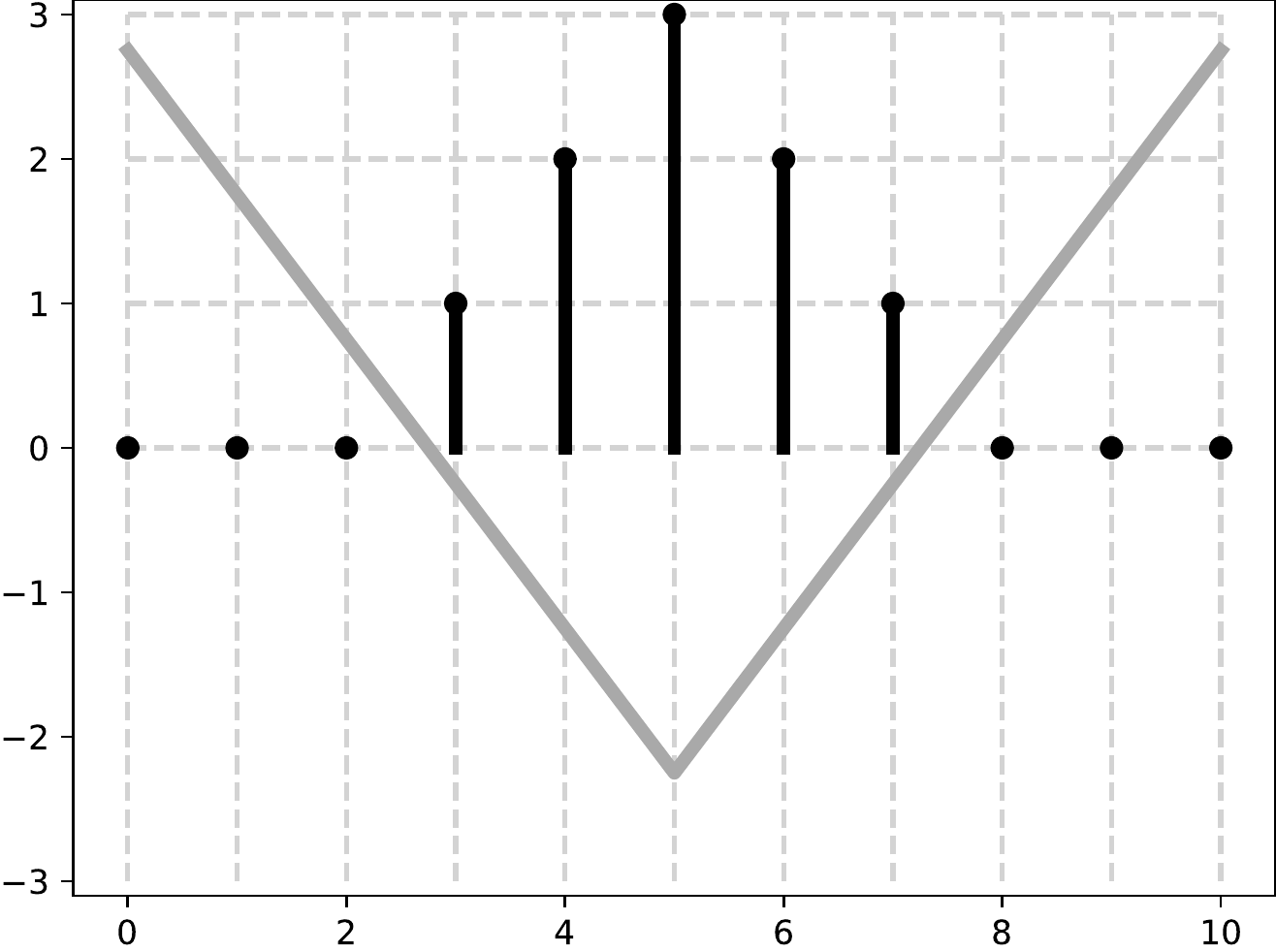}%
      \label{fig:sdt:inside}
    } &
    \subfloat[{$d(nh, A)$}]{
      \includegraphics[width=0.45\linewidth]{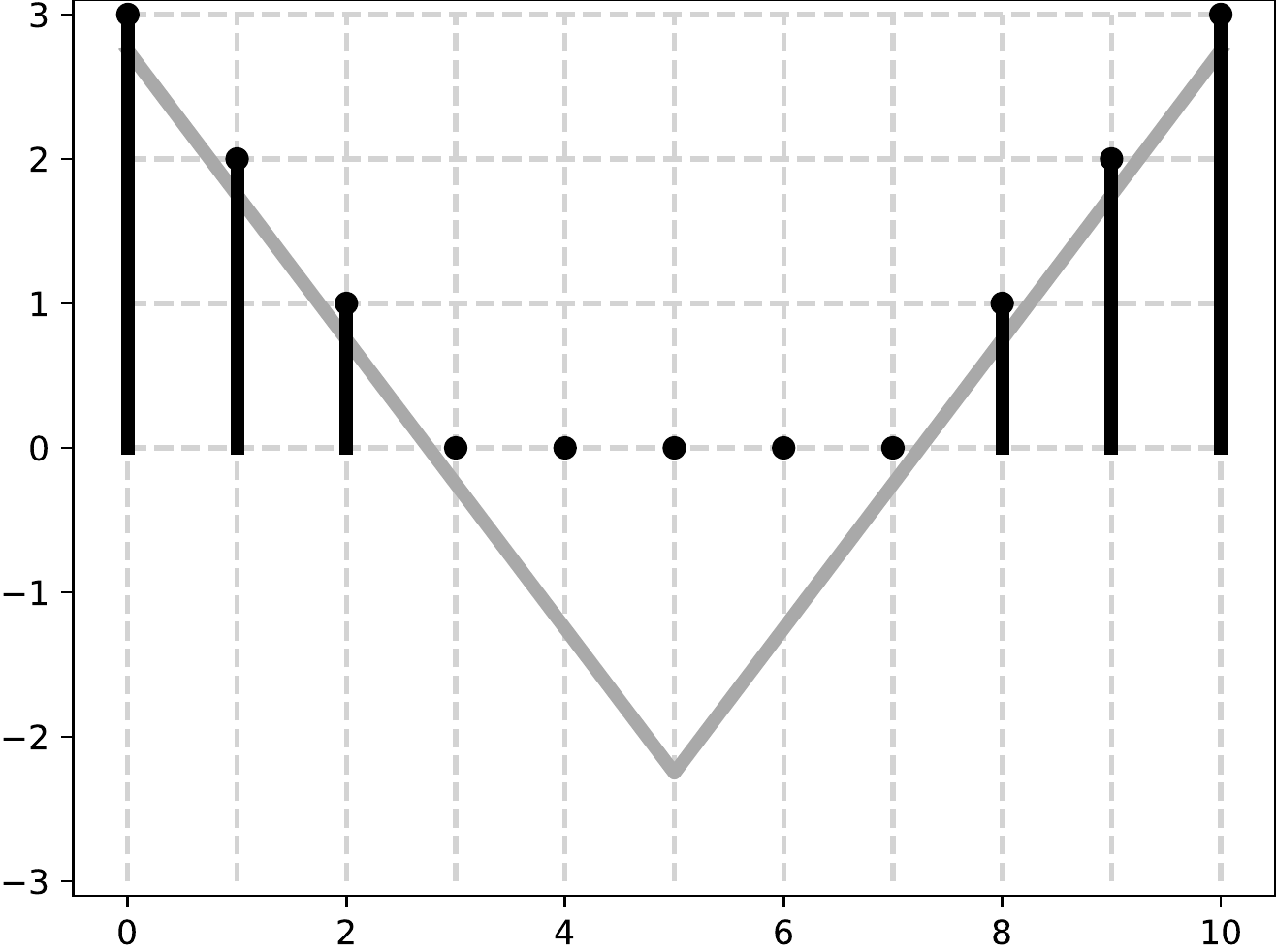}%
      \label{fig:sdt:outside}
    } \\
    \subfloat[Eqn~\ref{eqn:sdt}]{
      \includegraphics[width=0.45\linewidth]{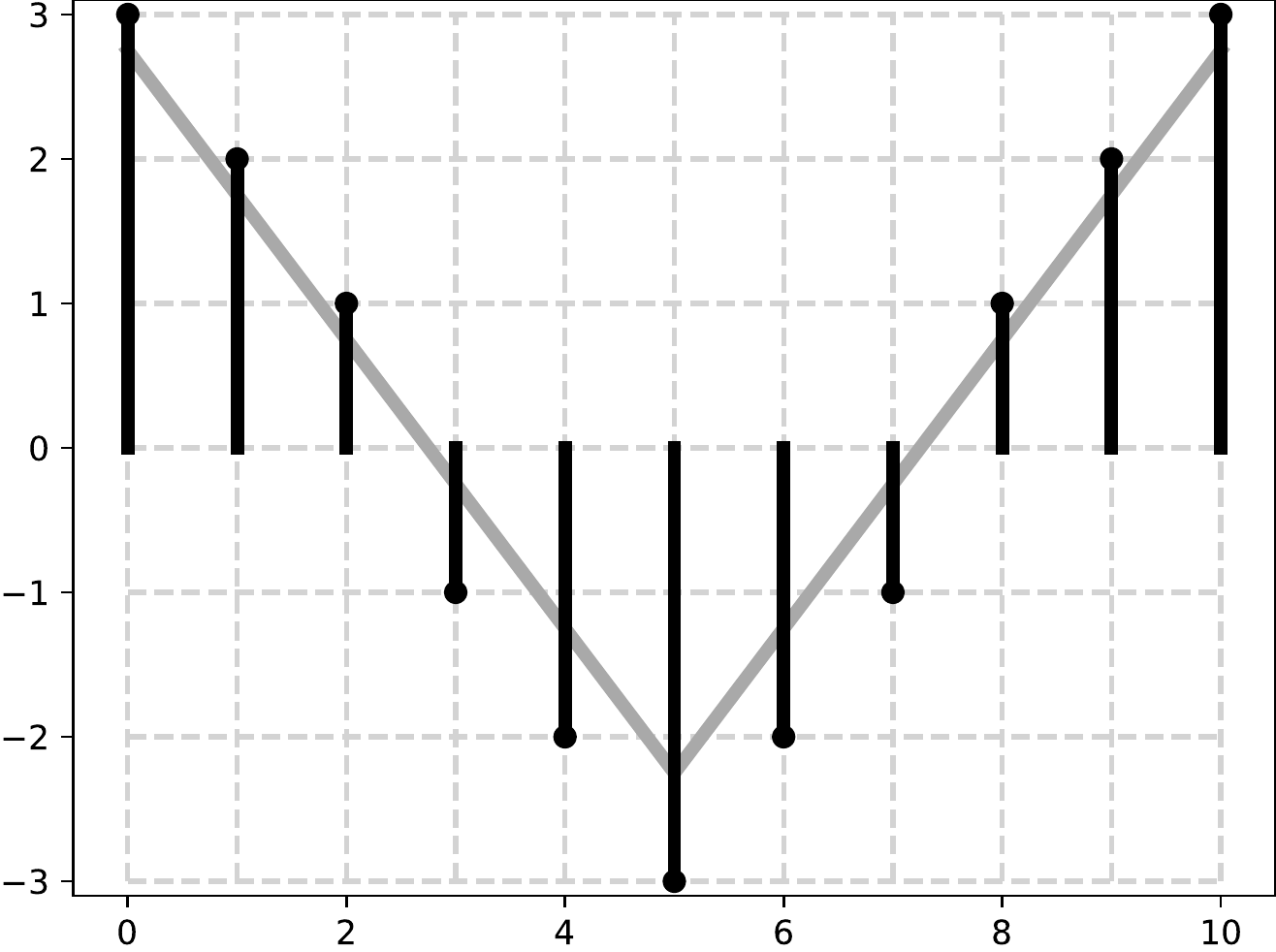}%
      \label{fig:sdt:standard}
    } &
    \subfloat[Eqn~\ref{eqn:sdt_corrected}]{
      \includegraphics[width=0.45\linewidth]{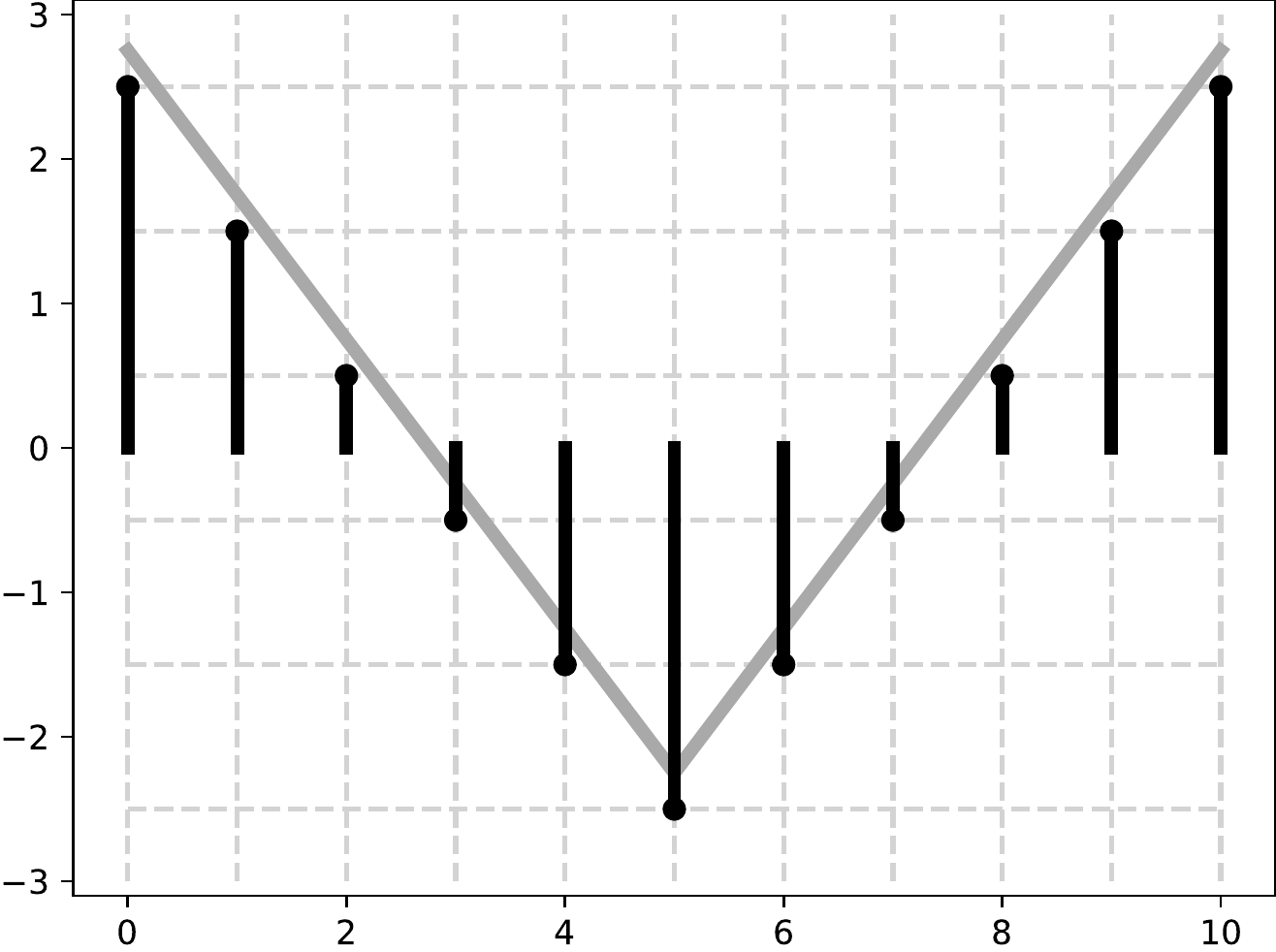}%
      \label{fig:sdt:corrected}
    } \\
  \end{tabular}
  \caption{Construction of the signed distance transform from two distance transforms. In~\ref{fig:sdt:standard}, the signal jumps from $+1$ to $-1$ at the zero crossing. This is resolved in~\ref{fig:sdt:corrected}.}
  \label{fig:sdt}
\end{figure}

Constructing a signed distance transform requires some details.
The authors prefer the following procedure:
\begin{equation}
  \label{eqn:sdt_corrected}
  \phi[nh] = \left\{
    \begin{matrix}
      -h/2 + d(nh, A) & \text{if } nh \in A^C \\
      +h/2 - d(nh, A^C) & \text{if } nh \in A
    \end{matrix}
    \right.
\end{equation}
Justification is given in Figure~\ref{fig:sdt}.
By adding half the sampling period to the embedding, the resulting signed transform crosses zero between the two edge samples.
Furthermore, it keeps the magnitude gradient equal to one at the edge.
This is why the quantization aligns on half-integers in Figure~\ref{fig:example1d:dt_Heaviside}.

\bibliographystyle{IEEEtran}
\bibliography{DT-Quantization.bib}

\end{document}